\documentclass[useAMS,usenatbib,referee]{biom}

\usepackage[figuresright]{rotating}
\usepackage{amsmath,subfigure,hyperref}
\usepackage{amssymb}
\usepackage{amsfonts}
\usepackage{multirow}
\usepackage{color} 
\usepackage{soul}
\usepackage{threeparttable}
\usepackage{xcolor}

\newtheorem{proposition}{Proposition}

\newcommand{\sumi}{\sum_{i=1}^n}

\newcommand{\mH}{\mathcal{H}}
\newcommand{\mF}{\mathcal{F}}

\newcommand{\mN}{\mathcal{N}}

\newcommand{\mK}{\mathcal{K}}
\newcommand{\NO}{\widetilde{N}}

\newcommand{\mA}{\mathcal{A}}
\newcommand{\mB}{\mathcal{B}}

\newcommand{\bZ}{\bm Z}
\newcommand{\bz}{\bm z}

\newcommand{\bbeta}{\bm \beta}
\newcommand{\tbeta}{\widetilde{\bbeta}}
\newcommand{\balpha}{\bm \alpha}
\newcommand{\bgamma}{\bm \gamma}
\newcommand{\btheta}{\bm \theta}

\newcommand{\hbeta}{\widehat{\bbeta}}
\newcommand{\halpha}{\widehat{\balpha}}

\newcommand{\SZ}{\mathcal{S}_Z}

\newcommand{\hshape}{h}

\newcommand{\gammaexp}{\widehat{\bgamma}_{\text{exp}}}
\newcommand{\gammamre}{\widehat{\bgamma}_{\text{mre}}}

\newcommand{\bxi}{\bm \xi}

\newcommand{\Ntr}{N_{\rm tr}}

\newcommand{\hc}{\widehat{c}}

\newcommand{\tr}{\widetilde{r}_h}

\newcommand{\tR}{\widetilde{R}_h}
\newcommand{\hF}{\widehat{F}}

\newcommand{\tu}{\widetilde{u}}
\newcommand{\hN}{\widehat{\mathcal{N}}}

\newcommand{\hf}{\widehat{f}_{1}}
\newcommand{\hff}{\widehat{f}_{2}}

\newcommand{\Rtr}{R^*}

\title[]{Statistical inference for counting processes under shape heterogeneity}

\author{
Yifei Sun$^{1,*}$ \email{ys3072@cumc.columbia.edu}
and 
Ying Sheng$^{2}$\\
$^{1}$ Department of Biostatistics, Mailman School of Public Health, Columbia University,\\ New York, NY 10032, U.S.A. \\
$^{2}$ Academy of Mathematics and Systems Sciences,  Chinese Academy of Sciences, \\ Beijing, Beijing 100190, China
}

\begin{document}

\begin{abstract} 
Proportional rate models are among the most popular methods for analyzing the rate function of counting processes. Although providing a straightforward rate-ratio interpretation of covariate effects, the proportional rate assumption implies that covariates do not modify the shape of the rate function. When such an assumption does not hold, we propose describing the relationship between the rate function and covariates through two indices: the shape index and the size index. The shape index allows the covariates to flexibly affect the shape of the rate function, and the size index retains the interpretability of covariate effects on the magnitude of the rate function. To overcome the challenges in simultaneously estimating the two sets of parameters, we propose a conditional pseudolikelihood approach to eliminate the size parameters in shape estimation and an event count projection approach for size estimation. The proposed estimators are asymptotically normal with a root-$n$ convergence rate. Simulation studies and an analysis of recurrent hospitalizations using SEER-Medicare data are conducted to illustrate the proposed methods.
\end{abstract}


\maketitle

\section{Introduction}

Statistical inferences of counting processes are often formulated based on their rate functions. The rate function is defined as the occurrence rate unconditional on the event history \citep{lawless1995some,lin2000semiparametric}, which is in contrast with the intensity function, defined as the occurrence rate conditional on history \citep{gail1980analysis,prentice1981regression,andersen1982cox}. The rate function has been commonly used for evaluating treatment effects or identifying risk factors. Denote by $\NO(t)$ the number of events occurring at or before time $t$ and by $\bZ$ a $p$-dimensional vector of covariates. Let $\mu(t\mid \bZ)$ be the conditional rate function of $\NO(\cdot)$ given $\bZ$, that is, $\mu(t\mid \bZ)dt = \operatorname{E}\{d\NO(t) \mid \bZ \}$. In the literature,  various models have been proposed to describe the effect of $\bZ$ on the rate function $\mu(t\mid \bZ)$; popular choices include multiplicative models \citep{pepe1993some,lawless1995some, lin2000semiparametric,wang2001analyzing}, additive models \citep{schaubel2006semiparametric}, scale-change models \citep{lin1998accelerated,ghosh2004accelerated,xu2017joint}, and transformation models \citep{lin2001semiparametric}. The availability of these models allows investigators to identify an appropriate tool to conduct inference for each specific application but requires the model to be specified a priori. Since the rate function is a curve that depends on time and is intrinsically infinite-dimensional, correctly specifying the form of $\mu(t\mid\bZ)$ can be challenging. Although graphical techniques and hypothesis testing have been commonly used for model checking, clear guidance on model selection and post-selection inference procedures are lacking. To avoid misleading conclusions drawn from misspecified models, one possible solution is to consider flexible models. On the other hand, models allowing full flexibility are often less favorable due to limited interpretability.

\begin{figure}
\includegraphics[height=20cm, width=17 cm]{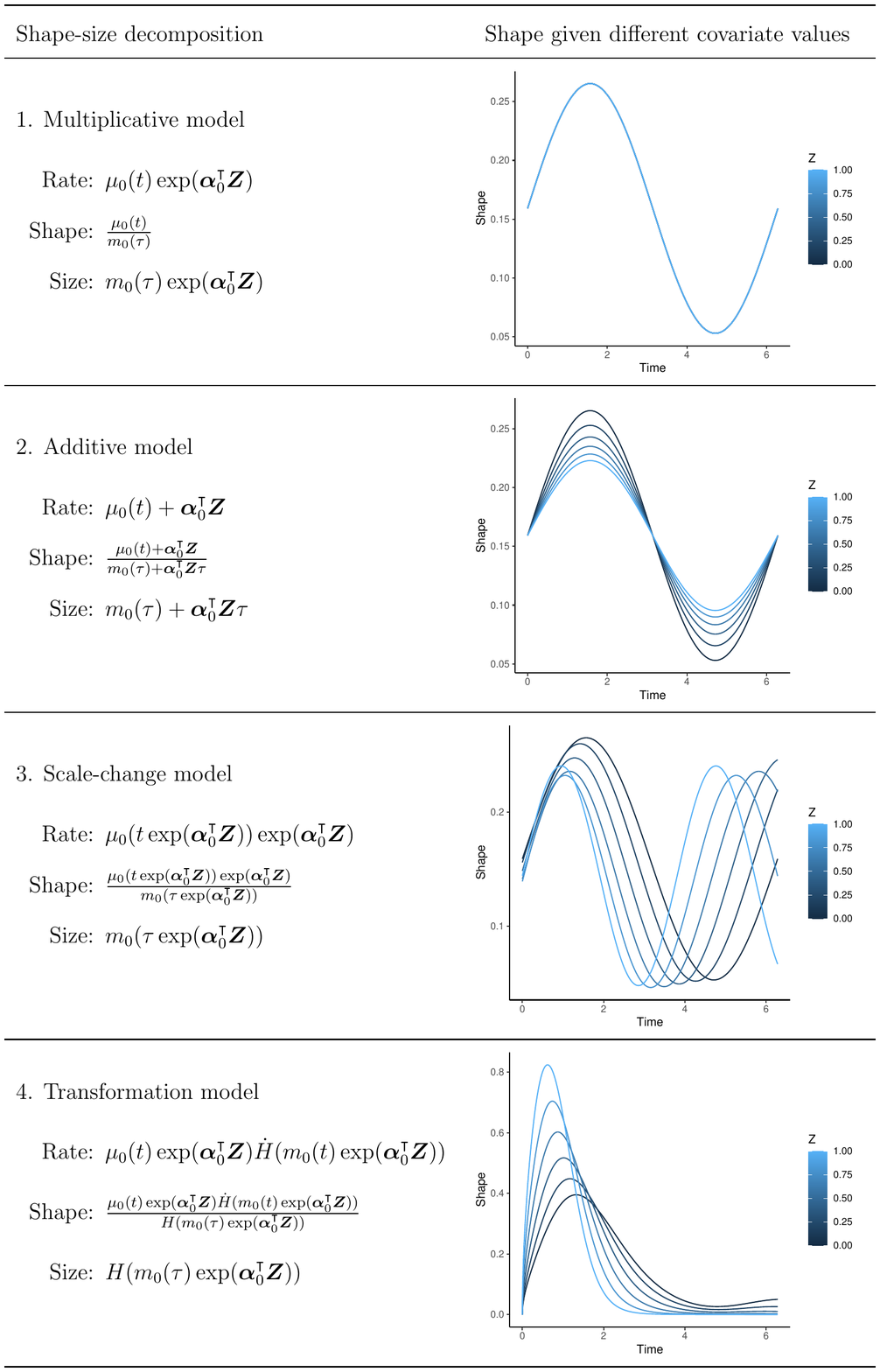}
\caption{\label{fig:shapesize} Shape and size decomposition of common recurrent event models. $\balpha_0$ is a vector of regression parameters; $\mu_0(t)$ and  $m_0(t) = \int_{0}^t \mu_0(u)du$ are the baseline rate and cumulative rate functions, respectively. For the transformation model, $H$ is an increasing function, and $\dot{H}$ is its first order derivative. The right column shows how covariates affect the shape function under a special case where $\bZ$ is one dimensional, $\mu_0(t) = \sin(t) + 1.5$, $\alpha_0 = 1$, and $H$ is the cumulative distribution function of the Weibull distribution with shape parameter 1.5 and scale parameter 5.}
\end{figure}

The goal of this paper is to develop a unified framework for modeling the rate function, which strikes a balance between flexibility and interpretability. Our framework is based on a decomposition of the rate function \citep{wang2014statistical}. Let $[0,\tau]$ be the time period of interest. Following \cite{wang2014statistical}, one can express the  rate function as the product of a \emph{size} component and a \emph{shape} component:  the size, defined as $\operatorname{E}\{\NO(\tau)\mid \bZ\}=\int_0^\tau \mu(t\mid\bZ)dt$, is the expected number of events over the entire study period; the shape, defined as $\mu(t\mid \bZ) / \operatorname{E}\{\NO(\tau)\mid \bZ\}$, characterizes the time-varying profile of the event rate standardized by its magnitude. The shape-size decomposition of the aforementioned rate-based models is presented in Figure \ref{fig:shapesize}. The figure demonstrates that shape is covariate-independent under the multiplicative model, while it depends on covariates in different ways under other models. Additionally, it is noted that existing approaches generally assume that size is monotonically increasing in a linear combination of covariates, allowing us to interpret the association between covariates and the total event count.

In this paper, we focus on the case of shape heterogeneity, where the shape of the rate function varies based on covariates. In our motivating SEER-Medicare data example \citep{warren2002overview,enewold2020updated}, the events of interest are repeated inpatient hospitalizations after the diagnosis of breast cancer. Younger patients tend to recover more quickly, and their hospitalizations are more likely to occur in the early phase after diagnosis. Older patients, on the other hand, have a more consistent rate of hospitalization over time. 
This implies that the shape of the hospitalization rate may depend on age at diagnosis, and the heterogeneity in shape should be taken into account when modeling the rate function. In this case, one may wonder how to draw definitive conclusions, such as whether a larger covariate is associated with an increased risk of events.
To solve this problem, we propose a {shape-size} model.
The model indexes the shape of the rate function using a linear predictor, which captures the relationship between the shape and multiple covariates; the size of the rate function is modeled through a separate index.
Our model extends the original single index model \citep[see, for example,][]{powell1989semiparametric,hardle1989investigating,hardle1993optimal,ichimura1993semiparametric} for a scalar outcome to a counting process outcome and offers a reasonable compromise between parametric and fully nonparametric modeling.

Single index models for the time to a single event under censoring have been widely studied in the literature. Researchers have considered proportional hazards regression where the hazard ratio depends on a single index \citep{huang2006polynomial,wang2009proportional}. 
Without the proportional hazards assumption, single index models for the density function \citep{bouaziz2010conditional,strzalkowska2013maximum,strzalkowska2014beran}, the mean \citep{lopez2009single,lopez2013single,huang2020estimation}, the hazard function \citep{ding2013local,chiang2018versatile}, and the quantile function \citep{kong2017uniform,christou2019single,bucher2021single} of the event time have also been studied. However, single index models for the counting processes of recurrent events have been less studied.   \cite{zhao2014sufficient} considered sufficient dimension reduction for the rate of nonstationary Poisson processes and assumed that covariates do not modify the shape of the rate function. \cite{bouaziz2015semiparametric} considered a single index model where the rate function is indexed by a linear combination of covariates but did not distinguish the effects on shape and size. 

The contribution of our work is twofold. First, we focus on the case where shape is covariate-dependent, and thus our method serves as a natural alternative when the commonly used proportional rate model (i.e., multiplicative model) does not hold. When assuming that size depends on covariates through an exponential link, we obtain comparable interpretations on the size component. Compared to other rate models where shape depends on covariates, our model is more flexible and allows covariates to have different effects on shape and size. Second, our model can involve up to two unspecified link functions, while existing single index models typically have one link function. To overcome the methodological challenges, we propose a two-step estimation procedure. In the first step, we estimate shape parameters through a conditional pseudolikelihood approach that eliminates size parameters. In the second step, we use the estimated shape function to project the event count from the follow-up period to the time interval of interest; regression analysis is then performed on the projected event count to estimate the size parameters.

This paper is organized as follows. In Section \ref{sec:model}, we introduce the model setup. In Section \ref{sec:shape}, we propose a conditional pseudolikelihood approach to estimate the shape parameters; in addition, we propose a simplified estimator with a lower computational cost and the same asymptotic variance. In Section \ref{sec:size}, we propose a unified approach for size estimation under different assumptions of link functions. In Section \ref{sect:simu}, we conduct simulation studies to evaluate the finite sample performance of the estimators. In Section \ref{sec:data}, the proposed methods are  applied to model the recurrent hospitalizations in SEER-Medicare breast cancer data. A discussion in Section \ref{sec:disc} concludes the paper. 

\section{Model setup}
\label{sec:model}

We propose the following model for the conditional rate function,
\begin{align}
\label{model}
  \mu(t \mid \bZ) = f(t, \bbeta_0^\intercal \bZ)  g(\bgamma_0^\intercal \bZ),~~ 0 \le t \le \tau,
\end{align}
where $\bbeta_0$ and $\bgamma_0$ are $p$-dimensional vectors of regression parameters, $f(t,\bbeta_0^\intercal \bZ)$ is the shape function indexed by $\bbeta_0^\intercal \bZ$, and $g(\cdot)$ is the size function indexed by $\bgamma_0^\intercal \bZ$. We assume that the shape function $f(\cdot,x)$ is an unknown function satisfying $\int_0^\tau f(t,x)dt = 1$ for any $x\in\mathbb{R}$. For identifiability, we assume that $\|\bbeta_0\|=1$ and the $p$th element of $\bbeta_0$ is positive, where $\|\cdot\|$ denotes the Euclidean norm. 
The size function $g(\cdot)$ can be either prespecified or unspecified. 

In practice, the size parameter $\bgamma_0$ is the focus of interest and assumptions on $g(\cdot)$ can facilitate the interpretation of $\bgamma_0$. For example, when evaluating the effect of a risk factor, imposing monotonic assumptions on $g(\cdot)$ allows a better understanding of whether the factor is associated with a higher event risk. As another example, since we have $\operatorname{E} \{\widetilde{N}(\tau)\mid \bZ\} = g(\bgamma_0^\intercal \bZ)$,  assuming $g(x) \propto \exp(x)$ yields a size ratio interpretation similar to a Poisson regression model for the total event count $\widetilde{N}(\tau)$. The interpretation of $\bbeta_0$ is of less interest, and shape is estimated to ensure accurate estimation of the covariate effects on size. When assuming that $g(\cdot)$ is monotonically increasing, Models 2-4 in Figure \ref{fig:shapesize} are special cases of Model \eqref{model} and force $\bbeta_0/\bgamma_0$ to be a constant. In other words, Models 2-4 assume all the covariates have consistent effects on the shape and size. However, in real applications, such an assumption is often violated. We note that \cite{sun2022statistical} considered a more restricted shape-size model, where monotone constraints are imposed on the shape and size functions. However, monotone constraints on shape generally do not hold under the models in Figure \ref{fig:shapesize} and are thus not imposed in our model. 
Additionally, to make interpretations more accessible to a broader audience, we incorporate a multiplicative size model as an extension of the proportional rate model; we also explored special cases where misspecified size models still yield consistent estimations of normalized covariate effects.

In practice, the process $\widetilde{N}(\cdot)$ may not be completely observed due to a limited follow-up period. Denote by $C = \min(C^*,\tau)$ the censoring time, where $C^*$ is the time to the end of follow-up. We assume that $C$ is independent of $\widetilde{N}(\cdot)$ conditioning on $\bZ$. The observed counting process is $N(t) = \widetilde{N}(\min(t,C))$, and we have $dN(t) = I(C\ge t)d\NO(t)$. The observed data, denoted by $\{(N_i(t), \bZ_i, C_i) , {t\in[0,\tau]},  i=1,\ldots,n \}$, are assumed to be independent and identically distributed (i.i.d.) replicates of $\{(N(t), \bZ, C), {t\in[0,\tau]} \}$.

\section{Estimation of the shape parameters}
\label{sec:shape}

\subsection{Conditional likelihood under a working model}
\label{sec:shape1}
The estimation of shape parameters can be derived from a conditional pseudolikelihood approach that eliminates the size component $g(\bgamma_0^\intercal \bZ)$. To motivate the estimation, we first consider a working assumption that $\widetilde{N}(\cdot)$ is a nonhomogeneous Poisson process. Later we will show that the estimator remains valid without the Poisson process assumption.  Let $m_i = \widetilde{N}_i(C_i)$ be the number of events observed on $[0,C_i]$. Conditional on $(C_i,\bZ_i, m_i)$, the event times, $T_{i1},\ldots, T_{im_i}$, are the order statistics of a set of i.i.d. random variables with the density function $f(t,\bbeta_0^\intercal \bZ_i)/F(C_i,\bbeta_0^\intercal \bZ_i )$ \citep{cox1966statistical, wang2001analyzing}, where $F(t,x) = \int_0^t f(u,x)du$ is the cumulative shape function. The conditional likelihood of event times $(T_{i1},\ldots, T_{im_i})$ given $(C_i,\bZ_i, m_i)$, $i=1,\ldots,n$, is proportional to
\begin{align}
\label{lik}
\prod_{i=1}^{n} \prod_{k=1}^{m_i} \frac{f(T_{ik},\bbeta^\intercal \bZ_i)}{F(C_i,\bbeta^\intercal \bZ_i )}  .
\end{align}
Notably, the expression in \eqref{lik} takes the form of likelihood for right-truncated data, where the truncation time for $T_{ik}$ is $C_i$. Define $r(t,x) = f(t,x)/F(t,x)$; then, we have $F(t,x) = \exp\{-\int_{t}^{\tau} r(u,x)du \}$. The log conditional likelihood, up to a constant, can be expressed as
\begin{align}
\label{lik_cont}
\frac{1}{n}\sum_{i=1}^{n}\sum_{k=1}^{m_i} \left\{\log r(T_{ik},\bbeta^\intercal \bZ_i)- \int_{T_{ik}}^{C_i}r(u,\bbeta^\intercal \bZ_i)du  
\right\} .
\end{align}  
To estimate $\bbeta_0$, a natural idea is to ``profile out'' the nuisance function $r(t,x)$ in \eqref{lik_cont}.

If $\bbeta$ is known, we can construct a local conditional likelihood to estimate $r(t,x)$ and its integral. For each $x$, define $R(t,x) = \int_t^\tau r(u,x)du$, which is a decreasing function in $t$. 
We consider a discrete version of $R(\cdot,x)$, which is a left-continuous step function and jumps down at the observed event times. Specifically, let $R\{ t, x\}$ be the decrease of $R(t, x)$ at $t$; then, the local log conditional likelihood can be written as
\begin{align}
\label{lik_disc}
\sum_{i=1}^{n}\sum_{k=1}^{m_i}K_h(\bbeta^\intercal \bZ_i - x) \left(\log {R\{T_{ik},x\}} - \sum_{(j,l):T_{ik}\le T_{jl}\le C_i} R\{T_{jl},x\}  \right),
\end{align}
{where $K_h(\cdot) = K(\cdot/h)/h$ is a scaled kernel function, with $K(\cdot)$ being a kernel function with a bounded support on $[-1,1]$ and $h$ being a bandwidth parameter.} In \eqref{lik_disc}, subjects whose shape indices $\bbeta^\intercal\bZ$ are in a small neighborhood of $x$ are used to estimate $R(t,x)$. 
Elementary calculus shows that \eqref{lik_disc} is maximized by
\begin{align*}
\widehat R\{t,x\} = \frac{\sum_{i=1}^n \sum_{k=1}^{m_i} K_h(\bbeta^\intercal \bZ_i - x)I(T_{ik}=t)}{\sum_{i=1}^n \sum_{k=1}^{m_i}K_h(\bbeta^\intercal \bZ_i - x)I(T_{ik}\le t \le C_i)}.
\end{align*}
Given $N_i(t) = \sum_{k=1}^{m_i}I(T_{ik}\le t)$, we construct the following estimator for $R(t,x)$,
\begin{align}
\label{Rhat}
 \widehat{R}_h(t,x, {\bbeta}) =\int_t^\tau \frac{\sum_{i=1}^n K_h(x-{\bbeta}^\intercal \bZ_i)dN_i(u)}{\sum_{i=1}^n K_h(x-{\bbeta}^\intercal \bZ_i)N_i(u)I(C_i\ge u)}.
\end{align}
The log conditional likelihood \eqref{lik_cont} also involves $r(t,x)$, which can be estimated by the following bivariate kernel estimator,
\begin{align*}
\widehat{r}_h(t,x, {\bbeta}) =  \frac{ \sum_{i=1}^n\sum_{k=1}^{m_i} K_h(x-{\bbeta}^\intercal \bZ_i)  K_{h}(t-T_{ik})}{\sum_{i=1}^n K_h(x-{\bbeta}^\intercal \bZ_i)N_i(t)I(C_i\ge t) }.
\end{align*}
We plug $\widehat r_h$ and $\widehat R_h$ into \eqref{lik_cont} and obtain the following objective function:
\begin{align}
\label{l1}
\frac{1}{n}\sum_{i=1}^{n}\int_0^\tau \left\{ \log \widehat{r}_h(t,\bbeta^\intercal \bZ_i,\bbeta) -  \widehat{R}_h(t,\bbeta^\intercal \bZ_i,\bbeta) +  \widehat{R}_h(C_i,\bbeta^\intercal \bZ_i,\bbeta)  \right\}dN_i(t).
\end{align}
\begin{remark}
Note that we replace $r(t,x)$ in \eqref{lik_cont} with $\widehat r_h(t,x)$ rather than the point mass $\widehat R\{t,x\}$. This is different from existing arguments of the profile likelihood for {the Cox model}, where the likelihood has a similar structure and the nuisance baseline hazard at event times is replaced by point masses \citep[see, for example,][]{murphy2000profile}. In our setting, replacing $r(t,x)$ in \eqref{lik_cont} with $\widehat R\{t,x\}$ does not yield a consistent estimator for $\bbeta_0$. 
Moreover, for each $x$, $r(t,x)$ is mathematically equivalent to a reverse time hazard function, and our method can be easily adapted to estimate a single index hazard model for time to a single event. 
This is noteworthy because the existing local profile likelihood method has been shown to fail when applied to a single index hazard model \citep{ding2013local}.
\end{remark}
\begin{remark}
The pseudo conditional likelihood approach can be applied when the form of the size function is misspecified, as the size function is eliminated in  \eqref{lik}. 
\end{remark}

\subsection{A simplified objective function}
\label{sec:simp}
The objective function \eqref{l1} can be further simplified to reduce the computational cost. Define $R(t,x,\bbeta) = \int_t^\tau  \operatorname{E}\{N(u)I(C\ge u)\mid\bbeta^\intercal\bZ = x\}^{-1}\operatorname{E}\{dN(u)\mid\bbeta^\intercal\bZ = x\}$ and we have $R(t,x,\bbeta_0) = R(t,x)$. The quantity $n^{-1}\sum_{i=1}^{n}\int_0^\tau \{   \widehat{R}_h(t,\bbeta^\intercal \bZ_i,\bbeta) -  \widehat{R}_h(C_i,\bbeta^\intercal \bZ_i,\bbeta)  \}dN_i(t)$ converges in probability to $\operatorname{E} \left[ \int_0^\tau  \{{R}(t,\bbeta^\intercal \bZ,\bbeta) -  {R}(C,\bbeta^\intercal \bZ,\bbeta)\} dN(t) \right]$ as $n\to \infty$. 
In Proposition \ref{p1}, we show that the limiting value does not depend on $\bbeta$. Moreover, the result holds for general counting processes under Model \eqref{model} and does not require the Poisson process assumption. The proof of Proposition \ref{p1} is given in the Supporting Information. 

\begin{proposition}
\label{p1}
For any $\bbeta\in\mathbb{R}^p$, we have
\begin{align*}
 \operatorname{E} \left[ \int_0^\tau  \{{R}(t,\bbeta^\intercal \bZ,\bbeta) -  {R}(C,\bbeta^\intercal \bZ,\bbeta)\} dN(t) \right] = \operatorname{E} \{N(\tau)\}.
\end{align*}
\end{proposition}

\begin{remark}
Assuming the left hand side of the above equation does not depend on $\bbeta$, we can obtain the expression $\operatorname{E} \{N(\tau)\}$ by considering a simple special case.
Under the working Poisson process assumption and conditioning on $(C_i,\bZ_i,m_i)$, the variables $\{R(T_{ik},\bbeta_0^\intercal \bZ_i, \bbeta_0)-R(C_i,\bbeta_0^\intercal \bZ_i, \bbeta_0 ), k = 1,\ldots, m_i\}$  are the order statistics of a set of i.i.d. random variables that follow the exponential distribution with mean 1. As a result, $\int_0^\tau\{R(t,\bbeta_0^\intercal \bZ_i, \bbeta_0)-R(C_i,\bbeta_0^\intercal \bZ_i, \bbeta_0 )\}dN_i(t)$ follows the Gamma distribution with shape parameter $m_i = N_i(\tau)$ and rate parameter $1$.
\end{remark}

Based on Proposition \ref{p1}, an alternative way is to maximize the following objective function, 
\begin{align}
\label{obj0}
\frac{1}{n}\sum_{i=1}^{n}\int_0^\tau \left\{ \log \widehat{r}_h(t,\bbeta^\intercal \bZ_i,\bbeta)   \right\} dN_i(t) .
\end{align}
Since the quantity $n^{-1}\sum_{i=1}^{n}\int_0^\tau \{   \widehat{R}_h(t,\bbeta^\intercal \bZ_i,\bbeta) -  \widehat{R}_h(C_i,\bbeta^\intercal \bZ_i,\bbeta)  \}dN_i(t)$ in \eqref{l1} involves double integrals, maximizing \eqref{l1} can be computationally expensive. When bootstrap is applied for variance estimation, the simplified objective function \eqref{obj0} significantly improves the computation speed.

\subsection{The proposed estimators}
\label{sec:est}

Based on the derivation presented above, the objective functions \eqref{l1} or \eqref{obj0} can be maximized to estimate the parameter $\bbeta_0$. To establish the large-sample properties of the proposed estimators, the objective functions are slightly modified by trimming regions of sparse data.
To make this modification, we define $\SZ$ to be a fixed subset of the support of $\bZ$ and focus on observations with $\bZ\in\SZ$; details of $\SZ$ are given in the regularity conditions.  Additionally, we confine the event times to a predetermined interval $[\tau_0,\tau_1]$ ($0 < \tau_0 < \tau_1 < \tau$) to circumvent the boundary problem encountered by the kernel estimator $\widehat r_h(t,x,\bbeta)$ near $t=0$ and $t=\tau$. 
We then adjust the objective function \eqref{l1} as follows:
$$
\ell(\bbeta) \!=\! \frac{1}{n} \sum_{i=1}^{n} \! \int_{\tau_0}^{\tau_1} \! \left\{ \log \widetilde{r}_h( t,\bbeta^\intercal \bZ_i,\bbeta) \!-\!  \widetilde{R}_h(t,\bbeta^\intercal \bZ_i,\bbeta) \!+ \! \widetilde{R}_h(C_i,\bbeta^\intercal \bZ_i,\bbeta)  \right\} \! I(\bZ_i \in \SZ)dN_i(t),$$ 
where for $t\in(\tau_0,\tau_1)$, we define the trimmed kernel estimators, 
$$ \widetilde{r}_h( t,x,\bbeta) = \frac{\sum_{i=1}^n\sum_{k=1}^{m_i} K_h(x-{\bbeta}^\intercal \bZ_i)  K_{h}(t-T_{ik}) I(\bZ_i \in \SZ)}{\sumi K_h(\bbeta^\intercal \bZ_i - x)I(\bZ_i \in \SZ,C_i\ge t)\{N_i(t)-N_i(\tau_0)\}}, $$
\begin{align*}
 \widetilde{R}_h(t,x, {\bbeta}) =\int_t^{\tau_1} \frac{\sum_{i=1}^n K_h(x-{\bbeta}^\intercal \bZ_i)I(\bZ_i \in \SZ)dN_i(u)}{\sum_{i=1}^n K_h(x-{\bbeta}^\intercal \bZ_i)I(\bZ_i \in \SZ,C_i\ge u)\{N_i(u)-N_i(\tau_0)\}}.
\end{align*}
Note that when $\bbeta_0$ is known, $\widetilde{r}_h( t,x,\bbeta_0)$ and $\widetilde{R}_h( t,x,\bbeta_0)$ estimate  $r^*(t,x) = f(t,x)/\int_{\tau_0}^{t} f(u,x)du$ and $R^*(t,x) {=} \int_t^{\tau_1} r^*(u,x)du$, respectively. The use of $\widetilde r_h$ and $\widetilde R_h$ in place of $\widehat r_h$ and $\widehat R_h$ ensures that the limiting objective function is maximized at $\bbeta_0$. 
{Define $\mathcal{B} =\{\bbeta = (\beta_1, \beta_2, \ldots, \beta_p)^\intercal: \|\bbeta\| = 1, \beta_p > 0\}$. We estimate $\bbeta_0$ by 
\begin{align}
\label{betahat}
\widehat{\bbeta} = {\arg\max}_{\bbeta\in\mathcal{B}} \ell(\bbeta).
\end{align}
{The constraint $\bbeta\in\mathcal{B}$ can be achieved by reparameterizing $\bbeta$ in the polyspherical coordinate system as a $(p-1)$-dimensional vector $\balpha = (\alpha_1,\alpha_2,\ldots,\alpha_{p-1})^\intercal\in \mathcal{A}$, where $\mathcal{A} = [0,\pi]^{(p-1)}$ is the $(p-1)$-ary Cartesian power of $[0,\pi]$. Specifically, set $\bbeta = \mathbb{S}(\balpha)$, where  $\mathbb{S}(\balpha) = (\cos(\alpha_1), \sin(\alpha_1)\cos(\alpha_2),\ldots,$ $ \prod_{k=1}^{p-2}\sin(\alpha_k)\cos(\alpha_{p-1}),  \prod_{k=1}^{p-1}\sin(\alpha_k))^\intercal$ maps $\mathcal{A}$ to $\mB$ \citep[see, for example,][]{balabdaoui2019score}. } 

Similar to Section \ref{sec:simp}, we can further simplify the trimmed objective function by removing the term that converges to a constant. 
By substituting $\int_0^t I(\bZ \in \SZ, \tau_0\le u\le \tau_1 ) d\widetilde{N}(u)$ for $\NO(t)$ and employing the arguments in Proposition \ref{p1}, it can be shown that  $n^{-1}\sum_{i=1}^{n}\int_{\tau_0}^{\tau_1} \left\{ -  \widetilde{R}_h(t,\bbeta^\intercal \bZ_i,\bbeta) +  \widetilde{R}_h(C_i,\bbeta^\intercal \bZ_i,\bbeta)  \right\} I(\bZ_i \in \SZ)dN_i(t)$ converges to a constant that does not depend on $\bbeta$. Further details are given in the Supporting Information. This leads to the objective function
$\ell'(\bbeta) 
=  n^{-1}\sum_{i=1}^{n}\int_{\tau_0}^{\tau_1} \left\{ \log \widetilde{r}_h( t,\bbeta^\intercal \bZ_i,\bbeta)  \right\}I(\bZ_i \in \SZ)dN_i(t). $
Then, an alternative estimator for $\bbeta_0$ is 
\begin{align}
\label{betatilde}
\widetilde{\bbeta} = {\arg\max}_{\bbeta\in\mathcal{B}} \ell'(\bbeta).
\end{align}

Although the objective functions are derived under the assumption of a Poisson process, we derive the large sample properties for $\widehat\bbeta$ and $\widetilde\bbeta$ without this assumption. 
Define $dM^*(t) = I(\bZ\in\SZ)dN(t) - I(C\ge t, \bZ\in\SZ) N(t)r^*(t,\bbeta_0^\intercal \bZ)dt$. Then $dM^*(t)$ has a similar form compared to the martingale residual in survival analysis, and we have $\operatorname{E}\{dM^*(t)\mid\bZ\} = 0$. For a vector $\bm v$, define $\bm v^{\otimes 0} = 1$, $\bm v^{\otimes 1} = \bm v$, and $\bm v^{\otimes 2} = \bm v \bm v^\intercal$. For $k = 0,1$, define $q^{(k)*}(t,x) = \operatorname{E} \{\bZ^{\otimes k} I(C\ge t, \bZ\in\SZ) g(\bgamma_0^\intercal \bZ)\mid\bbeta_0^\intercal\bZ = x\}$.
We use $\dot r^*(t,x)$ to denote the partial derivative of $r^*(t,x)$ with respect to $x$ and use $^-$ to denote the Penrose-Moore inverse of a matrix. Theorems \ref{thm:hatbeta} and \ref{thm:tildebeta} state the asymptotic normality of $\widehat\bbeta$ and $\widetilde\bbeta$, with proofs given in the Supporting Information. 

\vspace{-0cm}
\begin{theorem}
\label{thm:hatbeta}
Under conditions (C1)-(C5) in the Supporting Information, $\sqrt{n}(\widehat{\bbeta}-\bbeta_0)$ converges in distribution to a zero-mean normal random vector with the variance-covariance matrix $V^{-}\Sigma V^{-}$ as $n\to \infty$, where we define $\Sigma = \operatorname{E} (\psi \psi^{\intercal})$,  $\psi = \int_{\tau_0}^{\tau_1}  \{r^*(t,\bbeta_0^\intercal\bZ)\}^{-1} \dot r^*(t,\bbeta_0^\intercal \bZ)$ $ \left\{ \bZ - q^{(1)*}(t,\bbeta_0^\intercal\bZ)/q^{(0)*}(t,\bbeta_0^\intercal\bZ)\right\} d {M^*}(t)$, and  $$V = - \operatorname{E} \left[\int_{\tau_0}^{\tau_1} \left\{\frac{\dot r^*(t,\bbeta_0^\intercal \bZ)} {r^*(t,\bbeta_0^\intercal\bZ)}\right\}^2 \left\{ \bZ - \frac{q^{(1)*}(t,\bbeta_0^\intercal\bZ)}{q^{(0)*}(t,\bbeta_0^\intercal\bZ)} \right\}^{\otimes 2}  I(\bZ \in\SZ) dN(t) \right].$$
\end{theorem}

\begin{remark}
Note that the asymptotic variance of $\widehat\bbeta$ is the same as if we were to maximize
$$
\frac{1}{n}\sum_{i=1}^{n}\int_{\tau_0}^{\tau_1} \left\{ \log {r^*}( t,\bbeta^\intercal \bZ_i,\bbeta) -  {R^*}(t,\bbeta^\intercal \bZ_i,\bbeta) +  {R^*}(C_i,\bbeta^\intercal \bZ_i,\bbeta)  \right\}I(\bZ_i \in \SZ)dN_i(t),$$ 
in the case where $R^*(t,x,\bbeta) = \int_t^{\tau_1}  \operatorname{E}[\{N(u)-N(\tau_0)\}I(C\ge u,\bZ \in\SZ)\mid\bbeta^\intercal\bZ = x]^{-1}\operatorname{E}\{I(\bZ \in\SZ) dN(u)\mid\bbeta^\intercal\bZ = x\}$ 
and $r^*(t,x,\bbeta) = - \partial R^*(t,x,\bbeta)/\partial t$ are known.
\end{remark}

\begin{theorem}
\label{thm:tildebeta}
Under conditions (C1)-(C5) in the Supporting Information, 
$\sqrt{n}(\widetilde{\bbeta}-\bbeta_0)$ converges in distribution to a zero-mean normal random vector with the variance-covariance matrix $V^{-}\Sigma V^{-}$ as $n\to \infty$.
\end{theorem}

According to Theorems \ref{thm:hatbeta} and \ref{thm:tildebeta},  $\widehat\bbeta$ and $\widetilde\bbeta$ have the same asymptotic variance. In practice, both $\widehat\bbeta$ and $\widetilde\bbeta$ can be used to estimate $\bbeta_0$, with the computation of $\tbeta$ being faster. When bootstrap is applied for variance estimation and computational resource is limited, the bootstrap standard error of $\widetilde\bbeta$ can serve as an approximation for the bootstrap standard error of $\widehat\bbeta$, leading to a significant improvement in computation speed.

\section{Estimation of size parameters}
\label{sec:size}

Since we have $\operatorname{E}\{\NO(\tau)\mid\bZ\} = g(\bgamma_0^\intercal \bZ)$, estimating $\bgamma_0$ in the absence of censoring is straightforward. 
In the presence of censoring, the process $\NO(\cdot)$ is only observed up to $C$. To deal with censoring, we ``project'' the observed event count in $[0, C]$ onto the interval $[0,\tau]$ and estimate $\bgamma_0$ by using the projected event count as the outcome. 
Under the assumption of conditional independent censoring, we derive the following equation, with proof provided in the Supporting Information:
\begin{align}
\label{P2}
\operatorname{E} \left\{ \frac{N(C)}{F(C,\bbeta_0^\intercal \bZ)}\;\Big|\;\bZ  \right\}  =  g(\bgamma_0^\intercal \bZ) = \operatorname{E} \{\NO(\tau)\mid \bZ\}.
\end{align}
According to Equation \eqref{P2},  $N(C)/F(C,\bbeta_0^\intercal \bZ)$ shares the same conditional expectation as $\NO(\tau)$. Equation \eqref{P2} extends the results in \cite{wang2001analyzing}, where the shape does not depend on covariates.
In the ideal case where the shape is known, $N(C)/F(C,\bbeta_0^\intercal \bZ)$ could be used as the outcome for regression modeling. Since $F(\cdot,\cdot)$ and $\bbeta_0$ are usually unknown, we replace them with the corresponding estimates.
The function $F(t, x)$ can be estimated by either $\widehat{F}_h(t,x,\hbeta)$ or $\widehat{F}_h(t,x,\tbeta)$, where $\widehat{F}_h(t,x,\bbeta) = \exp\{ - \widehat{R}_h(t,x,\bbeta)\}$. The formulas for $\hbeta$, $\tbeta$, and $\widehat{R}_h$ are in Equations \eqref{betahat}, \eqref{betatilde}, and \eqref{Rhat} in Section \ref{sec:shape}.  Then the total event count, $\NO_i(\tau)$, can be replaced with the projected event count, ${N_i(C_{i})}{\widehat F_h(C_{i},\widehat\bbeta^\intercal \bZ_i,\widehat\bbeta)^{-1}}$. In what follows, we consider scenarios with different assumptions imposed on the size function $g(\cdot)$.

\subsection{A multiplicative model for size}
\label{sec:exp}

A simple way of analyzing recurrent events is to count the number of events observed within the interval $[0,\tau]$ and apply a Poisson regression model with the event count $\NO(\tau)$ as the outcome. Following the canonical Poisson regression specification, we assume that $g(a) \propto \exp(a)$. In Section \ref{sec:exp}, no constraint is imposed on the norm of $\bgamma_0$. Then, model \eqref{model} can be viewed as an extension to the proportional rate model in the sense that covariates not only have multiplicative effects on the size of the rate function but also modify the shape. In the absence of censoring, one can solve the estimating equation, 
\begin{align}
\label{ee_exp_uc}
\sumi \widetilde \bZ_i \{ \NO_i(\tau) - \exp(c+ \bgamma^\intercal \bZ_i) \} = \bm 0,
\end{align}
to estimate the size parameters, where $\widetilde \bZ_i = (1,\bZ_i^\intercal)^\intercal$, and $c$ corresponds to an intercept in the regression model. 
Denote by $\gammaexp^{0}$ the solution of Equation \eqref{ee_exp_uc} with respect to $\bgamma$. The validity of the estimator generally relies on the correct specification of the link function $g$, and misspecification of the link function generally results in biased estimation. However, if $\bZ$ follows an elliptically symmetric distribution, such as the multivariate normal distribution, the relative covariate effects can be consistently estimated even under a misspecified link. The result is summarized in Proposition \ref{thm3}.

\begin{proposition}
\label{thm3}
If $\bZ$ has an elliptically symmetric distribution and the size function $g$ is monotonically increasing,  $\gammaexp^{0}/\|\gammaexp^{0}\|$ converges in probability to $\bgamma_0/\|\bgamma_0\|$ as $n\to \infty$.
\end{proposition}

\begin{remark}
Note that $\gammaexp^{0}$ can also be obtained by fitting the proportional rate model with complete data $\{\bZ_i, \NO_i(t), 0\le t\le\tau,  i = 1,\ldots, n\}$. Therefore, under the assumptions in Proposition \ref{thm3}, the directions of covariate effects can be correctly identified by the proportional rate model, even if the shape of the rate function depends on covariates and/or the size function is misspecified. However, for censored data, the normalized coefficients from the proportional rate model are generally biased under shape heterogeneity.
\end{remark}

In the presence of censoring, we propose to replace $\NO_i(\tau)$ in \eqref{ee_exp_uc} with the projected event count, ${N_i(C_{i})}{\widehat F_h(C_{i},\widehat\bbeta^\intercal \bZ_i,\widehat\bbeta)^{-1}}$, and solve the following estimating equation:
\begin{align}
\label{ee:size1}
    \sumi \widetilde \bZ_i  I(\bZ_i \in\SZ) \left\{ \frac{N_i(C_{i})}{\widehat F_h(C_{i},\widehat\bbeta^\intercal \bZ_i, \widehat\bbeta)} - \exp(c+ \bgamma^\intercal \bZ_i) \right\} = \bm 0.
\end{align}
The solution of the above estimating equation is denoted by $\gammaexp$. In Equation \eqref{ee:size1}, the indicator $I(\bZ_i \in\SZ)$ is added to ensure that the density of $\bbeta_0^\intercal \bZ$ is bounded away from zero, while trimming is not required for the kernel estimator $\widehat F_h$.

\begin{remark}
\label{rmk6}
When $\bZ$ follows a zero-mean normal distribution and $g$ is an increasing function (not necessarily exponential), we consider a trimming indicator $I(| \widehat\bbeta_{\parallel\widetilde\bgamma}^\intercal\bZ_i|\le a, |\widehat\bbeta_{\perp\widetilde\bgamma}^\intercal\bZ_i|\le a)$, where $a$ is a pre-specified positive constant, $\widetilde\bgamma$ is a consistent initial estimator of $\bgamma_0/\|\bgamma_0\|$ (e.g., the estimator presented in Section \ref{sec:unknown}), and $\widehat\bbeta_{\parallel\widetilde\bgamma}$ and $\widehat\bbeta_{\perp\widetilde\bgamma}$ denote the projection of $\hbeta$ on $\widetilde\bgamma$ and the corresponding rejection, respectively. Following the arguments in Proposition \ref{thm3}, it can be shown that $\gammaexp /\|\gammaexp \|$ consistently estimates $\bgamma_0/\|\bgamma_0\|$ even when the size function $g$ is misspecified. In simulation studies, we observed that the normalized estimator performed reasonably well in estimating $\bgamma_0/\|\bgamma_0\|$ with misspecified size functions. 
\end{remark}

We then establish the large-sample property of $\gammaexp$, assuming that the exponential link is correctly specified. 
Theorem \ref{thm:exp} summarizes the asymptotic normality of $\gammaexp$. 

\begin{theorem}
\label{thm:exp}
Assume $(c_0,\bgamma_0^\intercal)^\intercal \in \Xi$, where $\Xi$ is a compact subset of $\mathbb{R}^{p+1}$.
Under conditions {(C1)-(C8)} in the Supporting Information, $\sqrt{n}((\widehat c,\gammaexp^\intercal)^\intercal-(c_0,\bgamma_0^\intercal)^\intercal)$ converges in distribution to a zero-mean normal random vector with the variance-covariance matrix $\Gamma_1^{-1}\Omega_1 \Gamma_1^{-1}$ as $n\to \infty$, {where $\Gamma_1$ and $\Omega_1$ are defined in the Supporting Information.} 
\end{theorem}

The proof of Theorem \ref{thm:exp} is given in the Supporting Information. Alternatively, one may use $\widehat F_h(t,\widetilde\bbeta^\intercal \bz, \widetilde\bbeta)$ to estimate $F(t,\bbeta_0^\intercal \bZ)$ and establish asymptotic normality for this estimator. The proof follows a similar argument and is therefore omitted.

\subsection{Unknown size function}
\label{sec:unknown}

We now consider the case where $g(\cdot)$ is unknown but monotonically increasing. The monotonicity constraint is imposed for the interpretation of the direction of the covariate effects and is satisfied in most existing semiparametric rate models. For identifiability, we assume that $\|\bgamma_0\|=1$.
In the literature, various methods for monotone single index models have been proposed.
Similar to \cite{sun2022statistical}, we consider the maximum rank estimation \citep{cavanagh1998rank}, which does not involve kernel smoothing and has attractive computational properties.
By applying Equation \eqref{P2}, we have $\operatorname{E}\left\{ \frac{N_i(C_i)}{F(C_i,\bbeta_0^\intercal \bZ_i)}\Big|\bZ_i,\bZ_j  \right\} \ge \operatorname{E}\left\{ \frac{N_j(C_j)}{F(C_j,\bbeta_0^\intercal \bZ_j)}\Big|\bZ_i,\bZ_j  \right\}$ whenever  $\bgamma_0^\intercal \bZ_i \ge \bgamma_0^\intercal \bZ_j$.
In the ideal case where $F(t,\bbeta_0^\intercal \bZ)$ is known, we estimate $\bgamma_0$ by maximizing $\sum_{i=1}^n \sum_{j=1}^n I(\bgamma^\intercal \bZ_i >  \bgamma^\intercal \bZ_j)N_i(C_i)/F(C_i,\bbeta_0^\intercal \bZ_i)$ with the unit norm constraint on $\bgamma$.
In practice, we replace the unknown quantities with their estimates and maximize the following objective function subject to the unit norm constraint on $\bgamma$:
\begin{align*}
\sum_{i=1}^n \sum_{j=1}^n I(\bgamma^\intercal \bZ_i >  \bgamma^\intercal \bZ_j, \bZ_i \in\SZ, \bZ_j \in\SZ)\frac{N_i(C_{i})}{ \widehat{F}_h(C_{i},\widehat{\bbeta}^\intercal \bZ_i,\widehat{\bbeta})}.
\end{align*} 
Denote by $\gammamre$ the maximizer of the above function over the set $\{\bgamma: \bgamma \in \mathbb{R}^p, \|\bgamma\| = 1 \}$. The asymptotic normality of the proposed estimator is stated in Theorem \ref{th_gamma1}. 

\begin{theorem}
\label{th_gamma1}
Under conditions (C1)-(C10) in the Supporting Information, $\sqrt{n}(\gammamre-\bgamma_0)$ converges in distribution to a zero-mean normal random vector with the variance-covariance matrix $\Gamma_2^-\Omega_2 \Gamma_2^-$ as $n\to \infty$, {where $\Gamma_2$ and $\Omega_2$ are defined in the Supporting Information.} 
\end{theorem}

\section{Simulation studies}
\label{sect:simu}

Simulations studies were conducted to evaluate the finite-sample performance of the proposed estimators. The recurrent event process $\NO(\cdot)$ was generated from a non-homogeneous Poisson process whose rate function depends on the covariates $\bZ = (Z_1, Z_2)^\intercal$. For each setting, the two covariates $Z_1$ and $Z_2$ were independently generated from the standard normal distribution. We also allow the rate function to depend on a subject-specific latent variable, denoted by $W$. We generated the recurrent events from the following rate functions:
\begin{description}
\item[(M1)] $\mu(t|\bZ, W) = W f(t,\bbeta_0^\intercal \bZ)\exp(\bgamma_0^\intercal \bZ)$, $t \in[0, 1]$, where $f(t,x) \propto t(1-t)^{\exp(x)}$ is the probability density function of a Beta distribution with shape parameters $2$ and $\exp(x)$;
\item[(M2)] $\mu(t|\bZ,W) = 3W\exp\{-t\exp(\bbeta_0^\intercal \bZ)\} \exp(\bbeta_0^\intercal \bZ)$, $t \in[0, 2]$;
\item[(M3)] 
$\mu(t|\bZ,W) = W(t^3 + \bbeta_0^\intercal \bZ)$, $t \in[0, 2]$.
\end{description}
We set the shape parameter as $\bbeta_0=(\beta_{01}, \beta_{02})^\intercal = (0.8, 0.6)^\intercal$. In ({M1}), we set the size parameter as $\bgamma_0=(\gamma_{01}, \gamma_{02})^\intercal =(0.6,0.8)^\intercal$.
In ({M2}) and ({M3}), we have $\bgamma_0 = \bbeta_0$. The censoring time $C$ was generated from an exponential distribution with rate parameter $0.1W$. For each scenario, we consider two cases, $W = 1$ and $W$ follows a Gamma distribution with mean 1 and variance 1/3. In the latter case, $C$ was correlated with the recurrent event process through both $\bZ$ and $W$, and thus censoring was informative. Although informative censoring is not discussed, arguments in Sections \ref{sec:shape} and \ref{sec:size} can be easily extended to show that the same estimators can be applied. In each simulation, we generated 1,000 simulated datasets with sample sizes of $200$ and $400$. 
The average number of recurrent events observed for ({M1}), ({M2}), and ({M3}) under conditional independent censoring (informative censoring) are 1.6 (1.5), 2.2 (2.2), and 3.5 (3.3), respectively.

For shape estimation, we used the fourth order kernel function, 
$K(u)=315(3-11u^2)(1-u^2)^3I(|u|\le 1)/512$, in objective functions $\ell(\bbeta)$ and $\ell'(\bbeta)$. The bandwidth parameter $h$ was set as {$h = n^{-2/15}$}. When the sample size is relatively small, there is a small probability that $\widetilde r_h$ is negative because the kernel function can take negative values. To solve this issue, we replaced the original kernel estimator $\widetilde r_h(t,x,\bbeta)$ with $\max\{\widetilde r_h(t,x,\bbeta), r_0\}$, where $r_0$ was set as $10^{-6}$ in our simulations. For size estimation, we used the second-order Epanechnikov kernel function, $K(u)=3(1-u^2)I(|u|\le 1)/4$, in ${ \widehat{F}_h(t,x,{\bbeta})}$. The bandwidth parameter $h$ was set as $h =  n^{-2/7}$.
For the unit norm constraint, we apply the transformation $(\cos(\alpha),\sin(\alpha))^\intercal$, $\alpha \in [0,2\pi]$.
The maximization of the objective functions is then with respect to $\alpha$. For $\hbeta$ and $\tbeta$, if the last element of the optimal solution is negative, we multiply the optimal solution by $-1$. As trimming is often skipped in practice \citep{hardle2004nonparametric}, we considered the untrimmed estimators (i.e., $\SZ = \mathbb{R}^p$, $\tau_0 = 0$, and $\tau_1 = \tau$). 
Nonparametric bootstrap with 200 iterations was used to obtain the standard error and 95\% confidence intervals of the proposed estimators. To reduce computational time, we applied bootstrap for $\tbeta$ and used the bootstrap standard error of $\tbeta$ to construct the confidence interval for $\hbeta$ based on the normal approximation. Finally, the size estimator $\gammaexp$ is valid under an exponential link function in size, but may not provide consistent estimation for the true parameters when the link function is 
misspecified (e.g., scenarios ({M2}) and ({M3})). In light of Remark \ref{rmk6} and the fact that the covariates are normally distributed, we evaluated the performance of $\gammaexp/\|\gammaexp\|$ in estimating  $\bgamma_0$.

The results of shape and size estimation are summarized in Tables \ref{tab1} and \ref{tab2}, respectively. As shown in both tables, the proposed estimators yield small biases. The average standard errors obtained using nonparametric bootstrap are close to the empirical standard errors, and the coverage probabilities of the 95\% confidence intervals approximate the nominal level. The absolute values of bias and standard errors decrease as the sample size increases. 
As expected, the empirical standard errors of $\hbeta$ and $\tbeta$ are close. Therefore, when shorter computational time is desired, the standard error of $\hbeta$ can be estimated by bootstrapping $\tbeta$. The standardized estimator, $\gammaexp/\|\gammaexp\|$, works reasonably well under misspecified models, which confirms Proposition \ref{thm3}. When the link function is correctly specified (i.e., ({M1})), $\gammaexp/\|\gammaexp\|$ has smaller or comparable empirical standard errors compared to $\gammamre$. When the link function is misspecified, $\gammamre$ has smaller empirical standard errors than $\gammaexp/\|\gammaexp\|$. The proposed estimators remain satisfactory when the correlation between recurrent events and the censoring time is additionally characterized by a frailty variable. In summary, the proposed approaches exhibit good finite-sample performance under different rate models.

\begin{table}
  \caption{\label{tab1} Summary of simulation results for shape estimation }
  \begin{threeparttable}
  \setlength{\tabcolsep}{0.8mm}{\begin{tabular}{rrrrrrrrrrrrrrrrrrrrrrr}
    \hline
    && \multicolumn{8}{c}{$\tbeta$} & \multicolumn{6}{c}{$\hbeta$} \\
    \hline
    && \multicolumn{4}{c}{$W=1$ } & \multicolumn{4}{c}{$W\sim$ Gamma } && \multicolumn{3}{c}{$W=1$ } & \multicolumn{3}{c}{$W\sim$ Gamma } \\
    \hline
    Scenario && Bias & ESE & ASE & CP & Bias & ESE & ASE & CP && Bias & ESE & CP & Bias & ESE & CP\\ 
    \hline
    $n= 200$\\
    {M1} &	$\beta_1$	&	-11	&	84	&	88	&	96.3	&	-16	&	89	&	94	&	95.6	&&	-11	&	84	&	95.8	&	-12	&	91	&	95.8	\\
&	$\beta_2$	&	-2	&	112	&	118	&	95.0	&	3	&	121	&	126	&	95.6	&&	-2	&	114	&	95.9	&	-8	&	143	&	95.7	\\

{M2} &	$\beta_1$	&	-9	&	94	&	98	&	96.4	&	-15	&	99	&	105	&	95.9	&&	-6	&	96	&	94.9	&	-14	&	102	&	96.3	\\
&	$\beta_2$	&	-9	&	127	&	126	&	94.2	&	-6	&	144	&	139	&	95.9	&&	-14	&	132	&	95.5	&	-11	&	156	&	95.9	\\
												
{M3} &	$\beta_1$	&	-5	&	60	&	64	&	95.8	&	-6	&	68	&	72	&	96.4	&&	-7	&	64	&	96.2	&	-9	&	73	&	96.1	\\
&	$\beta_2$	&	-2	&	80	&	85	&	95.6	&	-3	&	90	&	96	&	95.7	&&	-2	&	95	&	95.8	&	-4	&	103	&	95.7	\\[1ex]
    
$n = 400$\\
{M1} &	$\beta_1$	&	-6	&	57	&	61	&	95.3	&	-5	&	60	&	64	&	95.4	&&	-2	&	61	&	95.3	&	-7	&	63	&	95.3	\\
&	$\beta_2$	&	0	&	78	&	82	&	95.4	&	-2	&	82	&	85	&	95.2	&&	-6	&	83	&	95.4	&	0	&	85	&	95.1	\\
											
{M2} &	$\beta_1$	&	-1	&	66	&	70	&	95.7	&	1	&	69	&	72	&	96.1	&&	-6	&	69	&	95.6	&	-3	&	72	&	95.1	\\
&	$\beta_2$	&	-10	&	92	&	95	&	94.6	&	-12	&	94	&	95	&	95.4	&&	-3	&	94	&	95.2	&	-9	&	97	&	95.2	\\
					
{M3} &	$\beta_1$	&	-8	&	41	&	43	&	96.1	&	-5	&	42	&	44	&	95.0	&&	-6	&	36	&	95.3	&	-7	&	45	&	95.3	\\
&	$\beta_2$	&	6	&	55	&	58	&	96.2	&	3	&	56	&	58	&	95.1	&&	5	&	53	&	94.9	&	5	&	63	&	95.3	\\

  \hline
  \end{tabular}}
  \begin{tablenotes}
    \item[] {\footnotesize Note: Bias is the empirical bias ($\times 1000$); ESE is the empirical standard error ($\times 1000$);  ASE is the average bootstrap standard error ($\times 1000$); and CP is the 95\% empirical coverage probability (\%).}
  \end{tablenotes} 
  \end{threeparttable}
\end{table}

\begin{table}
  \caption{  \label{tab2} Summary of simulation results for size estimation}
  \begin{threeparttable}
  \setlength{\tabcolsep}{0.6mm}{
  \begin{tabular}{rrrrrrrrrrrrrrrrrrrrrrr}
    \hline
    && \multicolumn{8}{c}{$\gammaexp/\|\gammaexp\|$} & \multicolumn{8}{c}{$\gammamre$} \\
    \hline
    && \multicolumn{4}{c}{$W=1$ } & \multicolumn{4}{c}{$W\sim$ Gamma } && \multicolumn{4}{c}{$W=1$ } & \multicolumn{4}{c}{$W\sim$ Gamma } \\
    \hline
    Scenario && Bias & ESE & ASE & CP & Bias & ESE & ASE & CP && Bias & ESE & ASE & CP & Bias & ESE & ASE & CP\\ 
    \hline
$n= 200$\\
{M1} &	$\gamma_1$	&	5	&	48	&	51	&	94.9	&	2	&	73	&	77	&	95.3	&&	5	&	59	&	63	&	95.1	&	2	&	73	&	78	&	94.4	\\
&	$\gamma_2$	&	-6	&	37	&	41	&	93.8	&	-7	&	58	&	61	&	96.1	&&	-7	&	44	&	48	&	94.4	&	-7	&	55	&	59	&	94.2	\\
{M2} &	$\gamma_1$	&	-12	&	122	&	126	&	95.4	&	-23	&	151	&	155	&	94.6	&&	-11	&	115	&	120	&	95.2	&	-24	&	152	&	155	&	94.8	\\
&	$\gamma_2$	&	-16	&	153	&	158	&	95.0	&	-23	&	202	&	205	&	96.0	&&	-17	&	153	&	157	&	94.5	&	-20	&	201	&	205	&	95.3	\\
{M3} &	$\gamma_1$	&	-10	&	84	&	87	&	96.2	&	-5	&	96	&	100	&	96.5	&&	-8	&	78	&	81	&	95.3	&	-7	&	91	&	95	&	94.7	\\
&	$\gamma_2$	&	-3	&	114	&	110	&	96.8	&	-17	&	137	&	142	&	96.4	&&	-5	&	110	&	107	&	96.5	&	-11	&	124	&	129	&	95.8	\\ [1ex]
    
$n = 400$\\
{M1} &	$\gamma_1$	&	5	&	34	&	36	&	95.5	&	0	&	51	&	54	&	96.2	&&	4	&	41	&	43	&	95.2	&	2	&	51	&	53	&	95.2	\\
&	$\gamma_2$	&	-5	&	26	&	29	&	94.9	&	-2	&	38	&	40	&	95.0	&&	-5	&	31	&	34	&	95.3	&	-4	&	39	&	42	&	94.2	\\
{M2} &	$\gamma_1$	&	-6	&	87	&	90	&	95.8	&	-16	&	108	&	110	&	94.8	&&	-3	&	81	&	84	&	95.5	&	-15	&	110	&	112	&	95.0	\\
&	$\gamma_2$	&	-9	&	112	&	114	&	95.0	&	-6	&	142	&	145	&	95.0	&&	-12	&	109	&	112	&	94.9	&	-6	&	142	&	145	&	94.8	\\
{M3} &	$\gamma_1$	&	-4	&	60	&	63	&	95.9	&	-3	&	72	&	75	&	95.6	&&	-3	&	56	&	58	&	95.4	&	-2	&	69	&	72	&	95.7	\\
&	$\gamma_2$	&	-2	&	76	&	78	&	94.7	&	-8	&	93	&	95	&	94.2	&&	-3	&	73	&	75	&	95.5	&	-7	&	90	&	92	&	94.8 \\  
\hline 
  \end{tabular}}
  \begin{tablenotes}
    \item[] {\footnotesize Note: Bias is the empirical bias ($\times 1000$); ESE is the empirical standard error ($\times 1000$); ASE is the average bootstrap standard error ($\times 1000$); and CP is the 95\% empirical coverage probability (\%). }
  \end{tablenotes} 
  \end{threeparttable}
\end{table}

\section{ Data example}
\label{sec:data}

The proposed methods are applied to SEER-Medicare data to analyze the recurrent inpatient hospitalizations in breast cancer patients \citep{warren2002overview}. In our analysis, we focus on females who were diagnosed with stage 0 breast cancer at or after the age of 65 during the period spanning from 2010 to 2017. The outcome is the counting process for inpatient hospitalizations after the diagnosis of breast cancer. For each inpatient hospitalization, its event time is defined as the time from breast cancer diagnosis to the date of admission. The end of follow-up is 12/01/2018 or death, whichever is earlier. We focus on subjects who were enrolled in Medicare Part A from the 12 months before diagnosis to the end of follow-up. The time interval of interest is from the breast cancer diagnosis to 8 years after diagnosis (i.e., $\tau = 8$ years). The covariates include age at diagnosis, tumor size, year of diagnosis, and race (white vs. other). Our analysis is based on a random subsample of 5000 subjects satisfying the above inclusion criteria. The median follow-up time was 4.67 years, and a total of 15,150 events were observed.  In our study sample, 4,330 subjects (86.6\%) were white. The mean age at diagnosis was 75.8  years old; the mean  tumor size was 1.4  cm.

We applied the proposed shape-size model to estimate the covariate effects on the rate of inpatient hospitalizations. The results are presented in Table \ref{tab:medicare}. The results of the proportional rate model are included for comparison. The two approaches proposed for shape estimation yield similar results. Age at diagnosis {and race} are significantly associated with the shape of the process, which suggests heterogeneity in the shape function due to age {and race}. 
For the size component, we considered both an unspecified monotone link function and the exponential link function. Different assumptions on the link function do not alter the direction of associations. Under the unspecified link function, older age, larger tumor size, and more recent year of diagnosis are significantly associated with larger size of the rate function and higher overall event rate. Under the exponential link, one can further make interpretations on the size ratio. Specifically, a one-year increase in age is associated with a 6.7\% increase in the size of the process (95\% confidence interval [CI], 5.7\%--7.7\%); a one-centimeter increase in tumor size is associated with a 37.4\% increase in the size of the process (95\% CI,  25.1\%--51.0\%); a one-year increase in year of diagnosis is associated with a {116.8\% increase} in the size of the process (95\% CI,  87.2\%--151.2\%). When a proportional rate model is applied, {the effects of tumor size, year of diagnosis, and age at diagnosis are attenuated}. Therefore, ignoring shape heterogeneity leads to different results on covariate effects.

\begin{table}
 \caption{\label{tab:medicare} Shape and size estimation for SEER-Medicare data }	
  \begin{threeparttable}
  \setlength{\tabcolsep}{4pt}{
  \begin{tabular}{ccccccccccc}
\hline
    & Tumor size   & Race (non-white)  & Year of diagnosis  & Age at diagnosis  \\
    \hline
    \multicolumn{5}{c}{Shape estimation}\\
    Coef ($\hbeta$) & -0.064 & -0.151 & 0.045 & 0.985   \\
    SE & ~0.038  & ~0.063 & 0.031  & 0.037 \\
    Coef ($\tbeta$) & -0.050 & -0.134 & 0.052 & 0.988 \\
    SE  & ~0.037 & ~0.054 & 0.035 & 0.046 \\
  \hline
    \multicolumn{5}{c}{Size estimation}\\ 
      \multicolumn{5}{c}{(i) shape is covariate-dependent, unspecified link}\\ 
       
    Coef & ~0.306 & 0.055 & 0.692 & 0.652 \\
    SE & ~0.062 & 0.105 & 0.044 & 0.062 \\
      
      \multicolumn{5}{c}{(ii) shape is covariate-dependent, exponential link}\\ 
    Coef & 0.318 & 0.166 & 0.774 & 0.646 \\
    SE & 0.048 & 0.121 & 0.075 & 0.047 \\

      \multicolumn{5}{c}{(iii) shape is covariate-independent, exponential link (proportional rate model)}\\ 
      Coef & 0.126 & 0.028 & 0.221 & 0.208\\
      SE & 0.023 & 0.045 & 0.032 & 0.021\\    
    
   \hline
  
  \end{tabular} }
   \begin{tablenotes}
    \item[] {\footnotesize Note: Coef is the estimated regression coefficient. SE is the bootstrap standard error based on 500 bootstrap samples. The unit of age at diagnosis is 10 years and the unit of tumor size is centimeters.}
  \end{tablenotes} 
  \end{threeparttable}
  
\end{table}

\section{Discussion}
\label{sec:disc}

We proposed a comprehensive framework for modeling counting processes based on flexible and interpretable rate functions. The rate functions depend on two sets of regression parameters, which capture the effect of covariates on the shape and size of the rate function. Our model is particularly useful when the shape of the rate function varies with covariates and includes many commonly used models as special cases.

In the literature, \cite{Sun2008ACO}  and \cite{xu2020generalized}  have considered modeling the rate function with two types of covariate effects: a scale-change covariate effect that modifies the time scale, and an additional multiplicative effect on the cumulative rate function. From a shape-size perspective, the shape functions of the above models only depend on the scale-change parameters, and the proposed method can yield a consistent estimation of the scale-change effects up to a normalization constant. The size function of these models generally depends on both the scale-change and multiplicative effects. This is different from our model, where the size depends on the size index, and the association between covariates and size can be easily interpreted.

In this paper, we considered an estimating equation approach and maximum rank estimation for size estimation. However, it is worth noting that other existing approaches for single index mean models may also be applied. For example, when the size function is completely unspecified and the monotonicity assumption is not imposed, one could apply the methods in \cite{ichimura1993semiparametric} or \cite{xia2006asymptotic} with the projected event count as the outcome. A more thorough study of the estimators will be conducted in our future research.

\backmatter

\section*{Acknowledgements}
This study used the linked SEER-Medicare database. The interpretation and reporting of these data are the sole responsibility of the authors. The authors acknowledge the efforts of the National Cancer Institute; the Office of Research, Development and Information, CMS; Information Management Services (IMS), Inc.; and the Surveillance, Epidemiology, and End Results (SEER) Program tumor registries in the creation of the SEER-Medicare database. The collection of cancer incidence data used in this study was supported by the California Department of Public Health pursuant to California Health and Safety Code Section 103885; Centers for Disease Control and Prevention’s (CDC) National Program of Cancer Registries, under cooperative agreement 1NU58DP007156; the National Cancer Institute’s Surveillance, Epidemiology and End Results Program under contract HHSN261201800032I awarded to the University of California, San Francisco, contract HHSN261201800015I awarded to the University of Southern California, and contract HHSN261201800009I awarded to the Public Health Institute. The ideas and opinions expressed herein are those of the author(s) and do not necessarily reflect the opinions of the State of California, Department of Public Health, the National Cancer Institute, and the Centers for Disease Control and Prevention or their Contractors and Subcontractors.


 
\bibliographystyle{biom}
\bibliography{SSIndex2.bib}

\begin{thebibliography}{}

\bibitem[\protect\citeauthoryear{Andersen and Gill}{Andersen and
  Gill}{1982}]{andersen1982cox}
Andersen, P.~K. and Gill, R.~D. (1982).
\newblock Cox's regression model for counting processes: A large sample study.
\newblock {\em The Annals of Statistics} {\bf 10,} 1100--1120.

\bibitem[\protect\citeauthoryear{Balabdaoui, Groeneboom, and
  Hendrickx}{Balabdaoui et~al.}{2019}]{balabdaoui2019score}
Balabdaoui, F., Groeneboom, P., and Hendrickx, K. (2019).
\newblock Score estimation in the monotone single-index model.
\newblock {\em Scandinavian Journal of Statistics} {\bf 46,} 517--544.

\bibitem[\protect\citeauthoryear{Bouaziz, Geffray, and Lopez}{Bouaziz
  et~al.}{2015}]{bouaziz2015semiparametric}
Bouaziz, O., Geffray, S., and Lopez, O. (2015).
\newblock Semiparametric inference for the recurrent events process by means of
  a single-index model.
\newblock {\em Statistics} {\bf 49,} 361--385.

\bibitem[\protect\citeauthoryear{Bouaziz and Lopez}{Bouaziz and
  Lopez}{2010}]{bouaziz2010conditional}
Bouaziz, O. and Lopez, O. (2010).
\newblock Conditional density estimation in a censored single-index regression
  model.
\newblock {\em Bernoulli} {\bf 16,} 514--542.

\bibitem[\protect\citeauthoryear{B{\"u}cher, El~Ghouch, and
  Van~Keilegom}{B{\"u}cher et~al.}{2021}]{bucher2021single}
B{\"u}cher, A., El~Ghouch, A., and Van~Keilegom, I. (2021).
\newblock Single-index quantile regression models for censored data.
\newblock In {\em Advances in Contemporary Statistics and Econometrics}, pages
  177--196. Springer.

\bibitem[\protect\citeauthoryear{Cavanagh and Sherman}{Cavanagh and
  Sherman}{1998}]{cavanagh1998rank}
Cavanagh, C. and Sherman, R.~P. (1998).
\newblock Rank estimators for monotonic index models.
\newblock {\em Econometrics} {\bf 84,} 351--381.

\bibitem[\protect\citeauthoryear{Chiang, Wang, and Huang}{Chiang
  et~al.}{2018}]{chiang2018versatile}
Chiang, C.-T., Wang, S.-H., and Huang, M.-Y. (2018).
\newblock Versatile estimation in censored single-index hazards regression.
\newblock {\em Annals of the Institute of Statistical Mathematics} {\bf 70,}
  523--551.

\bibitem[\protect\citeauthoryear{Christou and Akritas}{Christou and
  Akritas}{2019}]{christou2019single}
Christou, E. and Akritas, M.~G. (2019).
\newblock Single index quantile regression for censored data.
\newblock {\em Statistical Methods \& Applications} {\bf 28,} 655--678.

\bibitem[\protect\citeauthoryear{Cox and Lewis}{Cox and
  Lewis}{1966}]{cox1966statistical}
Cox, D.~R. and Lewis, P.~A. (1966).
\newblock {\em The Statistical Analysis of Series of Events}.
\newblock Methuen \& Co., Ltd., London; John Wiley \& Sons, Inc., New York.

\bibitem[\protect\citeauthoryear{Ding, Kosorok, and Zeng}{Ding
  et~al.}{2013}]{ding2013local}
Ding, K., Kosorok, M.~R., and Zeng, D. (2013).
\newblock On the local and stratified likelihood approaches in single-index
  hazards model.
\newblock {\em Communications in Mathematics and Statistics} {\bf 1,} 115--132.

\bibitem[\protect\citeauthoryear{Einmahl and Mason}{Einmahl and
  Mason}{2005}]{einmahl2005uniform}
Einmahl, U. and Mason, D.~M. (2005).
\newblock Uniform in bandwidth consistency of kernel-type function estimators.
\newblock {\em The Annals of Statistics} {\bf 33,} 1380--1403.

\bibitem[\protect\citeauthoryear{Enewold, Parsons, Zhao, Bott, Rivera, Barrett,
  Virnig, and Warren}{Enewold et~al.}{2020}]{enewold2020updated}
Enewold, L., Parsons, H., Zhao, L., Bott, D., Rivera, D.~R., Barrett, M.~J.,
  Virnig, B.~A., and Warren, J.~L. (2020).
\newblock Updated overview of the {SEER}-medicare data: Enhanced content and
  applications.
\newblock {\em Journal of the National Cancer Institute Monographs} {\bf 2020,}
  3--13.

\bibitem[\protect\citeauthoryear{Gail, Santner, and Brown}{Gail
  et~al.}{1980}]{gail1980analysis}
Gail, M., Santner, T., and Brown, C. (1980).
\newblock An analysis of comparative carcinogenesis experiments based on
  multiple times to tumor.
\newblock {\em Biometrics} {\bf 36,} 255--266.

\bibitem[\protect\citeauthoryear{Ghosh}{Ghosh}{2004}]{ghosh2004accelerated}
Ghosh, D. (2004).
\newblock Accelerated rates regression models for recurrent failure time data.
\newblock {\em Lifetime Data Analysis} {\bf 10,} 247--261.

\bibitem[\protect\citeauthoryear{H{\"a}rdle, Hall, and Ichimura}{H{\"a}rdle
  et~al.}{1993}]{hardle1993optimal}
H{\"a}rdle, W., Hall, P., and Ichimura, H. (1993).
\newblock Optimal smoothing in single-index models.
\newblock {\em The Annals of Statistics} {\bf 21,} 157--178.

\bibitem[\protect\citeauthoryear{H\"{a}rdle, M\"{u}ller, Sperlich, and
  Werwatz}{H\"{a}rdle et~al.}{2004}]{hardle2004nonparametric}
H\"{a}rdle, W., M\"{u}ller, M., Sperlich, S., and Werwatz, A. (2004).
\newblock {\em Nonparametric and semiparametric models}.
\newblock Springer Series in Statistics. Springer-Verlag, New York.

\bibitem[\protect\citeauthoryear{H{\"a}rdle and Stoker}{H{\"a}rdle and
  Stoker}{1989}]{hardle1989investigating}
H{\"a}rdle, W. and Stoker, T.~M. (1989).
\newblock Investigating smooth multiple regression by the method of average
  derivatives.
\newblock {\em Journal of the American Statistical Association} {\bf 84,}
  986--995.

\bibitem[\protect\citeauthoryear{Huang, Li, Liang, and Tang}{Huang
  et~al.}{2020}]{huang2020estimation}
Huang, H., Li, Y., Liang, H., and Tang, Y. (2020).
\newblock Estimation of single-index models with fixed censored responses.
\newblock {\em Statistica Sinica} {\bf 30,} 829--843.

\bibitem[\protect\citeauthoryear{Huang and Liu}{Huang and
  Liu}{2006}]{huang2006polynomial}
Huang, J.~Z. and Liu, L. (2006).
\newblock Polynomial spline estimation and inference of proportional hazards
  regression models with flexible relative risk form.
\newblock {\em Biometrics} {\bf 62,} 793--802.

\bibitem[\protect\citeauthoryear{Ichimura}{Ichimura}{1993}]{ichimura1993semiparametric}
Ichimura, H. (1993).
\newblock Semiparametric least squares ({SLS}) and weighted {SLS} estimation of
  single-index models.
\newblock {\em Journal of Econometrics} {\bf 58,} 71--120.

\bibitem[\protect\citeauthoryear{Kong and Xia}{Kong and
  Xia}{2017}]{kong2017uniform}
Kong, E. and Xia, Y. (2017).
\newblock Uniform bahadur representation for nonparametric censored quantile
  regression: A redistribution-of-mass approach.
\newblock {\em Econometric Theory} {\bf 33,} 242--261.

\bibitem[\protect\citeauthoryear{Lawless and Nadeau}{Lawless and
  Nadeau}{1995}]{lawless1995some}
Lawless, J.~F. and Nadeau, C. (1995).
\newblock Some simple robust methods for the analysis of recurrent events.
\newblock {\em Technometrics} {\bf 37,} 158--168.

\bibitem[\protect\citeauthoryear{Lin, Wei, and Ying}{Lin
  et~al.}{1998}]{lin1998accelerated}
Lin, D., Wei, L., and Ying, Z. (1998).
\newblock Accelerated failure time models for counting processes.
\newblock {\em Biometrika} {\bf 85,} 605--618.

\bibitem[\protect\citeauthoryear{Lin, Wei, and Ying}{Lin
  et~al.}{2001}]{lin2001semiparametric}
Lin, D., Wei, L., and Ying, Z. (2001).
\newblock Semiparametric transformation models for point processes.
\newblock {\em Journal of the American Statistical Association} {\bf 96,}
  620--628.

\bibitem[\protect\citeauthoryear{Lin, Wei, Yang, and Ying}{Lin
  et~al.}{2000}]{lin2000semiparametric}
Lin, D.~Y., Wei, L.-J., Yang, I., and Ying, Z. (2000).
\newblock Semiparametric regression for the mean and rate functions of
  recurrent events.
\newblock {\em Journal of the Royal Statistical Society: Series B} {\bf 62,}
  711--730.

\bibitem[\protect\citeauthoryear{Lopez}{Lopez}{2009}]{lopez2009single}
Lopez, O. (2009).
\newblock Single-index regression models with right-censored responses.
\newblock {\em Journal of Statistical Planning and Inference} {\bf 139,}
  1082--1097.

\bibitem[\protect\citeauthoryear{Lopez, Patilea, and Van~Keilegom}{Lopez
  et~al.}{2013}]{lopez2013single}
Lopez, O., Patilea, V., and Van~Keilegom, I. (2013).
\newblock Single index regression models in the presence of censoring depending
  on the covariates.
\newblock {\em Bernoulli} {\bf 19,} 721--747.

\bibitem[\protect\citeauthoryear{Murphy and Van~der Vaart}{Murphy and Van~der
  Vaart}{2000}]{murphy2000profile}
Murphy, S.~A. and Van~der Vaart, A.~W. (2000).
\newblock On profile likelihood.
\newblock {\em Journal of the American Statistical Association} {\bf 95,}
  449--465.

\bibitem[\protect\citeauthoryear{Pepe and Cai}{Pepe and
  Cai}{1993}]{pepe1993some}
Pepe, M.~S. and Cai, J. (1993).
\newblock Some graphical displays and marginal regression analyses for
  recurrent failure times and time dependent covariates.
\newblock {\em Journal of the American Statistical Association} {\bf 88,}
  811--820.

\bibitem[\protect\citeauthoryear{Powell, Stock, and Stoker}{Powell
  et~al.}{1989}]{powell1989semiparametric}
Powell, J.~L., Stock, J.~H., and Stoker, T.~M. (1989).
\newblock Semiparametric estimation of index coefficients.
\newblock {\em Econometrica} {\bf 57,} 1403--1430.

\bibitem[\protect\citeauthoryear{Prentice, Williams, and Peterson}{Prentice
  et~al.}{1981}]{prentice1981regression}
Prentice, R.~L., Williams, B.~J., and Peterson, A.~V. (1981).
\newblock On the regression analysis of multivariate failure time data.
\newblock {\em Biometrika} {\bf 68,} 373--379.

\bibitem[\protect\citeauthoryear{Schaubel, Zeng, and Cai}{Schaubel
  et~al.}{2006}]{schaubel2006semiparametric}
Schaubel, D.~E., Zeng, D., and Cai, J. (2006).
\newblock A semiparametric additive rates model for recurrent event data.
\newblock {\em Lifetime Data Analysis} {\bf 12,} 389--406.

\bibitem[\protect\citeauthoryear{Sherman}{Sherman}{1993}]{sherman1993limiting}
Sherman, R.~P. (1993).
\newblock The limiting distribution of the maximum rank correlation estimator.
\newblock {\em Econometrica} pages 123--137.

\bibitem[\protect\citeauthoryear{Strzalkowska-Kominiak and
  Cao}{Strzalkowska-Kominiak and Cao}{2013}]{strzalkowska2013maximum}
Strzalkowska-Kominiak, E. and Cao, R. (2013).
\newblock Maximum likelihood estimation for conditional distribution
  single-index models under censoring.
\newblock {\em Journal of Multivariate Analysis} {\bf 114,} 74--98.

\bibitem[\protect\citeauthoryear{Strzalkowska-Kominiak and
  Cao}{Strzalkowska-Kominiak and Cao}{2014}]{strzalkowska2014beran}
Strzalkowska-Kominiak, E. and Cao, R. (2014).
\newblock Beran-based approach for single-index models under censoring.
\newblock {\em Computational Statistics} {\bf 29,} 1243--1261.

\bibitem[\protect\citeauthoryear{Sun and Su}{Sun and Su}{2008}]{Sun2008ACO}
Sun, L. and Su, B. (2008).
\newblock A class of accelerated means regression models for recurrent event
  data.
\newblock {\em Lifetime Data Analysis} {\bf 14,} 357--375.

\bibitem[\protect\citeauthoryear{Sun, Chiou, Marr, and Huang}{Sun
  et~al.}{2022}]{sun2022statistical}
Sun, Y., Chiou, S.~H., Marr, K.~A., and Huang, C.-Y. (2022).
\newblock Statistical inference on shape-and size-indexes for counting
  processes.
\newblock {\em Biometrika} {\bf 109,} 195--208.

\bibitem[\protect\citeauthoryear{{van der Vaart}}{{van der
  Vaart}}{2000}]{van2000Asymptotic}
{van der Vaart}, A.~W. (2000).
\newblock {\em Asymptotic Statistics}.
\newblock Cambridge University Press.

\bibitem[\protect\citeauthoryear{Wang and Huang}{Wang and
  Huang}{2014}]{wang2014statistical}
Wang, M.-C. and Huang, C.-Y. (2014).
\newblock Statistical inference methods for recurrent event processes with
  shape and size parameters.
\newblock {\em Biometrika} {\bf 101,} 553--566.

\bibitem[\protect\citeauthoryear{Wang, Qin, and Chiang}{Wang
  et~al.}{2001}]{wang2001analyzing}
Wang, M.-C., Qin, J., and Chiang, C.-T. (2001).
\newblock Analyzing recurrent event data with informative censoring.
\newblock {\em Journal of the American Statistical Association} {\bf 96,}
  1057--1065.

\bibitem[\protect\citeauthoryear{Wang, Wang, and Wang}{Wang
  et~al.}{2009}]{wang2009proportional}
Wang, W., Wang, J.-L., and Wang, Q. (2009).
\newblock Proportional hazards regression with unknown link function.
\newblock {\em Lecture Notes-Monograph Series} {\bf 57,} 47--66.

\bibitem[\protect\citeauthoryear{Warren, Klabunde, Schrag, Bach, and
  Riley}{Warren et~al.}{2002}]{warren2002overview}
Warren, J.~L., Klabunde, C.~N., Schrag, D., Bach, P.~B., and Riley, G.~F.
  (2002).
\newblock Overview of the {SEER}-medicare data: Content, research applications,
  and generalizability to the united states elderly population.
\newblock {\em Medical Care} pages IV3--IV18.

\bibitem[\protect\citeauthoryear{Xia}{Xia}{2006}]{xia2006asymptotic}
Xia, Y. (2006).
\newblock Asymptotic distributions for two estimators of the single-index
  model.
\newblock {\em Econometric Theory} {\bf 22,} 1112--1137.

\bibitem[\protect\citeauthoryear{Xu, Chiou, Huang, Wang, and Yan}{Xu
  et~al.}{2017}]{xu2017joint}
Xu, G., Chiou, S.~H., Huang, C.-Y., Wang, M.-C., and Yan, J. (2017).
\newblock Joint scale-change models for recurrent events and failure time.
\newblock {\em Journal of the American Statistical Association} {\bf 112,}
  794--805.

\bibitem[\protect\citeauthoryear{Xu, Chiou, Yan, Marr, and Huang}{Xu
  et~al.}{2020}]{xu2020generalized}
Xu, G., Chiou, S.~H., Yan, J., Marr, K., and Huang, C.-Y. (2020).
\newblock Generalized scale-change models for recurrent event processes under
  informative censoring.
\newblock {\em Statistica Sinica} {\bf 30,} 1773--1795.

\bibitem[\protect\citeauthoryear{Zhao and Zhou}{Zhao and
  Zhou}{2014}]{zhao2014sufficient}
Zhao, X. and Zhou, X. (2014).
\newblock Sufficient dimension reduction on the mean and rate functions of
  recurrent events.
\newblock {\em Statistics in Medicine} {\bf 33,} 3693--3709.

\end{thebibliography}

\section*{Supporting Information}
The Supporting Information contains the regularity conditions and proofs of Theorems 1--4 and Propositions 1--2.

\newpage

\begin{centering}
\LARGE 
Supporting Information for ``Statistical inference for counting processes under shape heterogeneity" by  Yifei Sun and Ying Sheng
\end{centering}

\setcounter{section}{0}

\section{Regularity conditions and notations}

\subsection{Regularity conditions and notations for Theorems 1 and 2}

Denote by $f_{\bbeta^\intercal\bZ}(\cdot)$ the density function of $\bbeta^\intercal \bZ$. For $t\in[\tau_0,\tau_1]$, define  $f^*_1(t,x,\bbeta)dt=\operatorname{E}\{I( {\bZ \in \SZ})d N(t)\mid \bbeta^\intercal \bZ=x\}{f_{\bbeta^\intercal\bZ}(x)}$ and $f^*_2(t,x,\bbeta)=\operatorname{E}[I(C\ge t, {\bZ \in \SZ}) {\{N(t)-N(\tau_0)\}}\mid \bbeta^\intercal \bZ=x] {f_{\bbeta^\intercal\bZ}(x)}$. 
Denote by $\nabla_2$ the second order partial derivative operator with respect to $\balpha$ ({treating $\bbeta = \mathbb S(\balpha)$ as a function of $\balpha$}), and denote by  $\|\cdot\|_F$ the Frobenius norm of matrices.

We impose the following regularity conditions for Theorems 1 and 2:
\begin{itemize}
    
    \item[(C1)] 
The set $\SZ$ is a compact subset of $\mathbb{R}^p$.
        
    \item[(C2)] The process $\widetilde N(\cdot)$ is bounded on $[0,\tau]$.
    
    \item[(C3)] The censoring time $C$ {is independent of $\widetilde N(\cdot)$ conditioning on $\bZ$} and satisfies $\inf_{\bz \in \SZ} P(C\ge \tau\mid \bZ = \bz)>0$. 
    
    \item[(C4)] $K(\cdot)$ is a twice differentiable and fourth order kernel function on $[-1,1]$. 
    Moreover, $h = a_1n^{-a_2}$, where $a_1$ and $a_2$ are constants that satisfy $a_1>0$ and $1/8<a_2<1/6$.

    \item[(C5)]For any $\tau_0^*\in(\tau_0,\tau_1)$, assume $\inf_{\bbeta\in\mathcal{B}, \bz \in \SZ, t\in[\tau_0^*,\tau_1]} f^*_2(t,\bbeta^\intercal \bz,\bbeta) > 0$. For $\balpha_1,\balpha_2 \in\mathcal{A}$, there exist {square-}integrable functions $m^*_k(\bz,t)$ on $\SZ\times [\tau_0,\tau_1]$ such that 
    $$
    \|\nabla_2 f^*_k(t,\mathbb S(\balpha_1)^\intercal\bz,\mathbb S(\balpha_1)) - \nabla_2 f^*_k(t,\mathbb S(\balpha_2)^\intercal\bz,\mathbb S(\balpha_2))\|_F \le m^*_k(\bz,t)\|\balpha_1-\balpha_2\|,\;\;k=1,2.
    $$   
\end{itemize}

\subsection{Regularity conditions and notations for Theorem 3}

For Theorem 3, we first define $\widetilde {\bZ}=(1,\bZ^\intercal)^\intercal$, $\mN = N(C)/F(C,\bbeta_0^\intercal \bZ)$, $dM(t) = dN(t) - I(C\ge t) N(t)r(t,\bbeta_0^\intercal \bZ)dt$,  and  $q^{(k)}(t,x) = \operatorname{E} \{\bZ^{\otimes k} I(C\ge t) g(\bgamma_0^\intercal \bZ)\mid\bbeta_0^\intercal\bZ = x\}$ for $k = 0,1$. We then define the matrices in the variance-covariance matrix formula:
\begin{align*}
    \Gamma_1 =& - E\{\exp(c_0 + \bgamma_0^\intercal\bZ)\widetilde{\bZ}\widetilde{\bZ}^\intercal  I(\bZ \in\SZ) \},\\
    \Omega_1 =& E(\psi_1\psi_1^\intercal),\\
    \psi_1 =& \widetilde{\bZ}I(\bZ \in\SZ)\left\{\mN - \exp(c_0 + \bgamma_0^\intercal\bZ)  \right\}\\
    & + \! \int_0^\tau \! { \operatorname{E} \{ \widetilde{\bZ}\mN I(\bZ \in\SZ, C\le t) \! \mid \! \bbeta_0^\intercal \bZ  \}} \{{q^{(0)}(t,\bbeta_0^\intercal \bZ)F(t, \bbeta_0^\intercal \bZ) }\}^{-1} \!{dM(t)}f_{\bbeta_0^\intercal \bZ}(\beta_0^\intercal \bZ)\\ 
    & - \operatorname{E}[\widetilde{\bZ}I(\bZ \in\SZ){\mN}\int_C^\tau \dot r(t,\bbeta_0^\intercal\bZ){\{\bZ -  q^{(1)}(t,\bbeta_0^\intercal \bZ)/q^{(0)}(t,\bbeta_0^\intercal \bZ)\}^\intercal} ]dtV^{-} \psi.
\end{align*}

We additionally impose the following regularity conditions. Here 
we define the functions $f_1(t,x,\bbeta)dt=\operatorname{E}\{d N(t)\mid \bbeta^\intercal \bZ=x\}{f_{\bbeta^\intercal\bZ}(x)}$ and $f_2(t,x,\bbeta)=\operatorname{E}[I(C\ge t) {N(t)}\mid \bbeta^\intercal \bZ=x] {f_{\bbeta^\intercal\bZ}(x)}$. Denote by $\dot r(t,x)$ the partial derivative of $r(t,x)$ with respect to $x$. It is worth noting that for size estimation, condition (C6) on the kernel function $K$ and the bandwidth parameter $h$ in $\widehat F_h$ differs from condition (C4) for shape estimation.

\begin{itemize}
     \item[(C6)]  The kernel function $K(\cdot)$ in $\widehat F_h(t,x,\bbeta)$ is a second order kernel function on $[-1,1]$. The bandwidth parameter  is of the form $\hshape =b_1 n^{-b_2}$, where $b_1$ and $b_2$ are constants that satisfy $b_1>0$ and $1/4<b_2<1/2$. 
    \item[(C7)] There exists a small positive constant $\tau'$ such that $\inf_{\bz \in \SZ} P(C\ge \tau'\mid \bZ = \bz) = 1$ and $\inf_{\bz \in \SZ} F(\tau',\bbeta_0^\intercal\bz) > 0$.
  \item[(C8)] For $\balpha_1,\balpha_2 \in\mathcal A$, there exist square-integrable functions $m_k(\bz,t)$ such that 
    $$
    \|\nabla_2 f_k(t,\mathbb S(\balpha_1)^\intercal\bz,\mathbb S(\balpha_1)) - \nabla_2 f_k(t,\mathbb S(\balpha_2)^\intercal\bz,\mathbb S(\balpha_2))\|_F \le m_k(\bz,t)\|\balpha_1-\balpha_2\|,\;\;k=1,2.
    $$
\end{itemize}

\subsection{Regularity conditions and notations for Theorem 4}

For Theorem 4, the definitions of the matrices in the variance-covariance formula are given as follows:
\begin{align*}
\Gamma_2 =& {-} \operatorname{E}[\{  \phi^{(2)}(\bgamma_0^\intercal \bZ)\phi^{(0)}(\bgamma_0^\intercal \bZ) - \phi^{(1)}(\bgamma_0^\intercal \bZ)^{\otimes 2} \} \dot{g}(\bgamma_0^\intercal  \bZ) f_{\bgamma_0^\intercal \bZ}(\bgamma_0^\intercal \bZ)  ],\\
\Omega_2 =& \operatorname{E}(\psi_2\psi_2^\intercal),\\
\psi_2
=& \delta(\bZ) \left\{ \mN - g(\bgamma_0^\intercal \bZ)   \right\}  \\
&-  \operatorname{E} \left[ \int_C^\tau   \dot{r}(t,\bbeta_0^\intercal \bZ) dt \cdot \delta(\bZ)\mN {\{\bZ -  q^{(1)}(t,\bbeta_0^\intercal \bZ)/q^{(0)}(t,\bbeta_0^\intercal \bZ)\}^\intercal}\right]  {V}^-  \psi\\
& +\int_0^\tau{ \operatorname{E} \left\{ \delta(\bZ)  \mN I(C\le t) \;\middle|\; \bbeta_0^\intercal \bZ \right\}} \{{q^{(0)}(t,\bbeta_0^\intercal \bZ)F(t, \bbeta_0^\intercal \bZ) }\}^{-1} {dM(t)}f_{\bbeta_0^\intercal \bZ}(\bbeta_0^\intercal \bZ),
\end{align*}
with $\delta(\bZ) =  f_{\bgamma_0^\intercal \bZ}(\bgamma_0^\intercal \bZ)\{\bZ \phi^{(0)}( \bgamma_0^\intercal \bZ  )- \phi^{(1)}(\bgamma_0^\intercal \bZ ) \}I(\bZ \in\SZ)$ and $\phi^{(k)}(\bgamma_0^\intercal\bZ) = \operatorname{E}\{\bZ^{\otimes k} I(\bZ\in\SZ)\mid \bgamma_0^\intercal\bZ\}$ for $k=0,1,2$.

The transformation $\mathbb{S}$ in Section 3.3 can be applied to reduce $\bgamma$ to a $(p-1)$-dimensional vector $\btheta$ that lies within $[0,\pi]^{(p-2)}\times[0,2\pi]$. Let $\btheta_0$ represent the parameter value that corresponds to the true value, i.e., $\bgamma_0 = \mathbb{S}(\btheta_0)$. Define 
$
\eta(O,\btheta)=\int_{\bz\in\SZ} I( \mathbb{S}(\btheta)^\intercal\bz<\mathbb{S}(\btheta)^\intercal\bZ,\bZ\in\SZ) \{ \mN - g(\bgamma_0^\intercal\bZ) \} F_{\bZ}(d\bz),
$
where $F_{\bZ}(\cdot)$ is the cumulative distribution function of $\bZ$, and $O=\{(\bZ,C,N(t)), 0\le t \le \tau\}$ denotes the observed data. 
Let $f_{\bgamma_0^\intercal \bZ}(\cdot)$ be the density of $\bgamma_0^\intercal \bZ$, and let $\dot g$ be the first order derivative of $g$. We additionally impose the following regularity conditions:

\begin{itemize}
    \item[(C9)] There exists a square-integrable function $m(\cdot)$ such that $$\|\nabla_2\eta(O,\btheta)-\nabla_2\eta(O,\btheta_0)\|_F\le m(O)\|\btheta-\btheta_0\|.$$

    \item[(C10)] The size function $g(\cdot)$ and its derivatives up to the third order are bounded. 
    
    \end{itemize}

\section{Proof of Proposition 1}

Given $R(t,x,\bbeta) = \int_t^\tau   \left[\operatorname{E}\{N(u)I(C\ge u)\mid\bbeta^\intercal\bZ = x\}\right]^{-1} \operatorname{E}\{dN(u)\mid\bbeta^\intercal\bZ = x\}$, we have
\begin{eqnarray*}
&& \operatorname{E}\left[ \int_0^\tau  \{{R}(t,\bbeta^\intercal \bZ,\bbeta) -  {R}(C,\bbeta^\intercal \bZ,\bbeta)\} dN(t) \right]\\
&&=  \operatorname{E}\left[ \int_0^\tau   \int_0^\tau I(t \le u \le C)   \frac{\operatorname{E}\{ dN(u)\mid \bbeta^\intercal \bZ\}dN(t)}{\operatorname{E}\{ N(u)I(C\ge u)\mid \bbeta^\intercal \bZ\}}   \right]\\
&&=  \operatorname{E}\left[ \int_0^\tau  N(u)I(C\ge u)   \frac{\operatorname{E}\{ dN(u)\mid \bbeta^\intercal \bZ\}}{\operatorname{E}\{ N(u)I(C\ge u)\mid \bbeta^\intercal \bZ\}}   \right]\\
&&=  \operatorname{E}\left[ \int_0^\tau  E\{ N(u)I(C\ge u)\mid \bbeta^\intercal \bZ\}   \frac{\operatorname{E}\{ dN(u)\mid \bbeta^\intercal \bZ\}}{\operatorname{E}\{ N(u)I(C\ge u)\mid \bbeta^\intercal \bZ\}}   \right]\\
&&=  \operatorname{E}\left[ \int_0^\tau     \operatorname{E}\{ dN(u)\mid \bbeta^\intercal \bZ\}\right]\\
&&=  \operatorname{E} \{N(\tau)\}.
\end{eqnarray*}
Therefore, we have proved the proposition.

The same argument can be applied to the trimmed estimator. 
Define $d\Ntr(t) = I(\bZ\in\SZ,\tau_0 \le t \le \tau_1)dN(t)$. For $t\in(\tau_0,\tau_1)$, the kernel estimator, 
\begin{align*}
 \widetilde{R}_h(t,x, {\bbeta}) =\int_t^{\tau_1} \frac{\sum_{i=1}^n K_h(x-{\bbeta}^\intercal \bZ_i)I(\bZ \in \SZ)dN_i(u)}{\sum_{i=1}^n K_h(x-{\bbeta}^\intercal \bZ_i)I(\bZ \in \SZ,C_i\ge u)\{N_i(u)-N_i(\tau_0)\}},  
\end{align*}
estimates the quantity 
$$
\Rtr(t,x,\bbeta) = \int_t^{\tau_1} \frac{\operatorname{E}\{d\Ntr(u)\mid\bbeta^\intercal\bZ = x\}}{\operatorname{E}\{I(C\ge u)\Ntr(u)\mid\bbeta^\intercal\bZ = x\}}.
$$
Then we have
\begin{eqnarray*}
 && \operatorname{E}\left[ \int_{\tau_0}^{\tau_1}  \{\Rtr(t,\bbeta^\intercal \bZ,\bbeta) -  \Rtr(C,\bbeta^\intercal \bZ,\bbeta)\} I(\bZ\in\SZ)dN(t) \right]\\
&&=  \operatorname{E}\left[ \int_{\tau_0}^{\tau_1}   \int_{\tau_0}^{\tau_1} I(t \le u \le C, \bZ\in\SZ)   \frac{\operatorname{E} \{I(\bZ\in\SZ) dN(u)\mid \bbeta^\intercal \bZ\}dN(t)}{\operatorname{E}\{ \Ntr(u)I(C\ge u)\mid \bbeta^\intercal \bZ\}}   \right]\\
&&= \operatorname{E} \left[ \int_{\tau_0}^{\tau_1} \Ntr(u)I(C\ge u)   \frac{\operatorname{E} \{I(\bZ\in\SZ) dN(u)\mid \bbeta^\intercal \bZ\}}{\operatorname{E} \{ \Ntr(u)I( C\ge u)\mid \bbeta^\intercal \bZ\}}   \right]\\
&&= \operatorname{E} \left[ \int_{\tau_0}^{\tau_1}     \operatorname{E}\{ I(\bZ\in\SZ)dN(u)\mid \bbeta^\intercal \bZ\}\right]\\
&&= \operatorname{E} \{ \Ntr(\tau_1)\}.
\end{eqnarray*}

\section{The large-sample property of $\widetilde r_h$}

In this section, we study the large-sample property of $\widetilde r_h(t,\bbeta^\intercal\bz,\bbeta)$. 
The result is summarized in Lemma \ref{lemmatr} and will be used when proving Theorems 1 and 2. The constraint $\bbeta\in\mB$ can be achieved by reparameterizing $\bbeta$ in the polyspherical coordinate system as a $(p-1)$-dimensional vector $\balpha = (\alpha_1,\alpha_2,\ldots,\alpha_{p-1})^\intercal \in \mathcal{A}$ with $\mathcal{A}=[0,\pi]^{(p-1)}$. Specifically, we set 
\begin{eqnarray}
\label{transformation}
\bbeta = \mathbb{S}(\balpha) = 
 \left(\cos(\alpha_1), \sin(\alpha_1)\cos(\alpha_2),\ldots, \prod_{k=1}^{p-2}\sin(\alpha_k)\cos(\alpha_{p-1}),  \prod_{k=1}^{p-1}\sin(\alpha_k)\right)^\intercal.
\end{eqnarray}
Define the $p\times(p-1)$ matrix $J(\balpha)=\partial \mathbb{S}(\balpha)/\partial\balpha$.
Define  $r^*(t,\bbeta^\intercal\bz,\bbeta)=f^*_1(t,\bbeta^\intercal\bz,\bbeta)/f^*_2(t,\bbeta^\intercal\bz,\bbeta)$, where $f^*_1(t,x,\bbeta)dt=\operatorname{E}\{I( {\bZ \in \SZ})d N(t)\mid \bbeta^\intercal \bZ=x\}{f_{\bbeta^\intercal\bZ}(x)}$ and $$f^*_2(t,x,\bbeta)=\operatorname{E}[I(C\ge t, {\bZ \in \SZ}) {\{N(t)-N(\tau_0)\}}\mid \bbeta^\intercal \bZ=x] {f_{\bbeta^\intercal\bZ}(x)}.$$ 
Let $\bm{a} = (a_1, a_2, \dots, a_p)^\intercal$ be a vector in $\mathbb{R}^p$. We define the notation $\bm{a} < {\infty}$ to mean that $\bm{a}$ is a finite vector with $a_j<\infty$ for $j = 1,\ldots,p$.
Similarly, for a matrix $\bm A$, $\bm A<\infty$ means all the elements are finite.  
Let $\mH_1=\{h: h =a_1 n^{-a_2}, a_1>0,  1/8<a_2<1/6\}$. 

{\bf Throughout the proof, we use $\bbeta$ and $\mathbb S(\balpha)$ interchangeably for ease of expression.}

\begin{lemma}
\label{lemmatr}
Under conditions (C1)--(C5), with probability 1, we have
\begin{eqnarray}
\label{tr}
\lim_{n\to \infty} \sup_{ t\in[\tau_0^*,\tau_1], \bz\in\SZ,  \balpha\in\mA, h\in\mH_1} \frac{|\tr(t,\bbeta^\intercal\bz,\bbeta)-r^*(t,\bbeta^\intercal\bz,\bbeta)|}{n^{-1/2}h^{-1}(\log n)^{1/2}} = M_1^* < \infty,
\end{eqnarray}
\begin{eqnarray}
\label{tr1}
\lim_{n\to \infty} \sup_{t\in[\tau_0^*,\tau_1], \bz\in\SZ,  \balpha\in\mA, h\in\mH_1} \frac{\big| \partial\tr(t,\bbeta^\intercal\bz,\bbeta)/\partial \balpha - \partial r^*(t,\bbeta^\intercal\bz,\bbeta)/\partial \balpha \big|}{n^{-1/2}h^{-2}(\log n)^{1/2}} = \bm M_2^* < \infty,
\end{eqnarray}
\begin{eqnarray}
\label{tr2}
\lim_{n\to \infty} \sup_{t\in[\tau_0^*,\tau_1], \bz\in\SZ,  \balpha\in\mA, h\in\mH_1} \frac{ \big| \partial^2\tr(t,\bbeta^\intercal\bz,\bbeta)/\partial \balpha \partial \balpha^\intercal - \partial^2 r^*(t,\bbeta^\intercal\bz,\bbeta)/\partial \balpha \partial \balpha^\intercal \big| }{n^{-1/2}h^{-3}(\log n)^{1/2}} = \bm M_3^* < \infty,
\end{eqnarray}
for any constant $\tau_0^*\in(\tau_0,\tau_1)$.
\end{lemma}

{\it Proof of Lemma \ref{lemmatr}.}
Define $F_1^*(\bz, t) = \operatorname{E} \{ I(\bZ \le \bz,\bZ \in \SZ) \int_0^t I(C\ge u)dN(u) \}$ and 
$F_2^*(\bz, t) = \operatorname{E} \{I(\bZ \le \bz, C\ge t)\Ntr(t) \}$, where $\bZ \le \bz$ means $Z_j\le z_j$ for $j=1,\ldots,p$ and $\Ntr(t) = \int_0^t I(\bZ\in\SZ,\tau_0 \le u \le \tau_1)dN(u)$.
Let $\hf^*(t,x,\bbeta) = \int K_h(x -\bbeta^\intercal \bz') K_h(t-t') d \hF_1^*(z',t')$ and $\hff^*(t,x,\bbeta) = \int K_h(x -\bbeta^\intercal \bz') d \hF_2^*(\bz',t)$,
where $\widehat{F}_1^*$ and $\widehat{F}_2^*$ are the empirical estimates of $F_1^*$ and $F_2^*$, respectively. 
Then we have $\tr(t,x,\bbeta) = \hf^*(t,x,\bbeta)/\hff^*(t,x,\bbeta)$.  
It is easy to see
\begin{eqnarray}
\label{diff-r}
&&|\tr(t,\bbeta^\intercal\bz,\bbeta)-r^*(t,\bbeta^\intercal\bz,\bbeta)| \\\nonumber
&=& \left| \frac{\hf^*(t,\bbeta^\intercal \bz,\bbeta)}{\hff^*(t,\bbeta^\intercal \bz,\bbeta)} - \frac{f_1^*(t,\bbeta^\intercal \bz,\bbeta)}{f_2^*(t,\bbeta^\intercal \bz,\bbeta)} \right| \\\nonumber
&\le&  \frac{|\hf^*(t,\bbeta^\intercal \bz,\bbeta) - f_1^*(t,\bbeta^\intercal \bz,\bbeta)|}{|\hff^*(t,\bbeta^\intercal \bz,\bbeta)|}  +  \frac{|f_1^*(t,\bbeta^\intercal \bz,\bbeta)||\hff^*(t,\bbeta^\intercal \bz,\bbeta) - f_2^*(t,\bbeta^\intercal \bz,\bbeta)|} {|f_2^*(t,\bbeta^\intercal \bz,\bbeta) \hff^*(t,\bbeta^\intercal \bz,\bbeta)|}.
\end{eqnarray}
Under condition (C5), the classes of functions $\mK_1$ and $\mK_2$ satisfy the bracketing number condition, where $\mK_1=\{ \int_0^\tau K((\bbeta^\intercal\bz - \bbeta^\intercal\bZ)/h) K((t-t')/h) I(\bZ\in\SZ) dN(t'): h \in\mH_1, \bz\in\SZ,  \balpha\in\mA, t \in[\tau_0^*,\tau_1]\}$ and 
$\mK_2=\{ K((\bbeta^\intercal\bz -\bbeta^\intercal\bZ)/h) I(
C\ge t) \Ntr(t): h \in\mH_1, \bz\in\SZ, \balpha\in\mA, t \in[\tau_0^*,\tau_1] \}$. This indicates that for $k=1,2$, there exist constants $v^*_k>0$ and $v_k>0$ such that $N_{[]}(\epsilon,\mK_k,L_2(P))\le v^*_k \epsilon^{-v_k}$, $0 < \epsilon <1$, where $N_{[]}(\epsilon,\mK_k,L_2(P))$ is the bracketing number. By Theorems 1 and 3 in \cite{einmahl2005uniform}, with probability 1, 
we can derive 
$$
\lim_{n\to \infty} \sup_{t\in[\tau_0^*,\tau_1], \bz\in\SZ, \balpha\in\mA, h\in\mH_1} \frac{|\widehat f_k^*(t,\bbeta^\intercal \bz,\bbeta) - \operatorname{E} \{\widehat f_k^*(t,\bbeta^\intercal \bz,\bbeta)\}|}{n^{-1/2}h^{-1} (\log n)^{1/2} } = m_k < \infty,\;\; k=1,2.
$$
For $k=1,2$, since $K(\cdot)$ is a fourth order kernel function, applying classical kernel smoothing arguments yields  $\sup_{t\in[\tau_0^*,\tau_1], \bz\in\SZ, \balpha\in\mA} | \operatorname{E}\{\widehat f_k^*(t,\bbeta^\intercal \bz,\bbeta)\} - f_k^*(t,\bbeta^\intercal \bz,\bbeta)| =O(h^4)=o(n^{-1/2})$ due  to $nh^8 \to 0$.
This leads to 
$$
\lim_{n\to \infty} \sup_{t\in[\tau_0^*,\tau_1], \bz\in\SZ, \balpha\in\mA, h\in\mH_1} \frac{|\widehat f_k^*(t,\bbeta^\intercal \bz,\bbeta) - f_k^*(t,\bbeta^\intercal \bz,\bbeta) |}{n^{-1/2}h^{-1} (\log n)^{1/2} } = m'_k < \infty,\;\; k=1,2.
$$
Moreover, we have $\inf_{t\in[\tau_0^*,\tau_1], \bz\in\SZ, \bbeta\in\mB} f^*_2(t,\bbeta^\intercal\bz,\bbeta)>0$ by condition (C5). 
Therefore, by the inequality in \eqref{diff-r}, we can derive 
$$
\lim_{n\to \infty} \sup_{t\in[\tau_0^*,\tau_1], \bz\in\SZ, \balpha\in\mA, h\in\mH_1} \frac{|\tr(t,\bbeta^\intercal\bz,\bbeta)-r^*(t,\bbeta^\intercal\bz,\bbeta)|}{n^{-1/2}h^{-1}(\log n)^{1/2}} = M_1^* < \infty.
$$
This completes the proof of Equation \eqref{tr}.

We then study the convergence rate of $\partial\tr(t,\bbeta^\intercal\bz,\bbeta)/\partial \balpha$. Straightforward calculations lead to 
$$
\frac{\partial\tr(t,\bbeta^\intercal\bz,\bbeta)}{\partial \balpha} = \frac{ \partial \hf^*(t,\bbeta^\intercal \bz,\bbeta)/\partial \balpha \cdot \hff^*(t,\bbeta^\intercal \bz,\bbeta) - \partial \hff^*(t,\bbeta^\intercal \bz,\bbeta)/\partial \balpha \cdot  \hf^*(t,\bbeta^\intercal \bz,\bbeta) }{\{\hff^*(t,\bbeta^\intercal \bz,\bbeta)\}^2},
$$
$$
\frac{\partial r^*(t,\bbeta^\intercal\bz,\bbeta)} {\partial \balpha} = \frac{ \partial f_1^*(t,\bbeta^\intercal \bz,\bbeta)/ \partial \balpha \cdot 
f_2^*(t,\bbeta^\intercal \bz,\bbeta) - \partial f_2^*(t,\bbeta^\intercal \bz,\bbeta)/\partial \balpha \cdot  f_1^*(t,\bbeta^\intercal \bz,\bbeta) }{\{f_2^*(t,\bbeta^\intercal \bz,\bbeta)\}^2}.
$$
Then, we have 
\begin{eqnarray}
 \label{diff-tr1}
\Big|\frac{\partial\tr(t,\bbeta^\intercal\bz,\bbeta)}{\partial \balpha} - \frac{\partial r^*(t,\bbeta^\intercal\bz,\bbeta)}{\partial \balpha}\Big| \leq I_1+I_2+I_3+I_4,   
\end{eqnarray}
where 
\begin{eqnarray*}
I_1 &=& \Big|\frac{ \partial f_2^*(t,\bbeta^\intercal \bz,\bbeta)/\partial \balpha}{\{\hff^*(t,\bbeta^\intercal \bz,\bbeta)\}^2}\Big| |\hf^*(t,\bbeta^\intercal \bz,\bbeta) - f_1^*(t,\bbeta^\intercal \bz,\bbeta)|,\\
I_2 &=& \Big| \frac{ f_1^*(t,\bbeta^\intercal \bz,\bbeta) \{f_2^*(t,\bbeta^\intercal \bz,\bbeta)+\hff^*(t,\bbeta^\intercal \bz,\bbeta)\} \partial f_2^*(t,\bbeta^\intercal \bz,\bbeta)/\partial \balpha }{\{f_2^*(t,\bbeta^\intercal \bz,\bbeta)\}^2 \{\hff^*(t,\bbeta^\intercal \bz,\bbeta)\}^2} - \frac{ \partial f_1^*(t,\bbeta^\intercal \bz,\bbeta)/\partial \balpha }{f_2^*(t,\bbeta^\intercal \bz,\bbeta) \hff^*(t,\bbeta^\intercal \bz,\bbeta)}\Big|\\
&& \cdot  |\hff^*(t,\bbeta^\intercal \bz,\bbeta) - f_2^*(t,\bbeta^\intercal \bz,\bbeta)|,\\
I_3 &=& \Big|\frac{1}{\hff^*(t,\bbeta^\intercal \bz,\bbeta)}\Big|  \Big| \frac{\partial \hf^*(t,\bbeta^\intercal \bz,\bbeta)}{\partial \balpha} - \frac{\partial f_1^*(t,\bbeta^\intercal \bz,\bbeta)}{\partial \balpha} \Big|, \\
I_4 &=& \Big|\frac{ \hf^*(t,\bbeta^\intercal \bz,\bbeta)}{\{\hff^*(t,\bbeta^\intercal \bz,\bbeta)\}^2}\Big| \Big| \frac{\partial \hff^*(t,\bbeta^\intercal \bz,\bbeta)}{\partial \balpha} - \frac{\partial f_2^*(t,\bbeta^\intercal \bz,\bbeta)}{\partial \balpha} \Big|. 
\end{eqnarray*}
For $k=1,2$, we have shown in the proof of Equation \eqref{tr} that 
$$
\lim_{n\to \infty}\sup_{t\in[ \tau_0^*,\tau_1], \bz \in \SZ, \balpha\in\mA, h\in\mH_1} \frac{ |\widehat f_k^*(t,\bbeta^\intercal \bz,\bbeta) - f_k^*(t,\bbeta^\intercal \bz,\bbeta)|}{n^{-1/2}h^{-1}(\log n)^{1/2}}=m_k'<\infty,
$$ 
and thus we have $$I_1+I_2=O_p\left(n^{-1/2}h^{-1}(\log n)^{1/2}\right).$$ Next, we study $I_3$ and $I_4$.
Define $\mK_3=\{ \int_0^\tau \dot K((\bbeta^\intercal\bz - \bbeta^\intercal\bZ)/h) K((t-t')/h) (\bz-\bZ)^\intercal J(\balpha) I(\bZ\in\SZ) dN(t'): h \in\mH_1, \bz\in\SZ, \balpha\in\mA, t \in[ \tau_0^*,\tau_1]\}$ and  
$\mK_4=\{\dot K((\bbeta^\intercal\bz -\bbeta^\intercal\bZ)/h) (\bz-\bZ)^\intercal J(\balpha) I( C\ge t) \Ntr(t): h \in\mH_1, \bz\in\SZ, \balpha\in\mA, t \in[ \tau_0^*,\tau_1] \}$.
For $k=3,4$, the bracketing number $N_{[]}(\epsilon,\mK_k,L_2(P))$ is bounded by a polynomial in $1/\epsilon$.  For $k=1,2$, applying arguments in \cite{einmahl2005uniform} yields
$$ \sup_{t\in[\tau_0^*,\tau_1], \bz \in \SZ, \balpha\in\mA}|\partial \widehat f_k^*(t,\bbeta^\intercal \bz,\bbeta)/\partial \balpha - \operatorname{E}\{\partial \widehat f_k^*(t,\bbeta^\intercal \bz,\bbeta)/\partial \balpha\}|=O_p\left(n^{-1/2}h^{-2}(\log n)^{1/2}\right).$$  Moreover, we have $\sup_{t\in[\tau_0^*,\tau_1], \bz \in \SZ, \balpha\in\mA}| \operatorname{E}\{\partial \widehat f_k(t,\bbeta^\intercal \bz,\bbeta)/\partial \balpha\} - \partial f_k(t,\bbeta^\intercal \bz,\bbeta)/\partial \balpha|=O_p(h^4)=o_p(n^{-1/2})$ due to $nh^8\to 0$. Consequently, we can derive $\sup_{t\in[\tau_0^*,\tau_1], \bz \in \SZ, \balpha\in\mA}|\partial \widehat f_k^*(t,\bbeta^\intercal \bz,\bbeta)/\partial \balpha - \partial f_k^*(t,\bbeta^\intercal \bz,\bbeta)/\partial \balpha|=O_p\left(n^{-1/2}h^{-2}(\log n)^{1/2}\right)$, which leads to $$I_3+I_4=O_p\left(n^{-1/2}h^{-2}(\log n)^{1/2}\right).$$
It follows from \eqref{diff-tr1} that
$$
\lim_{n\to \infty} \sup_{t\in[\tau_0^*,\tau_1], \bz \in \SZ, \balpha\in\mA, h\in\mH_1} \frac{\Big| \partial\tr(t,\bbeta^\intercal\bz,\bbeta)/\partial \balpha - \partial r^*(t,\bbeta^\intercal\bz,\bbeta)/\partial \balpha \Big|}{n^{-1/2}h^{-2}(\log n)^{1/2}} =\bm M_2^*<\infty .
$$
This completes the proof of Equation \eqref{tr1}.

Equation \eqref{tr2} can be obtained along the same line and thus the proof is omitted.

\section{ Lemmas for Theorems 1 and 2}

When establishing the asymptotic normality of $\hbeta$ and $\tbeta$, the following lemmas are used to derive the first order derivatives of the objective functions with respect to $\balpha$.

\begin{lemma}
\label{lemma_deriv1}
Define 
$q^{(k)*}(t,x) = \operatorname{E} \{\bZ^{\otimes k} I(C\ge t, \bZ\in\SZ) g(\bgamma_0^\intercal \bZ)\mid\bbeta_0^\intercal\bZ = x\}$ for $k=0,1$. Under the conditions specified in Theorem 1, for $t\in[\tau_0^*,\tau_1]$, we have
$$
\frac{\partial r^*(t,\bbeta^\intercal\bz,\bbeta)}{\partial \balpha}\bigg|_{\balpha = \balpha_0} = \dot r^*(t,\bbeta_0^\intercal\bz)
\left\{\bz -  \frac{q^{(1)*}(t,\bbeta_0^\intercal\bz)}{q^{(0)*}(t,\bbeta_0^\intercal\bz)}\right\}^\intercal J(\balpha_0),
$$
where $\tau_0^*\in(\tau_0,\tau_1)$ is a constant, and $\dot r^*(t,x)$ denotes the partial derivative of $r^*(t,x)$ with respect to $x$.
\end{lemma} 

{\it Proof of Lemma \ref{lemma_deriv1}.} 
Define $\Ntr(t) = \int_0^t I(\bZ\in\SZ,\tau_0 \le t \le \tau_1)dN(u)$ and we have $r^*(t,\bbeta^\intercal\bz,\bbeta) dt =\operatorname{E}\{d\Ntr(t)  \mid \bbeta^\intercal \bZ = \bbeta^\intercal \bz  \}/\operatorname{E}\{\Ntr(t)I(C\ge t)\mid   \bbeta^\intercal \bZ = \bbeta^\intercal \bz \}$.
We first calculate $\partial \operatorname{E}\{\Ntr(t)  I(C\ge t, \bbeta^\intercal \bZ \le \bbeta^\intercal \bz + \delta)\}/\partial\balpha$, where $\delta$ is a constant. Under Model (1), we have $\operatorname{E}\{\Ntr(t)I(C\ge t)\mid\bZ \} = S_C(t\mid\bZ)I(\bZ\in\SZ) S_C(t\mid \bZ)g(\bgamma_0^\intercal \bZ) F^*(t,\bbeta_0^\intercal\bZ)$, where $F^*(t,x) = F(t,x) - F(\tau_0,x)$ and $S_C(t\mid\bZ) = P(C\ge t\mid\bZ)$.
Following the argument of \cite{sherman1993limiting}, it can be shown that
\begin{align*}
& \frac{\partial \operatorname{E}\{\Ntr(t)  I(C\ge t,\bbeta^\intercal \bZ \le \bbeta^\intercal \bz + \delta)\}}{\partial\balpha} \Big|_{\balpha = \balpha_0}   \\
= &  \operatorname{E}\{ (\bz -  \bZ)^\intercal I(\bZ\in\SZ) S_C(t\mid \bZ)g(\bgamma_0^\intercal \bZ) F^*(t,\bbeta_0^\intercal\bZ)   \mid \bbeta_0^\intercal\bZ = \bbeta_0^\intercal\bz + \delta  \}f_{\bbeta_0^\intercal\bZ}(\bbeta_0^\intercal \bz + \delta)J(\balpha_0).
\end{align*}
Then we have
\begin{eqnarray}
\label{l3:e1}
&& \hspace{0.2 in}\frac{\partial[ E\{\Ntr(t)  I(C\ge t)\mid\bbeta^\intercal\bZ = \bbeta^\intercal\bz \}f_{\bbeta^\intercal\bZ}(\bbeta^\intercal\bz)]}{\partial\balpha}\bigg|_{\balpha = \balpha_0}\\\nonumber
&= & \lim_{\delta\rightarrow 0}\frac{1}{\delta}\frac{\partial[ \operatorname{E}\{\Ntr(t)  I(C\ge t,\bbeta^\intercal \bZ \le \bbeta^\intercal \bz + \delta)\} - \operatorname{E}\{\Ntr(t)  I(C\ge t,\bbeta^\intercal \bZ \le \bbeta^\intercal \bz )\}]}{\partial\balpha} \bigg|_{\balpha = \balpha_0}   \\\nonumber
&=&  \{\eta^{*\intercal}(t,\bz)F^*(t,\bbeta_0^\intercal\bz)\dot f_{\bbeta_0^\intercal\bZ}(\bbeta_0^\intercal \bz  ) +  \dot \eta^{*\intercal}(t,\bz)F^*(t,\bbeta_0^\intercal\bz)f_{\bbeta_0^\intercal\bZ}(\bbeta_0^\intercal \bz  )\} J(\balpha_0)\\\nonumber
&&+ \eta^{*\intercal}(t,\bz)\dot F^*(t,\bbeta_0^\intercal\bz)f_{\bbeta_0^\intercal\bZ}(\bbeta_0^\intercal \bz  ) J(\balpha_0)
\end{eqnarray}  
where $\eta^*(t,\bz) = \bz q^{(0)*}(t,\bbeta_0^\intercal\bz) - q^{(1)*}(t,\bbeta_0^\intercal\bz) $, $\dot{f}_{\bbeta_0^\intercal\bZ}(x)$ denotes the derivative of ${f}_{\bbeta_0^\intercal\bZ}(x)$,  $\dot F^*(t,x)$ denotes the partial derivative of $F^*(t,x)$ with respect to $x$, $ \dot{\eta}^*(t,\bz) = \bz \dot q^{(0)*}(t,\bbeta_0^\intercal\bz) - \dot q^{(1)*}(t,\bbeta_0^\intercal\bz)$, and $\dot q^{(k)*}(t,x)$ is the partial derivative of $q^{(k)*}(t,x)$ with respect to $x$, $k = 0,1$. Similarly, we can also show that the result holds when $\Ntr(t)I(C\ge t)$ is replaced by $d\Ntr(t)$ and $F^*(t,\bbeta_0^\intercal\bz)$ is replaced by $f(t,\bbeta_0^\intercal\bz)dt$.

Under Model (1), we have
\begin{align}
\label{l3:e2}
 \operatorname{E}\{\Ntr(t)I(C\ge t)\mid   \bbeta_0^\intercal \bZ =& \bbeta_0^\intercal \bz \} = q^{(0)*}(t,\bbeta_0^\intercal\bz) F^*(t,\bbeta_0^\intercal\bz),\\
 \label{l3:e3}
\operatorname{E}\{d\Ntr(t) \mid   \bbeta_0^\intercal \bZ =& \bbeta_0^\intercal \bz \} = q^{(0)*}(t,\bbeta_0^\intercal\bz) f(t,\bbeta_0^\intercal\bz)dt. 
\end{align}
Applying \eqref{l3:e1}, \eqref{l3:e2}, \eqref{l3:e3}, and the chain rule, we have
{\small
\begin{eqnarray*}
&& {\frac{\partial [\operatorname{E}\{d\Ntr(t)  \mid \bbeta^\intercal \bZ = \bbeta^\intercal \bz  \}/\operatorname{E}\{\Ntr(t)I(C\ge t)\mid   \bbeta^\intercal \bZ = \bbeta^\intercal \bz \}]}{\partial\balpha} \bigg|_{\balpha = \balpha_0}}\\
&&=\;  \frac{\dot f(t,\bbeta_0^\intercal\bz) \eta^{*\intercal}(t,\bz) f_{\bbeta_0^\intercal\bZ}(\bbeta_0^\intercal \bz  )  + f(t,\bbeta_0^\intercal\bz)\dot\eta^{*\intercal}(t,\bz) f_{\bbeta_0^\intercal\bZ}(\bbeta_0^\intercal \bz  ) + f(t,\bbeta_0^\intercal\bz)\eta^{*\intercal}(t,\bz)\dot f_{\bbeta_0^\intercal\bZ}(\bbeta_0^\intercal \bz  ) }{F^*(t,\bbeta_0^\intercal\bz) q^{(0)*}(t,\bbeta_0^\intercal\bz) f_{\bbeta_0^\intercal\bZ}(\bbeta_0^\intercal \bz  ) }dt J(\balpha_0)\\
&&\;\;- f(t,\bbeta_0^\intercal\bz)\frac{\dot F^*(t,\bbeta_0^\intercal\bz) \eta^{*\intercal}(t,\bz) f_{\bbeta_0^\intercal\bZ}(\bbeta_0^\intercal \bz  )  + F^*(t,\bbeta_0^\intercal\bz)\dot\eta^{*\intercal}(t,\bz) f_{\bbeta_0^\intercal\bZ}(\bbeta_0^\intercal \bz  ) 
}{F^*(t,\bbeta_0^\intercal\bz)^2 q^{(0)*}(t,\bbeta_0^\intercal\bz) f_{\bbeta_0^\intercal\bZ}(\bbeta_0^\intercal \bz  )}dt J(\balpha_0)\\
&&\;\; - f(t,\bbeta_0^\intercal\bz)\frac{  F^*(t,\bbeta_0\bz)\eta^{*\intercal}(t,\bz) \dot f_{\bbeta_0^\intercal\bZ}(\bbeta_0^\intercal \bz  ) }{F^*(t,\bbeta_0^\intercal\bz)^2 q^{(0)*}(t,\bbeta_0^\intercal\bz) f_{\bbeta_0^\intercal\bZ}(\bbeta_0^\intercal \bz  )}dt J(\balpha_0)\\
&&=\; \dot r^*(t,\bbeta_0^\intercal\bz)
\left\{\bz -  \frac{q^{(1)*}(t,\bbeta_0^\intercal\bz)}{q^{(0)*}(t,\bbeta_0^\intercal\bz)}\right\}^\intercal J(\balpha_0)dt.
\end{eqnarray*}
}
This completes the proof of Lemma \ref{lemma_deriv1}.

\begin{lemma}
\label{lemma-gradient}
Under the conditions specified in Theorem 1, we have
\begin{align*}
 \frac{\partial l(\bbeta,\tr,\tR)}{\partial \balpha}\Big|_{\balpha=\balpha_0} =  \frac{1}{\sqrt{n}} W_n^\intercal J(\balpha_0) + o_p(n^{-1/2}),
\end{align*}
where
$$
\ell(\bbeta,\tr,\tR) = \frac{1}{n}\sum_{i=1}^{n}\int_{\tau_0}^{\tau_1} \left\{ \log \widetilde{r}_h( t,\bbeta^\intercal \bZ_i,\bbeta) -  \widetilde{R}_h(t,\bbeta^\intercal \bZ_i,\bbeta) +  \widetilde{R}_h(C_i,\bbeta^\intercal \bZ_i,\bbeta)  \right\}I(\bZ_i \in \SZ)dN_i(t), 
$$
$$
W_n = \frac{1}{\sqrt{n}} \sumi \int_{\tau_0}^{\tau_1} \frac{\dot{r}^*(t,\bbeta_0^\intercal \bZ_i)}{r^*(t,\bbeta_0^\intercal \bZ_i)} \left \{\bZ_i-  \frac{q^{(1)*}(t,\bbeta_0^\intercal \bZ_i)}{q^{(0)*}(t,\bbeta_0^\intercal \bZ_i)} \right\} d M_i^*(t),
$$
with $dM_i^*(t)=  I(\bZ_i\in\SZ) dN_i(t) - I(C_i\ge t, \bZ_i\in\SZ) N_i(t)r^*(t,\bbeta_0^\intercal \bZ_i)dt $.
\end{lemma}

{\it Proof of Lemma \ref{lemma-gradient}.} 
Define 
$$
\ell(\bbeta,r^*,R^*) = \frac{1}{n}\sum_{i=1}^{n}\int_{\tau_0}^{\tau_1} \left\{ \log r^*( t,\bbeta^\intercal \bZ_i,\bbeta) -  R^*(t,\bbeta^\intercal \bZ_i,\bbeta) +  R^*(C_i,\bbeta^\intercal \bZ_i,\bbeta)  \right\}I(\bZ_i \in \SZ)dN_i(t).
$$
By Lemma \ref{lemma_deriv1}, we have 
$\partial l(\bbeta,r^*,R^*)/\partial \balpha|_{\balpha=\balpha_0} =  n^{-1/2} W_n^\intercal J(\balpha_0).$
Therefore, Lemma \ref{lemma-gradient} can be proved by proving 
$\partial l(\bbeta,\tr,\tR)/\partial \balpha|_{\balpha=\balpha_0} - \partial l(\bbeta,r^*,R^*)/\partial \balpha|_{\balpha=\balpha_0}=o_p(n^{-1/2}).$
It is easy to see  
$$
\frac{\partial \ell(\bbeta,\tr,\tR)}{\partial \balpha} \Big|_{\balpha=\balpha_0} = \frac{\partial \ell_1(\bbeta,\tr)}{\partial \balpha} \Big|_{\balpha=\balpha_0} + \frac{\partial \ell_2(\bbeta,\tR)}{\partial \balpha} \Big|_{\balpha=\balpha_0},
$$
where 
$$
\ell_1(\bbeta,\tr) = \frac{1}{n}\sum_{i=1}^{n}\int_{\tau_0}^{\tau_1}  \log \widetilde{r}_h( t,\bbeta^\intercal \bZ_i,\bbeta)  I(\bZ_i \in \SZ)dN_i(t),
$$
$$
\ell_2(\bbeta,\tR) = \frac{1}{n}\sum_{i=1}^{n}\int_{\tau_0}^{\tau_1} \left\{ -  \widetilde{R}_h(t,\bbeta^\intercal \bZ_i,\bbeta) +  \widetilde{R}_h(C_i,\bbeta^\intercal \bZ_i,\bbeta)  \right\}I(\bZ_i \in \SZ)dN_i(t).
$$
Similarly, we have 
$$
\frac{\partial \ell(\bbeta,r^*,R^*)}{\partial \balpha} \Big|_{\balpha=\balpha_0} = \frac{\partial \ell_1(\bbeta,r^*)}{\partial \balpha} \Big|_{\balpha=\balpha_0} + \frac{\partial \ell_2(\bbeta,R^*)}{\partial \balpha} \Big|_{\balpha=\balpha_0},
$$
where $\ell_1(\bbeta,r^*)$ is defined by replacing $\tr$ with $r^*$ in $\ell_1(\bbeta,\tr)$, and $\ell_2(\bbeta,R^*)$ is defined by replacing $\tR$ with $R^*$ in $\ell_2(\bbeta,\tR)$.
We first study $\partial \ell_1(\bbeta,\tr)/\partial \balpha|_{\balpha=\balpha_0}-\partial \ell_1(\bbeta,r^*)/\partial \balpha|_{\balpha=\balpha_0}$. 
Note that
$$
\tr(t,\bbeta^\intercal\bZ,\bbeta)=\frac{\hf^*(t,\bbeta^\intercal\bZ,\bbeta)}{\hff^*(t,\bbeta^\intercal\bZ,\bbeta)},\;\; r^*(t,\bbeta^\intercal\bZ,\bbeta)=\frac{f_1^*(t,\bbeta^\intercal\bZ,\bbeta)}{f_2^*(t,\bbeta^\intercal\bZ,\bbeta)},
$$
where $\widehat f_k^*$ and $f_k^*$, $k=1,2$, are defined in the proof of Lemma \ref{lemmatr}. Therefore, we have 
\begin{eqnarray}
\label{Delta1234}
\frac{\partial \ell_1(\bbeta,\tr)}{\partial \balpha}\Big|_{\balpha=\balpha_0}-\frac{\partial \ell_1(\bbeta,r^*)}{\partial \balpha}\Big|_{\balpha=\balpha_0} = \Delta_1 - \Delta_2 + \Delta_3 - \Delta_4 + o_p(n^{-1/2}),  
\end{eqnarray}
where 
\begin{eqnarray*}
\Delta_1 &=& \frac{1}{n} \sumi \int_{\tau_0}^{\tau_1} \frac{\partial \hf^*(t,\bbeta^\intercal\bZ_i,\bbeta)/\partial\balpha|_{\balpha=\balpha_0} }{f_1^*(t,\bbeta_0^\intercal\bZ_i)} I(\bZ_i\in\SZ) dN_i(t),
\\
\Delta_2 &=& \frac{1}{n} \sumi \int_{\tau_0}^{\tau_1} \frac{\partial f_1^*(t,\bbeta^\intercal\bZ_i,\bbeta)/\partial\balpha|_{\balpha=\balpha_0} \cdot \hf^*(t,\bbeta_0^\intercal\bZ_i) }{\{f_1^*(t,\bbeta_0^\intercal\bZ_i)\}^2}  I(\bZ_i\in\SZ) dN_i(t),
\\
\Delta_3 &=& \frac{1}{n} \sumi \int_{\tau_0}^{\tau_1} \frac{\partial \hff^*(t,\bbeta^\intercal\bZ_i,\bbeta)/\partial\balpha|_{\balpha=\balpha_0} }{f_2^*(t,\bbeta_0^\intercal\bZ_i)} I(\bZ_i\in\SZ) dN_i(t),
\\
\Delta_4 &=& \frac{1}{n} \sumi \int_{\tau_0}^{\tau_1} \frac{\partial f_2^*(t,\bbeta^\intercal\bZ_i,\bbeta)/\partial\balpha|_{\balpha=\balpha_0} \hff^*(t,\bbeta_0^\intercal\bZ_i)}{\{f_2^*(t,\bbeta_0^\intercal\bZ_i)\}^2} I(\bZ_i\in\SZ) dN_i(t).
\end{eqnarray*}
Define 
\begin{eqnarray*}
w_1 = \frac{1}{n}\sumi \int_{\tau_0}^{\tau_1} \left\{ \bZ_i - \frac{q^{(1)*}(t,\bbeta_0^\intercal\bZ_i)}{q^{(0)*}(t,\bbeta_0^\intercal\bZ_i)}\right\}^\intercal \frac{\dot f_{\bbeta_0^\intercal\bZ}(\bbeta_0^\intercal\bZ_i)}{f_{\bbeta_0^\intercal\bZ}(\bbeta_0^\intercal\bZ_i)} I(\bZ_i\in\SZ) dN_i(t) J(\balpha_0),
\end{eqnarray*}
\begin{eqnarray*}
w_2 = \frac{1}{n}\sumi \int_{\tau_0}^{\tau_1} \left\{ \bZ_i - \frac{q^{(1)*}(t,\bbeta_0^\intercal\bZ_i)}{q^{(0)*}(t,\bbeta_0^\intercal\bZ_i)}\right\}^\intercal \frac{\dot f(t,\bbeta_0^\intercal\bZ_i)}{f(t,\bbeta_0^\intercal\bZ_i)} I(\bZ_i\in\SZ) dN_i(t) J(\balpha_0),
\end{eqnarray*}
\begin{eqnarray*}
w_3 = \frac{1}{n}\sumi \int_{\tau_0}^{\tau_1} \frac{\{\bZ_i \dot q^{(0)*}(t,\bbeta_0^\intercal\bZ_i)- \dot q^{(1)*}(t,\bbeta_0^\intercal\bZ_i)\}^\intercal}{ q^{(0)*}(t,\bbeta_0^\intercal\bZ_i)} I(\bZ_i\in\SZ) dN_i(t) J(\balpha_0),
\end{eqnarray*}
\begin{eqnarray*}
w_4 = \frac{1}{n}\sumi \int_{\tau_0}^{\tau_1} \dot{\mathcal E}(t,\bbeta_0^\intercal\bZ_i)^\intercal I(\bZ_i\in\SZ) dN_i(t) J(\balpha_0),
\end{eqnarray*}
\begin{eqnarray*}
w_5 = \frac{1}{n}\sumi \int_{\tau_0}^{\tau_1} \dot{\mathcal E}(t,\bbeta_0^\intercal\bZ_i)^\intercal r^*(t,\bbeta_0^\intercal\bZ_i) I(C_i\ge t, \bZ_i\in\SZ) N_i(t) dt J(\balpha_0),
\end{eqnarray*}
\begin{eqnarray*}
w_6 = \frac{1}{n}\sumi \int_{\tau_0}^{\tau_1} \left\{ \bZ_i - \frac{q^{(1)*}(t,\bbeta_0^\intercal\bZ_i)}{q^{(0)*}(t,\bbeta_0^\intercal\bZ_i)}\right\}^\intercal \frac{\dot F^*(t,\bbeta_0^\intercal\bZ_i)}{F^*(t,\bbeta_0^\intercal\bZ_i)} I(\bZ_i\in\SZ) dN_i(t) J(\balpha_0),
\end{eqnarray*}
\begin{eqnarray*}
w_7 = \frac{1}{n} \sumi \int_{\tau_0}^{\tau_1} \frac{\dot r^*(t,\bbeta_0^\intercal \bZ_i)}{r^*(t,\bbeta_0^\intercal\bZ_i)}  \left\{ \bZ_i - \frac{q^{(1)*}(t,\bbeta_0^\intercal\bZ_i)}{q^{(0)*}(t,\bbeta_0^\intercal\bZ_i)} \right\}^\intercal I(C_i\ge t, \bZ_i\in\SZ) N_i(t)r^*(t,\bbeta_0^\intercal \bZ_i)dt J(\balpha_0),
\end{eqnarray*}
where $\dot{\mathcal E}(t,x)$ denotes the partial derivative of $q^{(1)*}(t,x)/q^{(0)*}(t,x)$ with respect to $x$.
Applying Lemma 3.1 in \cite{powell1989semiparametric} yields
\begin{eqnarray*}
\Delta_1 &=& w_1+w_2+w_3+w_4+o_p(n^{-1/2}), \\
\Delta_2 &=& w_1+w_2+w_3+w_4+o_p(n^{-1/2}),\\
\Delta_3 &=& w_1 -w_3 +w_5+w_6-w_7+o_p(n^{-1/2}),\\
\Delta_4 &=& w_1 -w_3 +w_5+w_6+o_p(n^{-1/2}).
\end{eqnarray*}
It follows from \eqref{Delta1234} that
\begin{eqnarray*}
&&\frac{\partial \ell_1(\bbeta,\tr)}{\partial \balpha}\Big|_{\balpha=\balpha_0}-\frac{\partial \ell_1(\bbeta,r^*)}{\partial \balpha}\Big|_{\balpha=\balpha_0} \\\nonumber
&&= -\frac{1}{n} \sumi \int_{\tau_0}^{\tau_1} \frac{\dot r^*(t,\bbeta_0^\intercal \bZ_i)}{r^*(t,\bbeta_0^\intercal\bZ_i)}  \left\{ \bZ_i - \frac{q^{(1)*}(t,\bbeta_0^\intercal\bZ_i)}{q^{(0)*}(t,\bbeta_0^\intercal\bZ_i)} \right\}^\intercal I(C_i\ge t, \bZ_i\in\SZ) N_i(t)r^*(t,\bbeta_0^\intercal \bZ_i)dt J(\balpha_0)\\\nonumber
&& + o_p(n^{-1/2}).
\end{eqnarray*}
In a similar way, we can prove that
\begin{eqnarray*}
&&\frac{\partial \ell_2(\bbeta,\tR)}{\partial \balpha}\Big|_{\balpha=\balpha_0}-\frac{\partial \ell_2(\bbeta,R^*)}{\partial \balpha}\Big|_{\balpha=\balpha_0} \\\nonumber
&&= \frac{1}{n} \sumi \int_{\tau_0}^{\tau_1} \frac{\dot r^*(t,\bbeta_0^\intercal \bZ_i)}{r^*(t,\bbeta_0^\intercal\bZ_i)}  \left\{ \bZ_i - \frac{q^{(1)*}(t,\bbeta_0^\intercal\bZ_i)}{q^{(0)*}(t,\bbeta_0^\intercal\bZ_i)} \right\}^\intercal I(C_i\ge t, \bZ_i\in\SZ) N_i(t)r^*(t,\bbeta_0^\intercal \bZ_i)dt J(\balpha_0)\\
&& \;\;+ o_p(n^{-1/2}).
\end{eqnarray*}
Therefore, we can derive
$$
\frac{\ell(\bbeta,\tr,\tR)}{\partial \balpha}\Big|_{\balpha=\balpha_0}-\frac{\partial \ell(\bbeta,r^*,R^*)}{\partial \balpha}\Big|_{\balpha=\balpha_0} = o_p(n^{-1/2}).
$$
This completes the proof of Lemma \ref{lemma-gradient}.

\section{Proof of Theorem 1}


Define $F_0(\bz,c,t)=\operatorname{E}\{I(C\le c) N(t) I(\bZ\le \bz)\}$, and let $\hF_0(\bz,c,t)$ denote the corresponding empirical estimate. Define 
$$
\ell_0(\bbeta,r^*,R^*,u)=\int_{u}^{\tau_1} \int_0^\tau \int_{\bz \in\SZ}  \left\{ \log r^*(t,\bbeta^\intercal\bz, \bbeta) -  R^*(t,\bbeta^\intercal \bz,\bbeta) + R^*(c,\bbeta^\intercal \bz,\bbeta) \right\} d F_0(\bz,c,t),
$$ 
\begin{eqnarray}
\label{objective-function}
 \ell (\bbeta,\tr,\tR,u) \! = \! \int_{u}^{\tau_1}\! \int_0^\tau \!\int_{\bz \in\SZ} \! \left\{ \log \tr( t,\bbeta^\intercal \bz,\bbeta) \!- \! \tR(t,\bbeta^\intercal \bz,\bbeta) + \tR(c,\bbeta^\intercal \bz,\bbeta)  \right\} d \hF_0(\bz,c,t).
\end{eqnarray}
Let $\ell (\bbeta,\tr,\tR)=\ell (\bbeta,\tr,\tR,\tau_0)$ and $ \ell_0 (\bbeta,r^*,R^*)=\ell_0 (\bbeta,r^*,R^*,\tau_0).$  We first prove the consistency of the estimator $\hbeta$. This 
can be proved by showing that
$
\halpha=\arg\max_{\balpha \in\mA} \ell (\bbeta,\tr,\tR)
$
is a consistent estimator of $\balpha_0=\arg\max_{\balpha \in\mA} \ell_0 (\bbeta,r^*,R^*)$. Define $dM^*(t) = I(\bZ\in\SZ)dN(t) - I(C\ge t, \bZ\in\SZ) N(t)r^*(t,\bbeta_0^\intercal \bZ)dt$. It is easy to see $\operatorname{E}\{dM^*(t)\mid\bZ\} = 0$. Therefore, applying Lemma \ref{lemma_deriv1} yields 
$$
\frac{\ell_0 (\bbeta,r^*,R^*)}{\partial \balpha} \Big |_{\balpha=\balpha_0}= 
\operatorname{E} \left[ \int_{\tau_0}^{\tau_1}  \frac{\dot{r}^*(t,\bbeta_0^\intercal \bZ)}{r^*(t,\bbeta_0^\intercal \bZ)} \left \{\bZ-  \frac{q^{(1)*}(t,\bbeta_0^\intercal \bZ)}{q^{(0)*}(t,\bbeta_0^\intercal \bZ)} \right\}^\intercal d M^*(t)\right] J(\balpha_0)=\bm 0.
$$
By Lemma \ref{lemma_deriv1}, it can be shown that 
$$
\frac{\partial^2 \ell_0 (\bbeta,r^*,R^*)}{\partial \balpha \partial \balpha^\intercal}\Big|_{\balpha=\balpha_0} = J(\balpha_0)^\intercal V J(\balpha_0),
$$
where
\begin{eqnarray}
\label{matrix-V}
V= -\operatorname{E} \left[\int_{\tau_0}^{\tau_1} \left\{\frac{\dot r^*(t,\bbeta_0^\intercal \bZ)} {r^*(t,\bbeta_0^\intercal\bZ)}\right\}^2 \left \{\bZ-  \frac{q^{(1)*}(t,\bbeta_0^\intercal \bZ)}{q^{(0)*}(t,\bbeta_0^\intercal \bZ)} \right\}^{\otimes 2} I(\bZ\in\SZ)   dN(t) \right].
\end{eqnarray}
Consequently, we have $\sup_{\balpha: \|\balpha-\balpha_0\|\geq \epsilon} \ell_0(\bbeta,r^*,R^*)<\ell_0(\bbeta_0,r^*,R^*)$ for $\epsilon>0$. 
Moreover, we need to show that $\sup_{\balpha \in \mA}| \ell(\bbeta,\tr,\tR) - \ell_0(\bbeta,r^*,R^*)|$ converges in probability to zero as $n\to \infty$.
Under condition (C5), following Example 19.7 and Theorem 19.4 of \cite{van2000Asymptotic}, it can be shown that the class of functions 
$\{\int_{\tau_0}^{\tau_1} \log r^*(t,\bbeta^\intercal \bZ,\bbeta) d N(t)-\int_{\tau_0}^{\tau_1} \int_0^\tau I(t\le u\le C)r^*(u,\bbeta^\intercal\bZ,\bbeta) dudN(t): \balpha \in \mA \}$ is Glivenko--Cantelli. As a consequence, $\sup_{\balpha \in \mA}|\ell(\bbeta,r^*,R^*) - \ell_0(\bbeta,r^*, R^*)|$ converges in probability to zero as $n\to \infty$, where $\ell(\bbeta,r^*,R^*)$ is defined by replacing $\tr$, $\tR$, and $u$ with $r^*$, $R^*$, and $\tau_0$ in  \eqref{objective-function}, respectively.  By Lemma \ref{lemmatr}, we can derive 
$
\sup_{t\in[\tau_0^*,\tau_1], \bz \in\SZ, h\in\mH_1, \balpha \in \mA } |\tr(t,\bbeta^\intercal\bz, \bbeta) -r^*(t,\bbeta^\intercal\bz, \bbeta)|=o_p(1).
$
In a similar way,  we have $
\sup_{t\in[\tau_0^*,\tau_1], \bz \in\SZ, h\in\mH_1, \balpha\in\mA } |\tR(t,\bbeta^\intercal\bz,\bbeta) -R^*(t,\bbeta^\intercal\bz,\bbeta)|=o_p(1).
$
Applying the Continuous Mapping Theorem yields
$
\sup_{\balpha \in \mA} |\ell(\bbeta,\tr,\tR,\tau_0^*) - \ell(\bbeta,r^*,R^*,\tau_0^*)| = o_p(1).
$
By letting $\tau_0^*\to \tau_0$, we can show $
\sup_{\balpha \in \mA} |\ell(\bbeta,\tr,\tR) - \ell(\bbeta,r^*,R^*)| = o_p(1).
$
Hence, we have
$\sup_{\balpha \in \mA}|\ell(\bbeta,\tr, \tR) - \ell_0(\bbeta,r^*,R^*)| 
 \le \sup_{\balpha \in \mA}|\ell(\bbeta,r^*, R^*) - \ell_0(\bbeta,r^*,R^*)| + \sup_{\balpha \in \mA}| \ell(\bbeta,\tr, \tR) - \ell(\bbeta,r^*,R^*)| 
=o_p(1).
$
The consistency of $\halpha$ is proved.

Next, we establish the asymptotic normality for $\halpha$ and thus the asymptotic normality for $\hbeta$ can be obtained.
If  $r^*$ and $R^*$ are known, uniformly over $o_p(1)$ neighborhoods of $\balpha_0$, we have  
\begin{eqnarray}
\label{elldiff1}
\ell(\bbeta,r^*,R^*,\tau_0^*) - \ell(\bbeta_0,r^*,R^*,\tau_0^*) &=& \ell_0(\bbeta,r^*,R^*,\tau_0^*) - \ell_0(\bbeta_0,r^*,R^*,\tau_0^*)\\\nonumber
&& + O_p\left( n^{-1/2}\|\balpha-\balpha_0\| \right) + o_p(\|\balpha-\balpha_0\|^2),
\end{eqnarray}
where $\ell(\bbeta,r^*,R^*,\tau_0^*)$ is defined by replacing $\tr$, $\tR$, and $u$ with $r^*$, $R^*$, and $\tau_0^*$ in  \eqref{objective-function}, respectively.
We now study the function $\ell(\bbeta,\tr,\tR,\tau_0^*)$, which is defined by \eqref{objective-function} with $u=\tau_0^*$. 
Our goal is to prove  
\begin{eqnarray*}
\ell(\bbeta,\tr,\tR,\tau_0^*) - \ell(\bbeta_0,\tr,\tR,\tau_0^*) &=&\ell(\bbeta,r^*,R^*,\tau_0^*)   + \ell(\bbeta_0,r^*,R^*,\tau_0^*) \\ 
&&+ O_p(n^{-1/2}\|\balpha-\balpha_0\|) + o_p(\|\balpha-\balpha_0\|^2).
\end{eqnarray*}
Applying Taylor’s expansion yields
\begin{eqnarray*}
\log \frac{\tr(t,\bbeta^\intercal\bz, \bbeta)}{r^*(t,\bbeta^\intercal\bz, \bbeta)} &=& \frac{\tr(t,\bbeta^\intercal\bz, \bbeta) - r^*(t,\bbeta^\intercal\bz, \bbeta) }{r^*(t,\bbeta^\intercal\bz, \bbeta)} - \frac{ \{\tr(t,\bbeta^\intercal\bz, \bbeta) - r^*(t,\bbeta^\intercal\bz, \bbeta)\}^2 } { 2 \{r^*(t,\bbeta^\intercal\bz, \bbeta)\}^2 }\\
&&+\frac{\{\tr(t,\bbeta^\intercal\bz, \bbeta) - r^*(t,\bbeta^\intercal\bz, \bbeta)\}^3}{3\bar{r}^3(t,\bbeta^\intercal\bz, \bbeta)},
\end{eqnarray*} 
where $\bar{r}(t,\bbeta^\intercal\bz, \bbeta)$ lies between $\tr(t,\bbeta^\intercal\bz, \bbeta)$ and $r^*(t,\bbeta^\intercal\bz, \bbeta)$. 
This leads to
\begin{eqnarray}
\label{diff-ell}
&& \ell(\bbeta,\tr,\tR,\tau_0^*) - \ell(\bbeta,r^*,R^*,\tau_0^*) 
= A_1(\balpha) - A_2(\balpha) + A_3(\balpha) - A_4(\balpha),
\end{eqnarray}
where 
\begin{eqnarray*}
A_1(\balpha) &=& \int_{\tau_0^*}^{\tau_1} \int_0^\tau \int_{\bz \in\SZ}    \frac{\tr(t,\bbeta^\intercal\bz, \bbeta) - r^*(t,\bbeta^\intercal\bz, \bbeta) }{r^*(t,\bbeta^\intercal\bz, \bbeta)} d\hF_0(\bz,c,t),\\
A_2(\balpha) &=& \int_{\tau_0^*}^{\tau_1} \int_0^\tau \int_{\bz \in\SZ} \frac{ \{\tr(t,\bbeta^\intercal\bz, \bbeta) - r^*(t,\bbeta^\intercal\bz, \bbeta)\}^2}{2 \{r^*(t,\bbeta^\intercal\bz, \bbeta)\}^2 }  d\hF_0(\bz,c,t),\\
A_3(\balpha) &=& \int_{\tau_0^*}^{\tau_1} \int_0^\tau \int_{\bz \in\SZ}  \frac{ \{\tr(t,\bbeta^\intercal\bz, \bbeta) - r^*(t,\bbeta^\intercal\bz, \bbeta)\}^3}{3 \bar{r}^{3}(t,\bbeta^\intercal\bz, \bbeta)} d\hF_0(\bz,c,t),\\
A_4(\balpha) &=&\int_{\tau_0^*}^{\tau_1} \int_0^\tau \int_{\bz \in\SZ} \{\tR(t,\bbeta^\intercal\bz, \bbeta) -R^*(t,\bbeta^\intercal\bz, \bbeta)\} d\hF_0(\bz,c,t)\\
&&+ \int_{\tau_0^*}^{\tau_1} \int_0^\tau \int_{\bz \in\SZ} \{\tR(c,\bbeta^\intercal\bz, \bbeta) -R^*(c,\bbeta^\intercal\bz, \bbeta)\} d\hF_0(\bz,c,t).
\end{eqnarray*}
Uniformly over a $o_p(1)$ neighborhood of $\balpha_0$, applying Taylor's expansion leads to
\begin{eqnarray}
\label{diffr}
\hspace{0.3 in}\tr(t,\bbeta^\intercal\bz, \bbeta) -r^*(t,\bbeta^\intercal\bz, \bbeta) 
&=& \tr(t,\bbeta_0^\intercal\bz) -r^*(t,\bbeta_0^\intercal\bz)  + B_1(\balpha_0)(\balpha-\balpha_0)\\\nonumber
&& + (\balpha-\balpha_0)^\intercal B_2(\balpha_0) (\balpha-\balpha_0) + o_p(\|\balpha-\balpha_0\|^2),
\end{eqnarray}
where 
\begin{eqnarray*}
B_1(\balpha_0) &=& \frac{\partial \tr(t,\bbeta^\intercal\bz, \bbeta)}{\partial \balpha}\big|_{\balpha=\balpha_0} - \frac{\partial r^*(t,\bbeta^\intercal\bz, \bbeta)} {\partial \balpha}\big|_{\balpha=\balpha_0},\\
B_2(\balpha_0) &=& \frac{1}{2} \left( \frac{\partial^2 \tr(t,\bbeta^\intercal\bz, \bbeta)}{\partial \balpha \partial \balpha^\intercal}\big|_{\balpha=\balpha_0} - \frac{\partial^2 r^*(t,\bbeta^\intercal\bz, \bbeta)} {\partial \balpha \partial \balpha^\intercal}\big|_{\balpha=\balpha_0}\right).
\end{eqnarray*}
We first study $A_{1}(\balpha)$. By \eqref{diffr}, we can derive
\begin{eqnarray}
\label{A1}
A_1(\balpha) &=& A_{10}(\balpha) + A_{11}(\balpha) (\balpha-\balpha_0) + (\balpha-\balpha_0)^\intercal A_{12}(\balpha) (\balpha-\balpha_0) \\\nonumber &&+ o_p(\|\balpha-\balpha_0\|^2),
\end{eqnarray} 
where
\begin{eqnarray*}
A_{10}(\balpha) &=&\int_{\tau_0^*}^{\tau_1} \int_0^\tau \int_{\bz \in\SZ}  \frac{ \tr(t,\bbeta_0^\intercal\bz) -r^*(t,\bbeta_0^\intercal\bz) }{r^*(t,\bbeta^\intercal\bz, \bbeta) } d\hF_0(\bz,c,t),\\
A_{11}(\balpha) &=& \int_{\tau_0^*}^{\tau_1} \int_0^\tau \int_{\bz \in\SZ}  \frac{B_1(\balpha_0)}{r^*(t,\bbeta^\intercal\bz, \bbeta) } d\hF_0(\bz,c,t),\\
A_{12}(\balpha) &=& \int_{\tau_0^*}^{\tau_1} \int_0^\tau \int_{\bz \in\SZ}  \frac{ B_2(\balpha_0)}{2 r^*(t,\bbeta^\intercal\bz, \bbeta) } d\hF_0(\bz,c,t).
\end{eqnarray*}
In the following, we study $A_{10}(\balpha)$, $A_{11}(\balpha)$, and $A_{12}(\balpha)$ in turn.
For $\balpha$ satisfying $\|\balpha-\balpha_0\|=o_p(1)$, applying Taylor's expansion of $A_{10}(\balpha)$ at $\balpha_0$ leads to
\begin{eqnarray*}
A_{10}(\balpha) 
&=& A_{10}(\balpha_0) + \frac{\partial A_{10}(\balpha)}{ \partial \balpha}\big|_{\balpha=\balpha_0} (\balpha-\balpha_0) + \frac{1}{2}(\balpha-\balpha_0)^\intercal \frac{\partial^2 A_{10}(\balpha)}{ \partial \balpha \partial \balpha^\intercal}\big|_{\balpha=\balpha_0} (\balpha-\balpha_0) \\ &&+  o_p(\|\balpha-\balpha_0\|^2).
\end{eqnarray*}
It follows from Lemma \ref{lemmatr} that
\begin{eqnarray*}
&&\frac{\partial A_{10}(\balpha)}{ \partial \balpha} \big|_{\balpha=\balpha_0}\\
&=& -\int_{\tau_0^*}^{\tau_1} \int_0^\tau \int_{\bz \in\SZ} \frac{\frac{ \partial r^*(t,\bbeta^\intercal\bz, \bbeta)}{ \partial \balpha}|_{\balpha=\balpha_0}  \left\{ \tr(t,\bbeta_0^\intercal\bz) -r^*(t,\bbeta_0^\intercal\bz) \right\}}{\{r^{*}(t,\bbeta_0^\intercal\bz)\}^2} d(\hF_0-F_0)(\bz,c,t) \\
&&-\int_{\tau_0^*}^{\tau_1} \int_0^\tau \int_{\bz \in\SZ}
\frac{\frac{\partial r^*(t,\bbeta^\intercal\bz, \bbeta)}{\partial \balpha}|_{\balpha=\balpha_0} \left\{ \tr(t,\bbeta_0^\intercal\bz) -r^*(t,\bbeta_0^\intercal\bz) \right\}}{\{r^{*}(t,\bbeta_0^\intercal\bz)\}^2 } dF_0(\bz,c,t)\\
&=& O_p(n^{-1/2}).
\end{eqnarray*}
This leads to $\partial A_{10}(\balpha)/\partial \balpha|_{\balpha=\balpha_0}(\balpha-\balpha_0)=O_p(n^{-1/2}\|\balpha-\balpha_0\|)$. In a similar way, we can prove that $(\balpha-\balpha_0)^\intercal \partial^2 A_{10}(\balpha)/ \partial \balpha \partial \balpha^\intercal\big|_{\balpha=\balpha_0} (\balpha-\balpha_0)=o_p(\|\balpha-\balpha_0\|^2)$. 
Therefore, it can be derived that 
\begin{eqnarray}
\label{A10}
A_{10}(\balpha) 
= A_{10}(\balpha_0) + O_p(n^{-1/2}\|\balpha-\balpha_0\|) + o_p(\|\balpha-\balpha_0\|^2).
\end{eqnarray}
Applying Taylor's expansion of $A_{11}(\balpha)$ at $\balpha_0$ yields
$$
A_{11}(\balpha) = A_{11}(\balpha_0) + \frac{\partial A_{11}(\balpha)}{ \partial \balpha} \Big|_{\balpha=\balpha_0} (\balpha-\balpha_0) + o_p(\|\balpha-\balpha_0\|).
$$
Applying the techniques in studying $A_{10}(\balpha)$ yields $A_{11}(\balpha_0)=O_p(n^{-1/2})$ and $ \|\partial A_{11}(\balpha)/ \partial \balpha |_{\balpha=\balpha_0}\| =o_p(1)$. This leads to
\begin{eqnarray}
\label{A11}
A_{11}(\balpha)(\balpha-\balpha_0) = O_p(n^{-1/2}\|\balpha-\balpha_0\|)+ o_p(\|\balpha-\balpha_0\|^2).
\end{eqnarray}
Moreover, it can be shown that $\|A_{12}(\balpha)\|=o_p(1)$ for $\balpha$ satisfying $\|\balpha-\balpha_0\|=o_p(1)$ due to Equation \eqref{tr2} in Lemma \ref{lemmatr}. 
It follows from \eqref{A1}, \eqref{A10}, and \eqref{A11} that
\begin{eqnarray}
\label{eq-A1}
A_1(\balpha) = A_{10}(\balpha_0) + O_p(n^{-1/2}\|\balpha-\balpha_0\|) + o_p(\|\balpha-\balpha_0\|^2).
\end{eqnarray}

We then study $A_2(\balpha)$. It follows from Lemma \ref{lemmatr} that $B_1(\balpha_0)=o_p(1)$ and $B_2(\balpha_0)=o_p(1)$. Therefore, by  \eqref{diffr}, we have
\begin{eqnarray*}
A_2(\balpha) = A_{20}(\balpha) + A_{21}(\balpha)(\balpha-\balpha_0)+ o_p(\|\balpha-\balpha_0\|^2),
\end{eqnarray*}
where 
\begin{eqnarray*}
A_{20}(\balpha) &=& \int_{\tau_0^*}^{\tau_1} \int_0^\tau \int_{\bz \in\SZ} \frac{ \{\tr(t,\bbeta_0^\intercal\bz) -r^*(t,\bbeta_0^\intercal\bz)\}^2}{2 \{r^*(t,\bbeta^\intercal\bz, \bbeta)\}^2 } d \hF_0(\bz,c,t),\\
A_{21}(\balpha) &=& \int_{\tau_0^*}^{\tau_1} \int_0^\tau \int_{\bz \in\SZ} \frac{\{\tr(t,\bbeta_0^\intercal\bz) -r^*(t,\bbeta_0^\intercal\bz)\}B_1(\balpha_0)} { \{r^*(t,\bbeta^\intercal\bz, \bbeta)\}^2 } d \hF_0(\bz,c,t).
\end{eqnarray*}
For $\balpha$ satisfying $\|\balpha-\balpha_0\|=o_p(1)$, applying similar techniques in studying $A_{1}(\balpha)$ yields
$
A_{20}(\balpha) = A_{20}(\balpha_0) + O_p(n^{-1/2}\|\balpha-\balpha_0\|) + o_p(\|\balpha-\balpha_0\|^2)
$ 
and
$
A_{21}(\balpha) = A_{21}(\balpha_0) + o_p(\|\balpha-\balpha_0\|) = O_p(n^{-1/2}) + o_p(\|\balpha-\balpha_0\|).
$
Therefore, it can be derived that
\begin{eqnarray}
\label{eq-A2}
A_2(\balpha) = A_{20}(\balpha_0) + O_p(n^{-1/2}\|\balpha-\balpha_0\|) + o_p(\|\balpha-\balpha_0\|^2).
\end{eqnarray}

Finally, we study $A_3(\balpha)$ and $A_4(\balpha)$. Based on Lemma \ref{lemmatr}, applying similar techniques in studying $A_1(\balpha)$ and $A_2(\balpha)$, we can derive
\begin{eqnarray}
\label{eq-A34}
A_k(\balpha) = A_k(\balpha_0) + O_p(n^{-1/2}\|\balpha-\balpha_0\|) + o_p(\|\balpha-\balpha_0\|^2),~k=3,4.
\end{eqnarray}

It follows from \eqref{diff-ell}, \eqref{eq-A1}, \eqref{eq-A2}, and \eqref{eq-A34} that
\begin{eqnarray*}
\ell(\bbeta,\tr, \tR,\tau_0^*) - \ell(\bbeta,r^*,R^*,\tau_0^*) &=& A_{10}(\balpha_0) -
 A_{20}(\balpha_0)+ A_3(\balpha_0) - A_{4}(\balpha_0) \\&& + O_p(n^{-1/2}\|\balpha-\balpha_0\|) + o_p(\|\balpha-\balpha_0\|^2).
\end{eqnarray*}
Since $A_{10}(\balpha_0) -
 A_{20}(\balpha_0)+ A_3(\balpha_0) - A_{4}(\balpha_0)=\ell(\bbeta_0,\tr,\tR,\tau_0^*) - \ell(\bbeta_0,r^*,R^*,\tau_0^*)$, we have
\begin{eqnarray}
\label{elldiff2}
\ell(\bbeta,\tr,\tR,\tau_0^*) - \ell(\bbeta,r^*,R^*,\tau_0^*) &=& \ell(\bbeta_0,\tr,\tR,\tau_0^*) - \ell(\bbeta_0,r^*,R^*,\tau_0^*) \\\nonumber 
&& + O_p(n^{-1/2}\|\balpha-\balpha_0\|) + o_p(\|\balpha-\balpha_0\|^2).
\end{eqnarray}
By \eqref{elldiff1} and \eqref{elldiff2}, uniformly over $o_p(1)$ neighborhoods of $\balpha_0$, we have
\begin{eqnarray*}
\ell(\bbeta,\tr,\tR,\tau_0^*) &=& \ell_0(\bbeta,r^*,R^*,\tau_0^*) - \ell_0(\bbeta_0,r^*,R^*,\tau_0^*) + \ell(\bbeta_0,\tr,\tR,\tau_0^*) 
\\\nonumber 
&& + O_p\left( n^{-1/2}\|\balpha-\balpha_0\| \right)  + o_p(\|\balpha-\balpha_0\|^2).
\end{eqnarray*}
By letting $\tau_0^*\to \tau_0$, we can derive
\begin{eqnarray}
\label{diff}
\ell(\bbeta,\tr,\tR) &=& \ell_0(\bbeta,r^*,R^*) - \ell_0(\bbeta_0,r^*,R^*) + \ell(\bbeta_0,\tr,\tR) 
\\\nonumber 
&& + O_p\left( n^{-1/2}\|\balpha-\balpha_0\| \right)  + o_p(\|\balpha-\balpha_0\|^2).
\end{eqnarray}
By Theorem 1 in \cite{sherman1993limiting}, $\halpha$ is $\sqrt{n}$--consistent for $\balpha_0$.


We now establish the asymptotic normality of $\sqrt{n}(\halpha-\balpha_0)$. By Lemma \ref{lemma-gradient}, we can derive
$\partial \ell (\bbeta,\tr, \tR)/\partial\balpha|_{\balpha=\balpha_0} =n^{-1/2} W_n^\intercal J(\balpha_0)+o_p(n^{-1/2})$,
where $W_n$ is defined in Lemma \ref{lemma-gradient}.
Moreover, applying Lemmas \ref{lemmatr} and \ref{lemma_deriv1} can yield
\begin{eqnarray*}
\frac{\partial^2 \ell(\balpha,\tr,\tR)}{ \partial \balpha \partial \balpha^\intercal}\Big|_{\balpha=\balpha_0} = J(\balpha_0)^\intercal V J(\balpha_0) +o_p(1),
\end{eqnarray*}
where $V$ is defined by \eqref{matrix-V}.
Uniformly over $O_p(n^{-1/2})$ neighborhoods of $\balpha_0$, applying Taylor's expansion yields
\begin{eqnarray*}
\ell(\bbeta,\tr,\tR) - \ell(\bbeta_0,\tr,\tR) &=& \frac{1}{2} (\balpha-\balpha_0)^\intercal J(\balpha_0)^\intercal V J(\balpha_0) (\balpha-\balpha_0) \\
&& + \frac{1}{\sqrt{n}} (\balpha-\balpha_0)^\intercal J(\balpha)^\intercal W_n +o_p(n^{-1}).
\end{eqnarray*}
By the Central Limit Theorem, $W_n$ converges in distribution to a zero mean normal distribution with the variance-covariance matrix
$
\Sigma= \operatorname{E}(\psi \psi ^\intercal)
$ as $n\to \infty$,
where
\begin{eqnarray*}
\psi = \int_{\tau_0}^{\tau_1}  \frac{\dot r^*(t,\bbeta_0^\intercal \bZ)}{r^*(t,\bbeta_0^\intercal\bZ)}  \left\{ \bZ - \frac{q^{(1)*}(t,\bbeta_0^\intercal\bZ)}{q^{(0)*}(t,\bbeta_0^\intercal\bZ)} \right\} d {M^*}(t).
\end{eqnarray*}
Following Theorem 2 in \cite{sherman1993limiting},  $\sqrt{n}(\halpha-\balpha_0)$ converges  in distribution to a zero mean normal random variable with the variance-covariance matrix $$
(J(\balpha_0)^\intercal VJ(\balpha_0))^{-} J(\balpha_0)^\intercal \Sigma J(\balpha_0) (J(\balpha_0)^\intercal VJ(\balpha_0))^{-}$$ as $n \to \infty$, where $^-$ denotes the Penrose-Moore inverse of a matrix. 
Under the transformation $\bbeta=\mathbb{S}(\balpha)$ defined by \eqref{transformation}, we can rewrite $\balpha$ as $\balpha=\mathbb{S}^*(\bbeta)$. 
Define $J(\bbeta)=\partial \mathbb{S}^*(\bbeta)/ \partial \balpha$ and it can be shown that $J(\balpha)J(\bbeta)=\mathrm{I}$. Applying the property of generalized inverse leads to $$(J(\balpha_0)^\intercal VJ(\balpha_0))^{-} J(\balpha_0)^\intercal \Sigma J(\balpha_0) (J(\balpha_0)^\intercal VJ(\balpha_0))^{-} = J(\bbeta_0) V^{-} \Sigma V^{-} J(\bbeta_0)^\intercal.$$ Note that $\hbeta=\mathbb{S}(\halpha)$ and $\bbeta_0=\mathbb{S}(\balpha_0)$. Applying the delta method, we can prove that $\sqrt{n}(\hbeta-\bbeta_0)$ converges in distribution to a zero mean normal random variable with the variance-covariance matrix $V^{-} \Sigma V^{-}$ as $n \to \infty$.


\section{Proof of Theorem 2}

Similar to the proof of Theorem 1, we can show that $\tbeta$ is $\sqrt{n}-$consistent for $\bbeta_0$. To establish the asymptotic normality of $\tbeta$, we calculate $\partial \ell'(\bbeta,\tr)/\partial \balpha$ and $\partial^2 \ell'(\bbeta,\tr)/\partial \balpha\partial \balpha^\intercal$, where
$$
\ell'(\bbeta,\tr) 
= \frac{1}{n} \sum_{i=1}^{n} \int_{\tau_0}^{\tau_1} \left\{ \log \widetilde{r}_h( t,\bbeta^\intercal \bZ_i,\bbeta)  \right\}I(\bZ_i \in \SZ)dN_i(t)
$$
is the objective function.
By Lemma \ref{lemma_deriv1}, we have 
$$
\frac{\partial \ell'(\bbeta,r^*)}{\partial \balpha}\Big|_{\balpha=\balpha_0} =  \frac{1}{n} \sumi \int_{\tau_0}^{\tau_1} \frac{\dot r^*(t,\bbeta_0^\intercal \bZ_i)}{r^*(t,\bbeta_0^\intercal\bZ_i)}  \left\{ \bZ_i - \frac{q^{(1)*}(t,\bbeta_0^\intercal\bZ_i)}{q^{(0)*}(t,\bbeta_0^\intercal\bZ_i)} \right\}^\intercal I( \bZ_i\in\SZ) dN_i(t) J(\balpha_0),
$$
where $\ell'(\bbeta,r^*)$ is defined by replacing $\tr$ with $r^*$ in $\ell'(\bbeta,\tr)$.
In the proof of Lemma \ref{lemma-gradient}, we have derived
\begin{eqnarray*}
&&\frac{\partial \ell'(\bbeta,\tr)}{\partial \balpha}\Big|_{\balpha=\balpha_0}-\frac{\partial \ell'(\bbeta,r^*)}{\partial \balpha}\Big|_{\balpha=\balpha_0} \\\nonumber
&&= -\frac{1}{n} \sumi \int_{\tau_0}^{\tau_1} \frac{\dot r^*(t,\bbeta_0^\intercal \bZ_i)}{r^*(t,\bbeta_0^\intercal\bZ_i)}  \left\{ \bZ_i - \frac{q^{(1)*}(t,\bbeta_0^\intercal\bZ_i)}{q^{(0)*}(t,\bbeta_0^\intercal\bZ_i)} \right\}^\intercal I(C_i\ge t, \bZ_i\in\SZ) N_i(t)r^*(t,\bbeta_0^\intercal \bZ_i)dt J(\balpha_0)\\
&& \;\;+ o_p(n^{-1/2}).
\end{eqnarray*}
Therefore, we have
$$
\frac{\partial \ell'(\bbeta,\tr)}{\partial \balpha}\Big|_{\balpha=\balpha_0} = \frac{1}{\sqrt{n}} W_n^\intercal J(\balpha_0)
+ o_p(n^{-1/2}),
$$
where $W_n$ is defined in Lemma \ref{lemma-gradient}.
Moreover, applying Lemmas \ref{lemmatr} and \ref{lemma_deriv1} yields 
$$
\frac{\partial^2 \ell'(\bbeta,\tr)}{\partial \balpha \partial \balpha^\intercal}\Big|_{\balpha=\balpha_0} = J(\balpha_0)^\intercal VJ(\balpha_0)+o_p(1),
$$
where $V$ is defined by \eqref{matrix-V}. Applying similar techniques in establishing the asymptotic normality of $\hbeta$, we can show that $\sqrt{n}(\widetilde{\bbeta}-\bbeta_0)$ converges in distribution to a zero-mean normal random vector with the variance-covariance matrix $V^{-}\Sigma V^{-}$ as $n\to \infty$, where $\Sigma$ is defined in the proof of Theorem 1.

\section{Proof of Equation (10).}
Under the proposed model, we have
\begin{align*}
& \operatorname{E}\left\{ \frac{N(C)}{F(C,\bbeta_0^\intercal \bZ)}\Big|\bZ  \right\}  \\
= & \operatorname{E}\left\{  \frac{\operatorname{E}\{N(C)\mid \bZ,C \}}{F(C,\bbeta_0^\intercal \bZ)}\Big|\bZ  \right\}\\
= & \operatorname{E}\left\{  \frac{\int_0^\tau I(C\ge t) \operatorname{E}\{d\widetilde N(t)\mid \bZ,C \}}{F(C,\bbeta_0^\intercal \bZ)}\Big|\bZ  \right\}\\
= & \operatorname{E}\left\{  \frac{\int_0^\tau I(C\ge t)f(t,\bbeta_0^\intercal \bZ) g(\bgamma_0^\intercal \bZ)dt}{F(C,\bbeta_0^\intercal \bZ)}\Big|\bZ  \right\}\\
= & g(\bgamma_0^\intercal \bZ).
\end{align*}
Then we complete the proof of Equation (10).


\section{Proof of Proposition 2}

To prove Proposition 2, we show that there exist constants $c_0$ and $\mu>0$ such that
\begin{align}
\label{eq1}
\operatorname{E} \{\widetilde\bZ  \mN - \widetilde\bZ  \exp(\mu\bgamma_0^\intercal \bZ+c_0)\} = \bm 0,
\end{align}
where $\widetilde\bZ =(1,\bZ^\intercal)^\intercal$ and $E(\mN\mid \bZ )= g(\bgamma_0^\intercal\bZ)$.
Without loss of generality, we assume $E(\bZ)=\bm 0$. Define $W = \bgamma_0^\intercal \bZ$ and thus we have $\operatorname{E}(W)=0$. 
Using the property of elliptically symmetric random variables, we have $\operatorname{E} (\bZ \mid W) = W\bm b_{\bm Z}$, where $\bm b_{\bm Z}$ is a $p\times 1$ deterministic vector such that $\bm b_{\bm Z}=\Sigma_{\bZ} \bgamma_0(\bgamma_0^\intercal\Sigma_{\bZ} \bgamma_0)^{-1}$ with $\Sigma_{\bZ}=\text{var}(\bZ)$. By some calculations, we obtain the following equations:
$$
\operatorname{E} \{\bZ\exp(c_0+\mu\bgamma_0^\intercal \bZ)\} = \operatorname{E}[\operatorname{E} \{\bZ\exp(c_0+\mu W) \mid W\}] = \operatorname{E} \{\exp(c_0+\mu W)W\}\bm b_{\bm Z},
$$
$$
\operatorname{E} (\bZ \mN) = \operatorname{E} \{ \operatorname{E} (\bZ \mN\mid W)\} = \operatorname{E} \{W g(W)\}\bm b_{\bm Z}.
$$
This leads to
\begin{align*}
\operatorname{E} \{\bZ \mN - \bZ\exp(\mu\bgamma_0^\intercal \bZ+c_0)\} = \operatorname{E} \{ g(W) W- \exp(c_0+\mu W)W\}\bm b_{\bm Z}.
\end{align*}
Similarly, we have $\operatorname{E} \{ \mN -  \exp(\mu\bgamma_0^\intercal \bZ+c_0)\} = \operatorname{E} \{ g(W)  - \exp(c_0+\mu W) \} $. Let $(c_0,\mu)$ be the solution of $\operatorname{E} [(1,W)\{g(W) - \exp(c_0+\mu W)\}] = \bm 0$. If we can show $\mu>0$, the solution $(c_0,\mu)$ satisfies Equation \eqref{eq1}, and thus the proposition can be proved.

In what follows, we show that $\mu>0$. Since $(c_0,\mu)$ is the solution of $\operatorname{E} [(1,W)\{g(W) - \exp(c_0+\mu W)\}] = \bm 0$,  we have
\begin{eqnarray}
\label{eq2}
\operatorname{E} \{W g(W)\}=\exp(c_0)\operatorname{E} \{W\exp(\mu W)\}.   
\end{eqnarray}
If $\mu=0$, we have 
$\operatorname{E}  \{W g(W)\}=\exp(c_0)\operatorname{E} (W)=0$ by \eqref{eq2}. This contradicts the fact that $\operatorname{E}  \{Wg(W)\} > 0$ when $W$ is symmetric and $g(W)$ is positive and monotonically increasing.
If $\mu<0$, it can be shown that $\operatorname{E} \{W\exp(\mu W)\} < 0$ since $W$ is symmetric and $\exp(\mu W)$ is monotonically decreasing. This contradicts the fact that $\operatorname{E} \{W\exp(\mu W)\}= \operatorname{E} \{W g(W)\}/\exp(c_0)>0$ due to  $\operatorname{E}  \{Wg(W)\} > 0$ and \eqref{eq2}. This completes the proof of $\mu>0$.

\section{Lemmas for Theorems 3 and 4} 

Define $\mN_i=N_i(C_i)/F(C_i,\bbeta_0^\intercal\bZ_i)$ and $\hN_i(\hbeta)=N_i(C_i)/\hF_{\hshape} (C_i,\hbeta^\intercal\bZ_i,\hbeta)$, where
\begin{align*}
\hF_{\hshape} (t,x,\bbeta) = \exp\left\{ - \int_{t}^{\tau}   \frac{\sum_{i=1}^n K_{\hshape}(x-\bbeta^\intercal \bZ_i)dN_i(u)}{\sum_{i=1}^n K_{\hshape}(x-\bbeta^\intercal \bZ_i) N_i(u) I(C_i\ge u)} \right\},
\end{align*}
with $K_{\hshape}(\cdot)$ satisfying condition (C6). 
Denote by $O=\{(\bZ, C, N(t)), 0\le t \le C\}$ the observed data.  Define 
\begin{eqnarray}
\label{function-d}
d(O,\bbeta_0) = \int_C^\tau  \left \{\bZ-  \frac{q^{(1)}(t,\bbeta_0^\intercal \bZ)}{q^{(0)}(t,\bbeta_0^\intercal \bZ)} \right\}  \dot{r}(t,\bbeta_0^\intercal\bZ) dt,
\end{eqnarray} 
where $\dot{r}(t,x)$ denotes the partial derivative of $r(t,x)$ with respect to $x$ and $q^{(k)}(t,x) = \operatorname{E} \{\bZ^{\otimes k} I(C\ge t) g(\bgamma_0^\intercal \bZ)\mid\bbeta_0^\intercal\bZ = x\}$ for $k = 0,1$. For $l=1,\ldots,n$, define $\rho(O_l,t,x) =  \int_t^\tau \{f_2(u,x)\}^{-1} \{d N_l(u)-I(C_l\ge u)N_l(u) r(u,x) du\}$, with $f_2(t,x) = \operatorname{E}\{I(C\ge t)N(t) \mid \bbeta_0^\intercal \bZ=x\}f_{\bbeta_0^\intercal\bZ}(x)$. 
 
\begin{lemma}
\label{lemma-N}
Let $m(\bZ) =\widetilde{\bZ}I(\bZ\in\SZ)$ with $\widetilde{\bZ}$ defined in Theorem 3, or $m(\bZ) =\delta(\bZ)$ with $\delta(\bZ)$ defined in Theorem 4, then we have
\begin{eqnarray*}
\frac{1}{n} \sumi m(\bZ_i) \{\hN_i(\hbeta)-\mN_i\} 
&=& \frac{1}{n^2} \sumi  \sum_{l=1}^n m(\bZ_i) \mN_i K_{\hshape}(\bbeta_0^\intercal\bZ_l-\bbeta_0^\intercal\bZ_i)  \rho(O_l,C_i,\bbeta_0^\intercal\bZ_i) \\
&&- \operatorname{E} \{m(\bZ)\mN d(O,\bbeta_0)^\intercal\} V^- \psi_i + o_p(n^{-1/2}).
\end{eqnarray*}
where $V$ is defined by \eqref{matrix-V} and
\begin{eqnarray}
   \label{psii} 
\psi_i = \int_{\tau_0}^{\tau_1}  \frac{\dot r^*(t,\bbeta_0^\intercal \bZ_i)}{r^*(t,\bbeta_0^\intercal\bZ_i)} \left\{ \bZ_i - \frac{q^{(1)*}(t,\bbeta_0^\intercal\bZ_i)}{q^{(0)*}(t,\bbeta_0^\intercal\bZ_i)} \right\} d M^*_i(t).
\end{eqnarray}

\end{lemma}

{\it Proof of Lemma \ref{lemma-N}.}
The functions $\mF_1(t,x,\bbeta)dt= \operatorname{E}\{N(t) \mid \bbeta^\intercal \bZ=x\}f_{\bbeta^\intercal\bZ}(x)$ and $f_2(t,x,\bbeta)= \operatorname{E}\{I(C\ge t)N(t) \mid \bbeta^\intercal \bZ=x\}f_{\bbeta^\intercal\bZ}(x)$ 
can be estimated by the kernel type estimators, $\widehat {\mF}_{1\hshape}(t,x,\bbeta)dt=n^{-1} \sum_{i=1}^n K_{\hshape}(x-\bbeta^\intercal \bZ_i) N_i(u)$ and $\widehat f_{2\hshape}(t,x,\bbeta)=n^{-1}\sum_{i=1}^n K_{\hshape}(x-\bbeta^\intercal \bZ_i) N_i(u) I(C_i\ge u)$, respectively. The kernel estimator for the shape function can  be expressed as 
$$
\hF_{\hshape} (t,x,\bbeta) = \exp\left[ -\int_t^\tau  \{\widehat f_{2\hshape}(u,x,\bbeta)\}^{-1} \widehat {\mF}_{1\hshape}(du,x,\bbeta) \right].
$$
Define the functions $\mF_{1\hshape}(t,x,\bbeta) = \operatorname{E}\{K_{\hshape}(x-\bbeta^\intercal\bZ)N(t)\}$, $f_{2\hshape}(t,x,\bbeta)= \operatorname{E}\{K_{\hshape}(x-\bbeta^\intercal\bZ)I(C\ge t)N(t)\}$, and $F_{\hshape}(t,x,\bbeta)=\exp[-\int_t^\tau \{f_{2\hshape}(u,x,\bbeta)\}^{-1} \mF_{1\hshape}(du,x,\bbeta)]$. 
It is easy to see that 
$$
\frac{1}{n} \sumi m(\bZ_i) \{\hN_i(\hbeta) - \mN_i\} = \frac{1}{n} \sumi  m(\bZ_i)(I_{1i}+I_{2i}+I_{3i}),
$$
where 
\begin{eqnarray*}
I_{1i} &=& \frac{N_i(C_i)}{\hF_{\hshape} (C_i,\bbeta_0^\intercal\bZ_i,\bbeta_0)} - \frac{N_i(C_i)}{F_{\hshape} (C_i,\bbeta_0^\intercal\bZ_i,\bbeta_0)},\\
I_{2i} &=& \frac{N_i(C_i)}{\hF_{\hshape} (C_i,\hbeta^\intercal\bZ_i,\hbeta)} - \frac{N_i(C_i)}{\hF_{\hshape} (C_i,\bbeta_0^\intercal\bZ_i,\bbeta_0)},\\
I_{3i} &=& \frac{N_i(C_i)}{F_{\hshape} (C_i,\bbeta_0^\intercal\bZ_i,\bbeta_0)} - \frac{N_i(C_i)}{F(C_i,\bbeta_0^\intercal\bZ_i,\bbeta_0)}.
\end{eqnarray*}
For $l=1,\ldots,n$, define
\begin{eqnarray}
\label{etai}
\rho(O_l,t,x,\bbeta) =  \int_t^\tau \{f_2(u,x,\bbeta)\}^{-1} dM_l(u,x,\bbeta),
\end{eqnarray}
with $dM_l(u,x,\bbeta)=d N_l(u)-I(C_l\ge u)N_l(u) r(u,x,\bbeta) du$.
Following \cite{sun2022statistical}, applying the functional delta method yields 
\begin{eqnarray*}
\frac{1}{n}\sumi m(\bZ_i) I_{1i} &=&\frac{1}{n^2}\sumi \sum_{l=1}^n m(\bZ_i) \mN_i K_{\hshape}(\bbeta_0^\intercal\bZ_l-\bbeta_0^\intercal\bZ_i)  \rho(O_l,C_i,\bbeta_0^\intercal\bZ_i)+R_n + O_p(\hshape^2),
\\
\frac{1}{n} \sumi m(\bZ_i) I_{2i} &=&\frac{1}{n} \sumi  m(\bZ_i) \mN_i d(O_i,\bbeta_0) ^\intercal (\hbeta-\bbeta_0) + o_p(n^{-1/2}),
\end{eqnarray*}
where $\rho(O_l,C_i,\bbeta_0^\intercal\bZ_i)=\rho(O_l,C_i,\bbeta_0^\intercal\bZ_i,\bbeta_0)$ with $\rho(O_l,t,x,\bbeta)$ defined by \eqref{etai}.
By condition (C6), we have  $O_p(\hshape^2)=o_p(n^{-1/2})$.
Moreover, following \cite{sun2022statistical}, we can show that $n^{-1} \sumi m(\bZ_i) I_{3i} +R_n = o_p(n^{-1/2}).$
Therefore, applying the results in Theorem 1 yields
\begin{eqnarray*}
\frac{1}{n} \sumi m(\bZ_i) \{\hN_i(\hbeta)-\mN_i\} &=& \frac{1}{n^2} \sumi \sum_{l=1}^n m(\bZ_i)  \mN_i  K_{\hshape}(\bbeta_0^\intercal\bZ_l-\bbeta_0^\intercal\bZ_i)  \rho(O_l,C_i,\bbeta_0^\intercal\bZ_i)\\
&&+ \frac{1}{n} \sumi m(\bZ_i) \mN_i d(O_i,\bbeta_0)^\intercal (\hbeta-\bbeta_0) + o_p(n^{-1/2})\\
&=& \frac{1}{n^2} \sumi  \sum_{l=1}^n m(\bZ_i) \mN_i K_{\hshape}(\bbeta_0^\intercal\bZ_l-\bbeta_0^\intercal\bZ_i)  \rho(O_l,C_i,\bbeta_0^\intercal\bZ_i) \\
&&- \operatorname{E} \{m(\bZ)\mN d(O,\bbeta_0)^\intercal\} V^- \psi_i + o_p(n^{-1/2}).
\end{eqnarray*}
This completes the proof of Lemma \ref{lemma-N}.



\begin{lemma}
\label{lemma-U}
Define $
u_h(O_i,O_l)= \{m(\bZ_i)\mN_i K_h(\bbeta_0^\intercal\bZ_i-\bbeta_0^\intercal\bZ_l) \rho(O_i,C_l,\bbeta_0^\intercal\bZ_l) + m(\bZ_l)\mN_l K_h(\bbeta_0^\intercal\bZ_l-\bbeta_0^\intercal\bZ_i) \rho(O_l, C_i,\bbeta_0^\intercal\bZ_i)\}/2,
$
where $m(\bZ) =\widetilde{\bZ}I(\bZ\in\SZ)$ with $\widetilde{\bZ}$ defined in Theorem 3, or $m(\bZ) =\delta(\bZ)$ with $\delta(\bZ)$ defined in Theorem 4. We have
\begin{eqnarray*}
&& \frac{1}{n(n-1)}\sum_{l\neq i}  u_h(O_i,O_l)- \operatorname{E}\{u_h(O_1,O_2)\} \\ &=&
\frac{1}{n}\sum_{i=1}^n 
\widetilde \rho(O_i)f_{\bbeta_0^\intercal \bZ}(\bbeta_0^\intercal \bZ_i) -2 \operatorname{E}\{u_h(O_1,O_2)\} + o_p(n^{-1/2}), 
\end{eqnarray*}
where $\widetilde\rho(o) = \operatorname{E}\left\{ m(\bZ_1)\mN_1 \rho(O_1, C_2, \bbeta_0^\intercal \bZ_2,\bbeta_0) \mid \bbeta_0^\intercal \bZ_2=\bbeta_0^\intercal \bz, O_1 = o \right\}$ with $\bz$ denoting the covariate component of $o$.

\end{lemma}

{\it Proof of Lemma \ref{lemma-U}.}
Under  conditions (C2) and (C4), we can show $\operatorname{E}\{\|u_h(O_1,O_2)\|^2\}=o(n)$.  
Since 
$
u_h(O_i,O_l)
$
is a symmetric function, applying Lemma 3.1 in \cite{powell1989semiparametric} yields
\begin{eqnarray*}
\frac{1}{n(n-1)}\sum_{ i\neq l} u_h(O_i,O_l)- \operatorname{E}\{u_h(O_1,O_2)\} =\frac{2}{n}\sum_{i=1}^n \tu_h(O_i) -2 \operatorname{E}\{u_h(O_1,O_2)\} + o_p(n^{-1/2}),   
\end{eqnarray*}
with $\tu_h(O_i)=\operatorname{E}\{u_h(O_i,O_l) \mid O_i\}$ for $i,l=1,\ldots,n$ and $i\neq l$. 
Moreover, we have
\begin{eqnarray*}
\frac{1}{n}\sum_{i=1}^n \tu_h(O_i)= \frac{1}{n}\sum_{i=1}^n \operatorname{E}\{u_h(O_i,O_l) \mid O_i\} = \frac{1}{2n}\sum_{i=1}^n \widetilde\rho(O_i) f_{\bbeta_0^\intercal \bZ}(\bbeta_0^\intercal \bZ_i) + o_p(n^{-1/2}).
\end{eqnarray*}
where $\widetilde\rho(o) = \operatorname{E}\left\{ m(\bZ_1)\mN_1 \rho(O_1, C_2, \bbeta_0^\intercal \bZ_2,\bbeta_0) \mid \bbeta_0^\intercal \bZ_2=\bbeta_0^\intercal \bz, O_1 = o \right\}$.
It follows that
\begin{eqnarray*}
&& \frac{1}{n(n-1)}\sum_{ i\neq l} u_h(O_i,O_l)- \operatorname{E}\{u_h(O_1,O_2)\} \\ &=&
\frac{1}{n}\sum_{i=1}^n \widetilde\rho(O_i) f_{\bbeta_0^\intercal \bZ}(\bbeta_0^\intercal \bZ_i) -2 \operatorname{E}\{u_h(O_1,O_2)\} + o_p(n^{-1/2}).
\end{eqnarray*}

To prove Theorem 4, the constraint on $\bgamma$ can be achieved by reparameterizing $\bgamma$ in the polyspherical coordinate system as a $(p-1)$-dimensional vector $\btheta = (\theta_1,\theta_2,\ldots,\theta_{p-1})^\intercal\in \Theta$ with $\Theta=[0,\pi]^{(p-2)}\times[0,2\pi]$. Specifically, we set 
\begin{eqnarray*}
\bgamma = \mathbb{S}(\btheta) = 
 \left(\cos(\theta_1), \sin(\theta_1)\cos(\theta_2),\ldots, \prod_{k=1}^{p-2}\sin(\theta_k)\cos(\theta_{p-1}),  \prod_{k=1}^{p-1}\sin(\theta_k)\right)^\intercal.
\end{eqnarray*}
Define the $p\times(p-1)$ matrix $J(\btheta)=\partial \mathbb{S}(\btheta)/\partial\btheta$.
When establishing the asymptotic normality of $\gammamre$, the following lemma is used to derive the first  and second order derivatives of the objective functions with respect to $\btheta$.

\begin{lemma}
\label{lemma_deriv2}
Under the conditions in Theorem 4, given $\operatorname{E}(\mN \mid \bZ) = {g}(\bgamma_0^\intercal  \bZ)$, we have

\begin{align*}
& \frac{\partial^2 \operatorname{E}\{(\mN_1 - \mN_2) I(\bZ_1\in\SZ,\bZ_2\in \SZ, \bgamma^\intercal \bZ_1 < \bgamma^\intercal \bZ_2  ) \}}{\partial\btheta\partial \btheta^\intercal} \bigg|_{\btheta = \btheta_0} \\
& =  2J(\btheta_0)^\intercal \operatorname{E} \left[ \left\{\phi^{(2)}(\bgamma_0^\intercal  \bZ)\phi^{(0)}(\bgamma_0^\intercal  \bZ) - \phi^{(1)}(\bgamma_0^\intercal  \bZ)^{\otimes 2} \right\} \dot{g}(\bgamma_0^\intercal  \bZ) f_{\bgamma_0^\intercal \bZ}(\bgamma_0^\intercal \bZ)  \right]J(\btheta_0),
\\\\
& \frac{\partial \operatorname{E}\{(y - \mN) I(\bZ\in\SZ,\bz\in \SZ, \bgamma^\intercal \bZ < \bgamma^\intercal \bz  ) \}}{\partial\btheta} \bigg|_{\btheta = \btheta_0} = \{ y - {g}(\bgamma_0^\intercal  \bz)  \} 
 \delta(\bz)^\intercal J(\btheta_0),
\end{align*}  
where $\delta(\bZ) =  f_{\bgamma_0^\intercal \bZ}(\bgamma_0^\intercal \bZ)\{\bZ \phi^{(0)}( \bgamma_0^\intercal \bZ  )- \phi^{(1)}(\bgamma_0^\intercal \bZ ) \}I(\bZ \in\SZ)$ and $\phi^{(k)}(\bgamma_0^\intercal\bZ) = E\{\bZ^{\otimes k}I(\bZ\in\SZ)\mid \bgamma_0^\intercal\bZ\}$ for $k=0,1,2$.
\end{lemma} 

Similar arguments to those used in the proof of Lemma \ref{lemma_deriv1} can be applied to prove Lemma \ref{lemma_deriv2} \cite[see, for example,][]{sherman1993limiting}, and thus the proof is omitted.

\section{Proof of Theorem 3}

The size parameter is estimated by solving the estimating equation $U(c,\bgamma,\hbeta)=\bm 0$, where
\begin{eqnarray}
\label{Ugamma}
U(c,\bgamma,\hbeta)= \frac{1}{n} \sumi \widetilde \bZ_i I(\bZ_i\in \SZ) \{ \hN_i(\hbeta) - \exp(c+ \bgamma^\intercal \bZ_i) \}
\end{eqnarray}
with $\widetilde \bZ_i=(1,\bZ_i^\intercal)^\intercal$. The solutions of $U(c,\bgamma,\hbeta)=\bm 0$ are denoted by $\hc$ and $\gammaexp$. Define $U_0(c,\bgamma) = \operatorname{E} [\widetilde{\bZ}I(\bZ\in\SZ) \{ \mathcal N - \exp(c+\bgamma^\intercal\bZ) \}]$ and let  $(c_0,\bgamma_0)$ be the solutions of $U_0(c,\bgamma)=\bm 0$. Let $\widehat{\bxi}=(\widehat c, \gammaexp^\intercal)^\intercal$ and $\bxi_0=(c_0,\bgamma_0^\intercal)^\intercal$. We first prove that $\widehat{\bxi}$ is a consistent estimator of $\bxi_0$. For convenience, rewrite $U(c,\bgamma,\hbeta)$ and $U_0(c,\bgamma)$ as $U(\bxi,\hbeta)$ and $U_0(\bxi)$, respectively. For $\epsilon>0$, it can be derived $\inf_{\bxi:\|\bxi-\bxi_0\| \geq \epsilon } \|U_0(\bxi)\|>0$ due to
\begin{eqnarray}
\label{matrix-Gamma1}
\frac{\partial U_0(\bxi)}{\partial \bxi}\Big|_{\bxi=\bxi_0} = \Gamma_1 = -\operatorname{E} \{\exp(c_0+ \bgamma_0^\intercal \bZ) \widetilde \bZ \widetilde \bZ^\intercal I(\bZ\in \SZ)\}.
\end{eqnarray}
Moreover, we need to show that
$\sup_{\bxi\in\Xi}\|U(\bxi,\hbeta)-U_0(\bxi)\|$ converges in probability to zero as $n\to \infty$, where $\Xi$ is a compact subset of $\mathbb{R}^{p+1}$. Let $U(\bxi,\mN)= n^{-1}\sumi \widetilde \bZ_i I(\bZ_i\in \SZ) \{ \mN_i - \exp(c+ \bgamma^\intercal \bZ_i) \}$. It is easy to see that $\sup_{\bxi\in\Xi}\|U(\bxi,\mN) -U_0(\bxi)\|=o_p(1)$. Moreover, applying Lemma \ref{lemma-N} leads to
$\sup_{\bxi\in\Xi} \|U(\bxi,\hbeta)-U(\bxi,\mN)\|=o_p(1)$.
Therefore, we have
$
\sup_{\bxi\in\Xi}\|U(\bxi,\hbeta)-U_0(\bxi)\| \leq \sup_{\bxi\in\Xi}\|U(\bxi,\mN)-U_0(\bxi)\|+ \sup_{\bxi\in\Xi}\|U(\bxi,\hbeta)-U(\bxi,\mN) \|=o_p(1)$.
The consistency of $\widehat{\bxi}=(\widehat c,\gammaexp^\intercal)^\intercal$ is proved.

We then prove the asymptotic normality of $\widehat{\bxi}=(\widehat c,\gammaexp^\intercal)^\intercal$. By Taylor's expansion of $U(\hc,\gammaexp,\hbeta)$ at the true values $(c_0,\bgamma_0)$, we have  
\begin{eqnarray*}
\bm 0 &=& 
U(c_0,\bgamma_0,\hbeta) -  \left\{ \frac{1}{n} \sumi \exp(c_0+ \bgamma_0^\intercal \bZ_i) \widetilde \bZ_i \widetilde \bZ_i^\intercal I(\bZ_i\in \SZ) \right \} \begin{pmatrix}
\hc-c_0\\
\gammaexp-\bgamma_0
\end{pmatrix} + \bm R_U,
\end{eqnarray*}
where the residual term $\bm R_U$ satisfies $\|\bm R_U\|=O_p(\|\widehat{\bxi}-\bxi_0\|^2)$.
Since $\widehat c$ and $\gammaexp$ are consistent, we have
\begin{eqnarray*}
\sqrt{n}\begin{pmatrix}
\hc-c_0\\
\gammaexp-\bgamma_0
\end{pmatrix} &=& \left\{ \frac{1}{n} \sumi \exp(c_0+ \bgamma_0^\intercal \bZ_i) \widetilde \bZ_i \widetilde \bZ_i^\intercal I(\bZ_i\in \SZ) \right \}^{-1} \sqrt{n} U(c_0,\bgamma_0,\hbeta) + o_p(1)\\
&=& -\Gamma_1^{-1} \sqrt{n} U(c_0,\bgamma_0,\hbeta) + o_p(1),
\end{eqnarray*}
where $\Gamma_1$ is defined by \eqref{matrix-Gamma1}. 
To establish the asymptotic normality for the proposed estimators $\hc$ and $\gammaexp$, we need to study $\sqrt{n} U(c_0,\bgamma_0,\hbeta)$. Applying Lemmas \ref{lemma-N} and \ref{lemma-U} yields
\begin{eqnarray}
\label{iid}
\sqrt{n} U(c_0,\bgamma_0,\hbeta) &=& \frac{1}{\sqrt{n}} \sumi  U_i(c_0,\bgamma_0) + o_p(1),
\end{eqnarray}
where
\begin{eqnarray*}
U_i(c_0,\bgamma_0)  &=& \widetilde \bZ_i I(\bZ_i\in \SZ) \left\{\mN_i - \exp(c_0+ \bgamma_0^\intercal \bZ_i) \right\} \\
&& + \int_0^\tau \frac{\operatorname{E} \{ \widetilde{\bZ}\mN I(\bZ \in\SZ, C\le t) \mid \bbeta_0^\intercal \bZ=\bbeta_0^\intercal \bZ_i  \}}{{q^{(0)}(t,\bbeta_0^\intercal \bZ_i)F(t, \bbeta_0^\intercal \bZ_i) }} {dM_i(t)}f_{\bbeta_0^\intercal \bZ}(\beta_0^\intercal \bZ_i) \\
&& - \operatorname{E}\left[\widetilde \bZ I(\bZ\in \SZ) \mN \int_C^\tau  \dot{r}(t,\bbeta_0^\intercal\bZ) \left \{\bZ-  \frac{q^{(1)}(t,\bbeta_0^\intercal \bZ)}{q^{(0)}(t,\bbeta_0^\intercal \bZ)} \right\}^\intercal  dt \right] V^- \psi_i
\end{eqnarray*}
with $V$ defined by \eqref{matrix-V} and $\psi_i$ defined in Lemma \ref{lemma-N}.
Define
\begin{align*}
    \psi_1 =& \widetilde{\bZ}I(\bZ \in\SZ)\left\{\mN - \exp(c_0 + \bgamma_0^\intercal\bZ)  \right\}\\
    & + \int_0^\tau{ \operatorname{E} \{ \widetilde{\bZ}\mN I(\bZ \in\SZ, C\le t) \mid \bbeta_0^\intercal \bZ  \}} \{{q^{(0)}(t,\bbeta_0^\intercal \bZ)F(t, \bbeta_0^\intercal \bZ) }\}^{-1} {dM(t)}f_{\bbeta_0^\intercal \bZ}(\beta_0^\intercal \bZ)\\ 
    & - \operatorname{E}[\widetilde{\bZ}I(\bZ \in\SZ){\mN}\int_C^\tau \dot r(t,\bbeta_0^\intercal\bZ){\{\bZ -  q^{(1)}(t,\bbeta_0^\intercal \bZ)/q^{(0)}(t,\bbeta_0^\intercal \bZ)\}^\intercal} ]dt V^{-} \psi,
\end{align*}
where $\psi$ is defined in Theorem 1.
It can be shown that
$
\text{var} \left\{ n^{-1/2} \sumi U_i(c_0,\bgamma_0) \right\} =  \Omega_1
$ with $\Omega_1 = \operatorname{E}(\psi_1\psi_1^\intercal)$.
Therefore, $\sqrt{n}((\widehat c,\gammaexp^\intercal)^\intercal-(c_0,\bgamma_0^\intercal)^\intercal)$ converges in distribution to a zero mean normal random variable with the variance-covariance matrix $\Gamma_1^{-1}\Omega_1 \Gamma_1^{-1}$ as $n\to \infty$.


\section{Proof of Theorem 4}


Define $\ell(\bgamma,\mN) = n^{-2} \sum_{i=1}^n \sum_{j=1}^n I(\bgamma^\intercal \bZ_i >  \bgamma^\intercal \bZ_j, \bZ_i \in\SZ, \bZ_j \in\SZ) \mN_i$ and rewrite the objective function as
$$
\ell(\bgamma,\widehat{\mN}) = \frac{1}{n^2} \sum_{i=1}^n \sum_{j=1}^n I(\bgamma^\intercal \bZ_i >  \bgamma^\intercal \bZ_j, \bZ_i \in\SZ, \bZ_j \in\SZ) \widehat{\mN}_i(\hbeta).
$$
Applying Lemma \ref{lemma_deriv2}, we have $\partial\ell(\bgamma,\mN) /\partial \btheta |_{\btheta=\btheta_0} = n^{-1} \sumi \{ \mN_i - g(\bgamma_0^\intercal \bZ_i) \} \delta(\bZ_i)^\intercal J(\btheta_0) +o_p(n^{-1/2}) $, with 
$\delta(\bZ) =  f_{\bgamma_0^\intercal \bZ}(\bgamma_0^\intercal \bZ)\{\bZ \phi^{(0)}( \bgamma_0^\intercal \bZ  )- \phi^{(1)}(\bgamma_0^\intercal \bZ ) \}I(\bZ \in\SZ)$.
Then by Lemma \ref{lemma-N}, we can show
$$
\frac{\partial\ell(\bgamma,\widehat {\mN}) }{\partial \btheta} \Big|_{\btheta=\btheta_0} = \frac{1}{n} \sumi \psi_{2i}^\intercal J(\btheta_0) + o_p(n^{-1/2}),
$$
where 
\begin{eqnarray*}
\psi_{2i} &=&   \delta(\bZ_i) \left\{ \mN_i - g(\bgamma_0^\intercal \bZ_i) \right\} \\&&- \operatorname{E} \left[ \int_C^\tau   \dot{r}(t,\bbeta_0^\intercal \bZ) dt \cdot \delta(\bZ)\mN {\{\bZ -  q^{(1)}(t,\bbeta_0^\intercal \bZ)/q^{(0)}(t,\bbeta_0^\intercal \bZ)\}^\intercal}\right]  {V}^-  \psi_i \\
&& +\int_0^\tau{ \operatorname{E} \left\{ \delta(\bZ)  \mN I(C\le t) \;\middle|\; \bbeta_0^\intercal \bZ = \bbeta_0^\intercal \bZ_i \right\}} \{{q^{(0)}(t,\bbeta_0^\intercal \bZ_i)F(t, \bbeta_0^\intercal \bZ_i) }\}^{-1} {dM_i(t)}f_{\bbeta_0^\intercal \bZ}(\bbeta_0^\intercal \bZ_i),
\end{eqnarray*}
with $V$ defined by \eqref{matrix-V}, and $\psi_i$ defined by \eqref{psii}.
Moreover, it follows from Lemmas \ref{lemma-N} and \ref{lemma_deriv2} that
$$
\frac{\partial^2\ell(\bgamma,\widehat{\mN}) }{\partial \btheta \partial \btheta^\intercal} \Big|_{\btheta=\btheta_0}  = J(\btheta_0)^\intercal\Gamma_2 J(\btheta_0) + o_p(1),
$$
with
$
\Gamma_2 = {-} \operatorname{E}[\{  \phi^{(2)}(\bgamma_0^\intercal \bZ)\phi^{(0)}(\bgamma_0^\intercal \bZ) - \phi^{(1)}(\bgamma_0^\intercal \bZ)^{\otimes 2} \} \dot{g}(\bgamma_0^\intercal  \bZ) f_{\bgamma_0^\intercal \bZ}(\bgamma_0^\intercal \bZ)  ].
$
Applying the technique in the proof of Theorem 2 in \cite{sun2022statistical}, we can prove that $\sqrt{n}(\gammamre-\bgamma_0)$ converges in distribution to a zero mean normal random variable with the variance-covariance matrix $\Gamma_2^-\Omega_2 \Gamma_2^-$ as $n \to \infty$, where $\Omega_2$ is defined in Theorem 4.

\end{document}